\documentstyle[12pt]{article}
\def\today{}
\renewcommand{\baselinestretch}{2}
\begin{document}
\title{Theory of $\tau$ mesonic decays}
\author{Bing An Li\\
Department of Physics and Astronomy, University of Kentucky\\
Lexington, KY 40506, USA}

\maketitle

\begin{abstract}
Studies of $\tau$ mesonic decays are presented.
A mechanism for axial-vector current at low energies
is proposed. The VMD is used to treat the vector current. All the
meson vertices of both normal parity and abnormal parity
(Wess-Zumino-Witten anomaly)
are obtained from an effective chiral theory of mesons.
$a_{1}$ dominance is found in
the decay modes of $\tau$ lepton: $3\pi$,
$f(1285)\pi$. Both the $\rho$ and the $a_{1}$ meson
contribute to the decay $\tau\rightarrow K^{*}K\nu$, it is found that
the vector current is dominant.
CVC is tested by studying $e^{+}e^{-}\rightarrow\pi^{+}\pi^{-}$.
The branching ratios of $\tau\rightarrow\omega\pi\nu$ and
$K\bar{K}\nu$ are calculated. In terms of similar mechanism
the \(\Delta s=1\) decay modes of $\tau$ lepton are studied
and $K_{a}$ dominance is found in $\tau\rightarrow K^{*}\pi\nu$
and $K^{*}\eta\nu$. The suppression of $\tau\rightarrow K\rho\nu$
is revealed.
The branching ratio of $\tau\rightarrow\eta K\nu$
is computed. As a test of this theory, the form factors of
$\pi\rightarrow e\gamma\nu$ and $K\rightarrow e\gamma\nu$ are
determined. Theoretical results agree with data reasonably well.
\end{abstract}

\newpage
\section{Introduction}
The $\tau$ mesonic
decays have been studied by Tsai[1] before the discovery of the
$\tau$ lepton.
All the hadrons in $\tau$ hadronic decays are mesons, therefore
the $\tau$ mesonic decays provide a test ground for all meson
theories. The mesons produced in $\tau$ hadronic decays are
made of the light quarks. Therefore, chiral symmtery play an
important role in studying $\tau$ mesonic decays. In Ref.[2]
the $\tau$ mesonic decays are associated with the chiral
dynamics. It is pointed out[2] that $\rho$-dominance is
necessary to be introduced and the chiral limits of the hadronic
matrix elements at low energies are set up. In Ref.[3]
the $\tau$ mesonic decays are studied by using
$SU(3)\times SU(3)$ chiral dynamics with the resonances
phenomenologically introduced. In Ref.[4] a Lagrangian for
psedoscalar and vector mesons has been constructed to investigate
$\tau$ physics. In Refs.[5] in studing three
pseudoscalar mesons decays of $\tau$ lepton the form factors of
these decays are constructed by chiral symmetry and dominated
by the lowest resonances. The abnormal decays have also been
studied[6,7,8,9,10]. The vector meson
dominance(VMD)[12]has been applyed to study the $\tau$ decays
in which the meson staes have even G-parity[2-10].

In Ref.[11] an effective Lagrangian of three nonets of
pseudoscalar, vector, and axial-vector mesons with $U(3)_{L}\times
U(3)_{R}$ is obtained. The chiral symmetry breaking scale $\Lambda$
determined in Ref.[11] is 1.6GeV, therefore, this theory is suitable
to be applyed to study $\tau$ mesonic decays.
The VMD is a natural result of this theory.
Wess-Zumino-Witten action[WZW][15]
is obtained from the leading terms of
the imaginary part of the effective Lagrangian.
This theory has been applied to study the form factors of $K_{l3}$,
$\tau\rightarrow\rho\nu$, $\tau\rightarrow K^{*}\nu$ and
theoretical results are in good agreements with data[11].
Based on this effective chiral theory of mesons[11],
a theory of $\tau$ mesonic decays is developed and a unified study
of $\tau$ mesonic decays is presented in this paper.

In the standard model the W bosons are coupled to
both the vector and
the axial-vector currents of the ordinary quarks.
The hadronization of the quark
currents is a problem of nonperturbative $QCD$.
In the dynamics of ordinary quarks both chiral symmetry and chiral
symmetry breaking are important.
The VMD and CVC are successful in studying
the matrix elements of vector current[16]. VMD is a natural result of
the effective chiral theory of mesons[11]. In this paper the VMD
is exploited to treat the matrix
elements of the vector currents of $\tau$ decays.
Before this paper the VMD
has been already exploited to study $\tau$ mesonic decays[2-10].
The difference between this paper and others in the case of two
flavors is that the coupling
of $\rho\pi\pi$, $f_{\rho\pi\pi}$, derived in Ref.[11] is no longer
a constant, but a function of $q^{2}$(q is the momentum of $\rho$
meson). It means that the vertex $\rho\pi\pi$ has a form factor.
Detailed discussion is presented in section 7.

It is well known that in the chiral limit,
the axial-vector currents and the vector currents
of ordinary quarks form a $SU(2)_{L}\times SU(2)_{R}$ algebra
which leads to Weinberg's sum rules[13]. On the other hand,
$a_{1}$ meson is the chiral partner of $\rho$ meson[14]. However,
$a_{1}$ meson is much heavier than $\rho$ meson. In Ref.[11] this
mass difference refers to the spontaneous chiral symmetry breaking
and this effect should be taken into account in bosonizing the
axial-vector currents. The effect of the spontaneous chiral
symmetry breaking on the bosonization of the axial-vector currents
at low energies is studied in this paper.
The axial-vector currents contribute to the
meson states of $\tau$ decays, which have negative G-parity.
The hadronic matrix elements of the axial-vector currents derived
in this paper are differnt from other studies. The Breit-Wigner
formula for $a_{1}$ resonance takes(see Eq.(35))
\[\frac{-g^{2}f^{2}_{a}m^{2}_{\rho}+i\sqrt{q^{2}}\Gamma_{a}(q^{2})}
{q^{2}-m^{2}_{a}+i\sqrt{q^{2}}\Gamma_{a}(q^{2})}.\]
The difference originates in the spontaneous chiral symmetry
breaking which is responsible for the mass difference of the $\rho$
and the $a_{1}$ mesons. Due to the chiral symmetry breaking,
at \(q^{2}=0\) this formula is equal to $g^{2}f^{2}_{a}{m^{2}_{\rho}
\over m^{2}_{a}}$ and the mass of the
$a_{1}$ meson is still there.

There are two kinds of meson vertices in $\tau$ mesonic decays
: the vertices of normal parity and the ones of
abnormal parity. The later are from the Wess-Zumino-Witten Lagrangian.
 In Ref.[11] it shows that the WZW Lagrangian is
the leading term of the imaginary part of the effective meson
Lagrangain and the fields in WZW Lagrangian are normalized to physical
meson fields. The normalization of the fileds of the WZW action is
very important. The normalization constants for the vector and
axial-vector fields are different[11] and due to the mixing effects
the axial-vector fields are always associated with $\partial_{\mu}P$,
where P is the corresponding pseudoscalar field.

All the meson vertices are obtained from Ref.[11] and
they are fixed completely and most of them have been tested
already[11]. It is necessary to point out that the vertices of VPP
obtained in ref.[11] are functions of momentum(see Eq.(37)
$f_{\rho\pi\pi}$ and see section 7 for details)
and the vertices of AVP depend on momentum strongly(see Eq.(27), for
example) and due to the cancellation in the vertices AVP the dependence
of momentum is very important in understanding $\tau\rightarrow
\nu(mesons)_{G=-}$.

In the chiral limit,
the theory used to study $\tau$ mesonic decays consists of three
parts: VMD for vector currents, a new expression of axial-vector
currents, and vertices of mesons. All these three parts are
determined by the effective chiral theory of mesons[11]. All parameters
have been fixed.

In many studies[5,6,8,17] besides
the $\rho$ meson the exicted $\rho$($\rho'$ and $\rho''$)
are taken part in. In this paper only the $\rho$ meson is included.
In the region of higher $q^{2}$ the effects of
the form factor of VPP vertex
and other decay channels of $\rho$(such as $K\bar{K}$, $KK^{*}$,...)
are taken into account in calculating the decay widths.
So far, theoretical results
agree with data reasonably well. In this paper only the lowest
resonances are taken into consideration.

It has been shown in Ref.[11] that the diagrams at tree level are at
order
of $N_{C}$ and the loop diagrams of mesons are at higher order in
large $N_{C}$ expansion. In Ref.[11] large $N_{C}$ expansion is implored
to argue the success of the effective theory. Following the same
argument, all calculations are done at tree level in this paper.

The paper is organized as
1.introduction; 2.general expression of the axial-vector currents;
3.determination of ${\cal L}^{V,A}$;
4.$a_{1}$ dominance in $\tau\rightarrow
\pi\pi\pi\nu$ decay; 5.$a_{1}$ dominance in
$\tau\rightarrow f_{1}(1285)\pi\nu$ decay;
6.$\tau\rightarrow K^{*}K\nu$;
7.CVC and $e^{+}e^{-}\rightarrow\pi^{+}\pi^{-}$;
8.$\tau\rightarrow\omega\pi\nu$; 9.$\tau\rightarrow K\bar{K}\nu$;
10.the form factors of $\pi\rightarrow
e\gamma\nu$;
11.effective Lagrangian of $\Delta s=1\) weak interactions;
12.$K_{a}$ dominance in $\tau\rightarrow K^{*}(892)\pi\nu$;
13.$\tau\rightarrow
K^{*}\eta\nu$; 14.$\tau\rightarrow\eta K\nu$;
15.the form factors of $K\rightarrow e\gamma\nu$;
16. conclusions.

\section{General expression of the axial-vector currents}
In the case of two flavors the expression of VMD[12]
is written as
\begin{equation}
{e\over 2f_{v}}\{-{1\over 2}F^{\mu\nu}(\partial_{\mu}v_
{\nu}-\partial_{\nu}v_{\mu})+A^{\mu}j^{v}_{\mu}\},
\end{equation}
where \(v=\rho^{0}, \omega, \phi\), $f_{v}$ is the decay constant
of these vector mesons respectively,
and $j^{v}_{\mu}$ are the
appropriate currents determined by the
substitution
\begin{equation}
v_{\mu}\rightarrow {e\over2f_{v}}A_{\mu}
\end{equation}
in the vertices involving neutral vector mesons.
CVC works very well in the weak interactions of hadrons and
in $\tau$
mesonic decays[16]. In the chiral limit,
the vector part of the weak
interaction of ordinary quarks is determined by CVC
\begin{equation}
{\cal L}^{V}={g_{W}\over 4}cos\theta_{C}
{1\over f_{\rho}}\{-{1\over 2}
(\partial_{\mu}A^{i}_{\nu}-\partial_{\nu}
A^{i}_{\mu})
(\partial_{\mu}\rho^{i}_
{\nu}-\partial_{\nu}\rho^{i}_{\mu})+A^{i}_{\mu}j^{i\mu}\},
\end{equation}
where \(i=1,2\) and $A^{i}_{\mu}$ are W boson fields. In the vector
part of the weak interaction there is $\rho$ dominance(two flavor case).
$j^{i}_{\mu}$ is derived by the substitution
\begin{equation}
\rho^{i}_{\mu}\rightarrow {g_{W}\over4f_{\rho}}cos\theta_{C}
A^{i}_{\mu}
\end{equation}
in the vertices involving $\rho$ mesons.
At low energies the matrix elements of the vector currents go back
to the chiral limit[2].

Chiral symmetry is one of major features of
$QCD$.
It is known for a long time that the $a_{1}$ meson is the chiral
partner of $\rho$ meson[14] and both are treated as nonabelian chiral
gauge fields[14]. On the other hand, it is well known that
in the chiral
limit, the vector and axial-vector currents form a $SU(2)_{L}\times
SU(2)_{R}$ algebra which leads to Weinberg's sum rules[13].
Based on the chiral symmetry it is reasonable to think
that in the axial-vector part of weak interaction of ordinary quarks
there is a term which is similar to VMD
\begin{equation}
-{g_{W}\over 4}cos\theta_{C}{1\over f_{a}}\{-{1\over 2}
(\partial_{\mu}A^{i}_{\nu}-\partial_{\nu}
A^{i}_{\mu})
(\partial_{\mu}a^{i}_
{\nu}-\partial_{\nu}a^{i}_{\mu})+A^{i\mu}j^{iW}_{\mu}\},
\end{equation}
where $a^{i}_{\mu}$ is the $a_{1}$ meson field,
$f_{a}$ is a constant,
and $j^{i,W}_{\mu}$ is the appropriate
current obtained by substituting
\begin{equation}
a^{i}_{\mu}\rightarrow -{g_{W}\over 4f_{a}}cos\theta_{C}A^{i}_{\mu}
\end{equation}
into the Lagrangian in which $a_{1}$ meson is involved.
On the other
hand, pion can couple to W boson directly. The second term
in the axial-vector part of weak interaction of the ordinary quarks
is
\begin{equation}
-{g_{W}\over4}cos\theta_{C}
f_{\pi}A^{i}_{\mu}\partial^{\mu}\pi^{i}.
\end{equation}

As a matter of fact, $a_{1}$ meson is much
heavier than $\rho$ meson. The spontaneous chiral
symmetry breaking is responsible for the mass
difference. Therefore, due to the effect of spontaneous chiral
symmetry breaking in the Lagrangian of meson theory there should be
an additional mass term for $a_{1}$ meson
\begin{equation}
{1\over 2}\Delta m^{2}f^{2}_{a}a^{i}_{\mu}a^{i\mu}.
\end{equation}
Using the substitution(6), a new coupling between W bosons and
$a_{1}$ mesons is revealed
\begin{equation}
-{g_{W}\over 4}cos\theta_{C}\Delta m^{2}f_{a}A^{i}_{\mu}a^{i,\mu}.
\end{equation}
Adding these three terms(5,7,9) together, the axial-vector part of the
effective Lagrangian of weak interaction of ordinary quarks is
obtained
\begin{eqnarray}
\lefteqn{{\cal L}^{A}=
-{g_{W}\over 4}cos\theta_{C}
{1\over f_{a}}\{-{1\over 2}(\partial_{\mu}A^{i}_{\nu}
-\partial_{\nu}A^{i}_{\mu})(\partial_{\mu}a^{i}_
{\nu}-\partial_{\nu}a^{i}_{\mu})+A^{i\mu}j^{iW}_{\mu}\}}\nonumber \\
&&-{g_{W}\over 4}cos\theta_{C}
\Delta m^{2}f_{a}A^{i}_{\mu}a^{i\mu}
-{g_{W}\over4}cos\theta_{C}
f_{\pi}A^{i}_{\mu}\partial^{\mu}\pi^{i}.
\end{eqnarray}
In Eq.(10) there are two parameters $f_{a}$ and $\Delta m^{2}$ which
are necessary to be determined. Weinberg's first sum rule
\begin{equation}
{g^{2}_{\rho}\over m^{2}_{\rho}}-{g^{2}_{a}\over m^{2}_{a}}=
f^{2}_{\pi}
\end{equation}
is derived by using $SU(2)_{L}\times SU(2)_{R}$ chiral symmetry,
current algebra, and VMD, where $g_{\rho}$ and $g_{a}$ are defined
by following formulas
\begin{equation}
<0|\bar{\psi}\tau_{i}\gamma_{\mu}\psi|\rho^{\lambda}_{j}>=g_{\rho}
\delta_{ij}\epsilon^{\lambda}_{\mu},\;\;\;
<0|\bar{\psi}\tau_{i}\gamma_{\mu}\gamma_{5}\psi|a^{\lambda}_{j}>
=g_{a}\delta_{ij}\epsilon^{\lambda}_{\mu}.
\end{equation}
Using Eqs.(3,10), we obtain
\begin{equation}
g_{\rho}=-{m^{2}_{\rho}\over f_{\rho}},\;\;\;
g_{a}=-{m^{2}_{a}\over f_{a}}+\Delta m^{2}f_{a}.
\end{equation}
It can be seen from Eqs.(4,6) that the $\rho$ fields are associated with
$f_{\rho}$ and $a_{1}$ with $f_{a}$. After spontaneous chiral symmetry
breaking the effective mass terms of $\rho$ and $a_{1}$ mesons
are written as
\begin{equation}
{1\over2}(\Delta m^{2}+m^{2}_{0})f^{2}_{a}a^{i}_{\mu}a^{i\mu}+
{1\over2}m^{2}_{0}f^{2}_{\rho}\rho^{i}_{\mu}\rho^{i\mu}
\end{equation}
and
\begin{equation}
m^{2}_{\rho}=m^{2}_{0}f^{2}_{\rho},\;\;\;m^{2}_{a}=f^{2}_{a}
(\Delta m^{2}+m^{2}_{0}).
\end{equation}
Eq.(15) leads to
\begin{equation}
\Delta m^{2}={m^{2}_{a}\over f^{2}_{a}}-{m^{2}_{\rho}\over
f^{2}_{\rho}}.
\end{equation}
From Eqs.(13,16) we obtain
\begin{equation}
g_{a}=-{f_{a}\over f^{2}_{\rho}}m^{2}_{\rho}.
\end{equation}
Substituting Eqs.(13,17) into Eq.(11), we determine
\begin{equation}
f^{2}_{a}=f^{2}_{\rho}(1-{f^{2}_{\pi}f^{2}_{\rho}\over m^{2}_{\rho}})
{m^{2}_{a}\over m^{2}_{\rho}},\;\;\;
\Delta m^{2}=f^{2}_{\pi}(1-{f^{2}_{\pi}f^{2}_{\rho}\over
m^{2}_{\rho}})^{-1}.
\end{equation}
The values of $f_{a}$ and $\Delta m^{2}$ are determined by
$f_{\pi}$, $f_{\rho}$, $m_{\rho}$, and $m_{a}$.
In general, \(f_{a}\neq f_{\rho}\).
Therefore, ${\cal L}^{A}$ is fixed.
The vector current is conserved in the limit of \(m_{q}=0\). The
axial-vector current derived from Eq.(10) must satisfy PCAC.
It will be shown that it is necessary to have
all the terms in Eq.(10) to satisfy PCAC.

\section{Determination of ${\cal L}^{V,A}$}
An effective chiral theory of pseudoscalar, vector, and axial-vector
mesons has been proposed[11]. In this theory both the physical processes
of normal parity and abnormal parity are studied by one Lagrangian.
Theoretical results agree with data well.
In the limit of \(m_{q}=0\), the Lagrangian is
\begin{eqnarray}
{\cal L}=\bar{\psi}(x)(i\gamma\cdot\partial+\gamma\cdot v
+\gamma\cdot a\gamma_{5}
-mu(x))\psi(x)\nonumber \\
+{1\over 2}m^{2}_{0}(\rho^{\mu}_{i}\rho_{\mu i}+
\omega^{\mu}\omega_{\mu}+a^{\mu}_{i}a_{\mu i}+f^{\mu}f_{\mu})
\nonumber \\
\end{eqnarray}
where \(a_{\mu}=\tau_{i}a^{i}_{\mu}+f_{\mu}\), \(v_{\mu}=\tau_{i}
\rho^{i}_{\mu}+\omega_{\mu}\),
, and \(u=exp\{i\gamma_{5}(\tau_{i}\pi_{i}+
\eta)\}\), $m$ is a parameter.
The fields in the Lagrangian(19) are not physical and the physical
meson fields have been defined in Ref.[11].
This Lagrangian is global $SU(2)_{L}\times SU(2)_{R}$ chiral symmetric.
On the other hand, the spontaneous chiral
symmetry breaking is revealed in this theory when
\(\pi^{i},\eta=0\) are taken.
This theory has both explicit chiral symmetry breaking(PCAC)
by adding the current quark masses $-\bar{\psi}M\psi$(M is the quark
matrix) to the Lagrangian
and dynamical chiral symmetry breaking(quark condensate)[11].

The explicit expression of VMD(1) has been found in Ref.[11].
The expressions of ${\cal L}^{V,A}$(3,10) have been derived by
this effective theory(see Eqs.(76,77,78) of Ref.[11]) too.
Following expressions are revealed from this effective chiral theory
\begin{eqnarray}
\lefteqn{g_{\rho}=-gm^{2}_{\rho},}\\
&&g_{a}=-g(1-{1\over 2\pi^{2}g^{2}})^{-{1\over 2}}m^{2}_{\rho},
\\
&&f_{\rho}=g^{-1},\\
&&f_{a}=g^{-1}(1-{1\over2\pi^{2}g^{2}})^{-{1\over2}},\\
&&(1-{1\over 2\pi^{2}g^{2}})m^{2}_{a}=6m^{2}+m^{2}_{\rho},\\
&&\Delta m^{2}=6m^{2}g^{2}=
f^{2}_{\pi}(1-{f^{2}_{\pi}\over g^{2}m^{2}_{\rho}})^{-1},
\end{eqnarray}
where g is a universal coupling constant and m is a parameter related
to quark condensate.
The term $6m^{2}$ in Eq.(24) and the factor
$(1-{1\over2\pi^{2}g^{2}})$ in Eqs.(23,24)
are from spontaneous chiral
symmetry breaking of this theory.
In this paper we choose \(g=0.39\).
Using this value, the
theoretical results obtained in Ref.[11]
agree with data well. For example, we obtain
\(\Gamma_{\rho}=142 MeV\) and \(m_{a}=1.20GeV\).
It is necessary to point out that Weinberg's first sum rule is
satisfied analytically in this effective chiral theory.
Eqs.(18) are satisfied too.
Therefore, all the parameters in the Lagrangians(${\cal L}^{V,A}
$) are fixed.

Besides the Lagrangians(3,10),
appropriate meson(pseudoscalar, vector, and
axial-vector)vertices are needed in
studying the $\tau$ mesonic decays and all these vertices can be
derived from the Lagrangian(19) and they are fixed[11]. The
most of these
vertices have been tested by calculating appropriate decay widths and
the results agree with data well.
Therefore, the Lagrangians of the weak interactions of mesons are
completely determined. There are no other undetermined parameters
in studying $\tau$ mesonic decays. This effective theory makes definite
predictions for $\tau$ mesonic decays.

\section{$a_{1}$ dominance in $\tau\rightarrow \pi\pi\pi\nu$}
The $a_{1}$ meson has a long history. In determining the parameters
of this meson the process of $a_{1}$ production in $\tau$ decays
play an important role[18,19]. In Ref.[19]
the flux tube quark model has been exploited to study
$\tau\rightarrow\pi\pi\pi\nu$. The decay rate of $\tau\rightarrow
a_{1}\nu$ has been calculated by the effective chiral theory[11].
However, the effect of wide resonance should be taken into account.
On the other hand,
the experimental observations[20-23,25,27] have
reported the $a_{1}$ dominance in $\tau\rightarrow3\pi\nu$.

Only the axial-vector part of the Lagrangian, ${\cal L}^{A}$,
takes part in this process.
There are five
diagrams contributing to this decay: $a_{1}$ couples to $\rho\pi$,
W-boson couples to $\rho\pi$ directly, $\pi$ couples
to $\rho\pi$, $a_{1}$ directly
couples to three pions, and $\pi$ directly couples to three pions.
The study done in Ref.[11]
indicates that the contribution of the four $\pi$ coupling
to $\pi\pi$ scattering
is smaller than the contribution of $\rho$ exchange by two orders of
magnitude. From Ref.[11] it is learned that the contribution of
the vertex of $a_{1}\pi\pi\pi$ to $a_{1}$ decay is very small and this
result agrees with data.
Therefore, we omit the contributions of both the vertices of
the four $\pi$ and the $a_{1}
\pi\pi\pi$. The remaining three diagrams
indicate the existence of
$\rho$ resonances in the final states of $\tau\rightarrow3\pi\nu$ and
this result is in agreement with data[26,27].
From these three vertices: $a_{1}\rho\pi$, $W\rho\pi$, and $\pi\rho\pi$
($\rho\pi\pi$ is included too)it is not obvious why
$a_{1}$
dominates this decay. This is a crucial test on this theory.
The vertices derived in Ref.[11]
contribute to $\tau\rightarrow
\pi\pi\pi\nu$
\begin{eqnarray}
\lefteqn{{\cal L}^{a_{1}\rho\pi}=\epsilon_{ijk}\{Aa^{i}_{\mu}
\rho^{j\mu}\pi^{k}-Ba^{i}_{\mu}\rho^{j}_{\nu}\partial^{\mu\nu}\pi^{k}
+Da^{i}_{\mu}\partial^{\mu}(\rho^{j}_{\nu}
\partial^{\nu}\pi^{k})\}},\\
&&A={2\over f_{\pi}}gf_{a}\{{m^{2}_{a}\over g^{2}f^{2}_{a}}
-m^{2}_{\rho}+p^{2}[{2c\over g}+{3\over4
\pi^{2}g^{2}}(1-{2c\over g})]\nonumber \\
&&+q^{2}[{1\over 2\pi^{2}g^{2}}-
{2c\over g}-{3\over4\pi^{2}g^{2}}(1-{2c\over g})]\},\\
&&c={f^{2}_{\pi}\over2gm^{2}_{\rho}},\\
&&B=-{2\over f_{\pi}}gf_{a}{1\over2\pi^{2}g^{2}}(1-{2c\over g}),\\
&&D=-{2\over f_{\pi}}f_{a}\{2c+{3\over2\pi^{2}g}(1-{2c\over g})\},\\
&&{\cal L}^{\rho\pi\pi}={2\over g}\epsilon_{ijk}\rho^{i}_{\mu}
\pi^{j}\partial^{\mu}\pi^{k}-{2\over \pi^{2}f^{2}_{\pi}g}
\{(1-{2c\over g})^{2}-4\pi^{2}c^{2}\}\epsilon
_{ijk}\rho^{i}_{\mu}\partial_{\nu}\pi^{j}\partial^{\mu\nu}\pi^{k}
\nonumber \\
&&-{1\over \pi^{2}f^{2}_{\pi}g}\{3(1-{2c\over g})^{2}
+1-{2c\over g}-8\pi^{2}c^{2}\}\epsilon_{ijk}\rho^{i}_{\mu}\pi_{j}
\partial^{2}\partial_{\mu}\pi_{k},
\end{eqnarray}
where p is the momentum of $\rho$ meson and q is
the momentum of $a_{1}$. Because the mesons of the vertices are
not necessary to be on mass-shell, hence,
in Eqs.(26,31) the divergence of $a_{\mu}$ and
$\partial^{2}\pi_{k}$
are kept. In the chiral limit,
these new terms do not contribute to the decays of $\rho$ or $a_{1}$,
however they are
important in keeping
the axial-vector current conserved in $\tau\rightarrow3\pi\nu$
in the chiral limit.
The vertex ${\cal L}^
{W\rho\pi}$ is derived by using the substitution(6) in
Eq.(26).

The two pions in $\tau\rightarrow3\pi\nu$ are from the decays of
$\rho$ meson, therefore, we only need to show that the
axial-vector current is conserved
(in the limit of \(m_{q}\rightarrow
0\)) in $\tau\rightarrow\rho\pi\nu$, then this conservation is
satisfied in $\tau\rightarrow3\pi\nu$.

Using ${\cal L}^{A}$(10) and three vertices ${\cal L}^{a_{1}
\rho\pi}$(26), ${\cal L}^{W\rho\pi}$, and ${\cal L}^{
\rho\pi\pi}$(31), the matrix element of the axial-vector current
of $\tau^{-}\rightarrow\rho^{0}\pi^{-}$ is obtained as
\begin{eqnarray}
\lefteqn{<\rho^{0}\pi^{-}|\bar{\psi}\tau_{+}\gamma_{\mu}\gamma_{5}
\psi|0>=\frac{i}{\sqrt{4\omega E}}\{\frac{1}
{f_{a}(q^{2}-m^{2}_{a})}(q_{\mu}q_{\nu}-q^{2}g_{\mu\nu})
(Ag_{\lambda\nu}+Bk_{\lambda}k_{\nu})\epsilon^{*\lambda}_{\sigma}}
\nonumber \\
&&-\frac{\Delta m^{2}f_{a}}{q^{2}-m^{2}_{a}}(\frac{q_{\mu}q_{\nu}}
{q^{2}}-g_{\mu\nu})
(Ag_{\lambda\nu}+Bk_{\lambda}k_{\nu})\epsilon^{*\lambda}_{\sigma}
-\frac{\Delta m^{2}f_{a}}{m^{2}_{a}}\frac{q_{\mu}}{q^{2}}
(A+k\cdot qB)k\cdot \epsilon^{*}_{\sigma}\nonumber \\
&&+{1\over f_{a}}
(Ag_{\mu\nu}+Bk_{\mu}k_{\nu})\epsilon^{*\nu}_{\sigma}
-({1\over f_{a}}-\frac{\Delta m^{2}f_{a}}{m^{2}_{a}})Dk\cdot
\epsilon^{*}_{\sigma}
q_{\mu}\nonumber \\
&&-{4f_{\pi}\over g}\frac{q_{\mu}}{q^{2}}\{1+\frac{p^{2}}{2\pi^{2}
f^{2}_{\pi}}[(1-{2c\over g})^{2}-4\pi^{2}c^{2}]
+\frac{q^{2}}{2\pi^{2}f^{2}_{\pi}}(1-{2c\over g})(1-{c\over g})\}
k\cdot \epsilon^{*}_{\sigma}\},
\end{eqnarray}
where k, p, and q are the momenta of pion, $\rho$, and $a_{1}$
respectively.
In the effective Lagrangian of mesons[11] derived from the Lagrangian
(19) there
is mass term of the $a_{1}$ meson, therefore, the propagator of
$a_{1}$ field is taken to be
\begin{equation}
{i\over(2\pi)^{4}}\frac{1}{q^{2}-m^{2}_{a}}(-g_{\mu\nu}+\frac{q_{\mu}
q_{\nu}}{m^{2}_{a}}).
\end{equation}
There are cancellations in Eq.(32). Using Eqs.(24,25,27,28,29,30),
it is proved
\begin{eqnarray}
\lefteqn{
-\frac{\Delta m^{2}f_{a}}{m^{2}_{a}}
(A+k\cdot qB)+{1\over f_{a}}(A+Bk\cdot q)
-({1\over f_{a}}-\frac{\Delta m^{2}f_{a}}{m^{2}_{a}})D
q^{2}}\nonumber \\
&&-{4f_{\pi}\over g}\{1+\frac{p^{2}}{2\pi^{2}
f^{2}_{\pi}}[(1-{2c\over g})^{2}-4\pi^{2}c^{2}]
+\frac{q^{2}}{2\pi^{2}f^{2}_{\pi}}(1-{2c\over g})(1-{c\over g})\}
=0.
\end{eqnarray}
Eq.(34) leads to the conservation of axial-vector current in the
limit of \(m_{q}=0\)
\[q^{\mu}<\rho^{0}\pi^{-}|\bar{\psi}\tau_{+}\gamma_{\mu}\gamma_{5}
\psi|0>=0.\]
From this discussion it is learned that in order to have the axial-
vector conserved(in the limit of \(m_{q}=0\)) the new term(9) is
necessary to be included in Eq.(10).
Using Eq.(34), the matrix element(32) is rewritten as
\begin{eqnarray}
\lefteqn{<\rho^{0}\pi^{-}|\bar{\psi}\tau_{-}\gamma_{\mu}\gamma_{5}
\psi|0>=\frac{i}{\sqrt{4\omega E}}(\frac{q_{\mu}q_{\nu}}{q^{2}}
-g_{\mu\nu})(Ag_{\nu\lambda}+Bk_{\nu}k_{\lambda})\epsilon^{*\nu}
_{\sigma}\frac{-\Delta m^{2}f_{a}+q^{2}f^{-1}_{a}}
{q^{2}-m^{2}_{a}+i\sqrt{q^{2}}\Gamma_{a}(q^{2})}}
\nonumber \\
&&+\frac{i}{\sqrt{4\omega E}}(\frac{q_{\mu}q_{\nu}}{q^{2}}
-g_{\mu\nu})(-
{1\over f_{a}})
(Ag_{\nu\lambda}+Bk_{\nu}k_{\lambda})\epsilon^{*\nu}_{
\sigma}\nonumber \\
&&=\frac{i}{\sqrt{4\omega E}}(\frac{q_{\mu}q_{\nu}}{q^{2}}
-g_{\mu\nu})(Ag_{\nu\lambda}+Bk_{\nu}k_{\lambda})\epsilon^{*\nu}
_{\sigma}\frac{g^{2}f_{a}m^{2}_{\rho}-if^{-1}_{a}\sqrt{q^{2}}\Gamma_{a}
(q^{2})}
{q^{2}-m^{2}_{a}+i\sqrt{q^{2}}\Gamma_{a}(q^{2})}
\end{eqnarray}
It is necessary to point out that the
Breit-Wigner formula of the
axial-vector meson, $a_{1}$, is new and
is different from the one of the vector
meson(see section 7). This difference is caused by the dynamical
chiral symmetry breaking. It is also important to notice that
the amplitude A(27) derived in Ref.[11] strongly depends on
the momentum. Due to the cancellation in Eq.(27) this dependence
is significant.

The decay width of $a_{1}$ meson has been introduced.
The $a_{1}$ dominance in $\tau\rightarrow \rho\pi\nu$
is revealed and the dominance is caused by
the cancellation(34) which leads to the axial-vector current
conservation.
Using the vertex ${\cal L}^{\rho\pi\pi}$(31),
the matrix
element is derived
\begin{eqnarray}
\lefteqn{<\pi^{+}\pi^{-}\pi^{-}|\bar{\psi}\tau_{+}\gamma_{\mu}\gamma_
{5}\psi|0>=\frac{i}{\sqrt{8\omega_{1}\omega_{2}\omega_{3}}}
(-g_{\mu\nu}
+\frac{q_{\mu}q_{\nu}}{q^{2}})\frac{g^{2}f_{a}m^{2}_{\rho}
-i\sqrt{q^{2}}f^{-1}_{a}\Gamma_{a}(q^{2})}{
q^{2}-m^{2}_{a}+i\sqrt{q^{2}}\Gamma_{a}(q^{2})}}\nonumber \\
&&\{\frac{f_{\rho\pi\pi}(k^{2})}{k^{2}-m^{2}_{\rho}
+ik\Gamma_{\rho}(k^{2})}
[A(k^{2})(k_{2}-k_{3})_{\nu}+Bk_{1\nu}
k_{1}\cdot(k_{2}-k_{3})]\nonumber \\
&&\frac{f_{\rho\pi\pi}(k^{'2})
}{k^{'2}-m^{2}_{\rho}+ik^{'}\Gamma_{\rho}
(k^{'2})}
[A(k^{'2})(k_{1}-k_{3})_{\nu}+Bk_{2\nu}
k_{2}\cdot(k_{1}-k_{3})]\},\\
&&f_{\rho\pi\pi}(k^{2})={2\over g}\{1+
{k^{2}\over 2\pi^{2}f^{2}_{\pi}}[(1-{2c\over g})^{2}-4\pi^{2}c^{2}]
\},\nonumber \\
&&\Gamma_{\rho}(k^{2})=\frac{f^{2}_{\rho\pi\pi}}{48\pi}{k^{2}\over
m_{\rho}}(1-4{m^{2}_{\pi}\over k^{2}})^{{3\over2}},
\end{eqnarray}
where $k_{i}$(\(i=1,2,3)\) are the momenta of $\pi^{-}$, $\pi^{-}$, and
$\pi^{+}$, \(k=k_{2}+k_{3}\), \(k'=k_{1}+k_{3}\), \(q=k_{1}+
k_{2}+k_{3}\). $A(k^{2})$($A(k^{'2})$) are obtained by taking \(p^{2}
=k^{2}(k^{'2})\) and \(p^{2}_{a}=q^{2}\) in Eq.(27)
, and $\Gamma_{a}$ is defined in Eq.(39).
The distribution of the decay width of $\tau^{-}\rightarrow\pi^{-}
\pi^{-}\pi^{+}\nu$ is derived
\begin{eqnarray}
\lefteqn{\frac{d\Gamma}{dq^{2}dk^{2}dk^{`2}}=
\frac{G^{2}}{(2\pi)^{5}}
\frac{cos^{2}\theta_{C}}
{3072m^{3}_{\tau}q^{6}}(m^{2}_{\tau}-q^{2})^{2}(m^{2}_{\tau}
+2q^{2})(q^{2}-k^{2})^{2}}\nonumber \\
&&\frac{g^{4}f^{2}_{a}m^{4}_{\rho}+q^{2}f^{-2}_{a}\Gamma^{2}_{a}
(q^{2})}
{(q^{2}-m^{2}_{a})^{2}+q^{2}\Gamma^{2}_{a}(q^{2})}F(k^{2},k^{'2})
,
\nonumber \\
&&G(k^{2},k^{'2})=\{\frac{f_{\rho\pi\pi}(k^{2})}{(k^{2}-m^{2}_{\rho})
^{2}+k^{2}\Gamma^{2}_{\rho}(k^{2})}(k^{2}-m^{2}_{\rho})
[A(k^{2})+k_{1}\cdot(k_{2}-k_{3})B]\nonumber \\
&&+\frac{f_{\rho\pi\pi}(k^{'2})}{(k^{'2}-m^{2}_{\rho})
^{2}+k^{'2}\Gamma^{2}_{\rho}(k^{'2})}(k^{'2}-m^{2}_{\rho})
2A(k^{'2})\}^{2}\nonumber \\
&&+\{\frac{f_{\rho\pi\pi}(k^{2})}{(k^{2}-m^{2}_{\rho})
^{2}+k^{2}\Gamma^{2}_{\rho}(k^{2})}\sqrt{k^{2}}\Gamma_{\rho}
(k^{2})
[A(k^{2})+k_{1}\cdot(k_{2}-k_{3})B]\nonumber \\
&&+\frac{f_{\rho\pi\pi}(k^{'2})}{(k^{'2}-m^{2}_{\rho})
^{2}+k^{'2}\Gamma^{2}_{\rho}(k^{'2})}\sqrt{k^{'2}}\Gamma_{\rho}
(k'^{2})2A(k^{'2})\}^{2}.\nonumber \\
&&F(k^{2},k^{'2})={1\over2}\{G(k^{2},k^{'2})+G(k^{'2},k^{2})\}.
\end{eqnarray}
$\Gamma_{a}$ is derived from the vertices(26,31)
\begin{eqnarray}
\lefteqn{
\Gamma_{a}(q^{2})=\frac{1}{192(2\pi)^{3}m_{a}q^{4}}\int dq^{2}_{1}
dq^{2}_{2}(q^{2}-q^{2}_{1})^{2}\{\frac{f_{\rho\pi\pi}(q^{2}_{1})}
{(q^{2}_{1}-m^{2}_{\rho})^{2}+q^{2}_{1}\Gamma_{\rho}(q^{2}_{1})}
(q^{2}_{1}-m^{2}_{\rho})}\nonumber \\
&&[A(q^{2}_{1})+{1\over2}
(q^{2}_{3}-q^{2}_{2})B]
+\frac{f_{\rho\pi\pi}(q^{2}_{2})}
{(q^{2}_{2}-m^{2}_{\rho})^{2}+q^{2}_{2}\Gamma_{\rho}(q^{2}_{2})}
(q^{2}_{2}-m^{2}_{\rho})2A(q^{2}_{2})]\}^{2}\nonumber \\
&&+\{\frac{f_{\rho\pi\pi}(q^{2}_{1})}
{(q^{2}_{1}-m^{2}_{\rho})^{2}+q^{2}_{1}\Gamma_{\rho}(q^{2}_{1})}
\sqrt{q^{2}_{1}}\Gamma_{\rho}(q^{2}_{1})
[A(q^{2}_{1})+{1\over2}
\nonumber \\
&&(q^{2}_{3}-q^{2}_{2})B]
+\frac{f_{\rho\pi\pi}(q^{2}_{2})}
{(q^{2}_{2}-m^{2}_{\rho})^{2}+q^{2}_{2}\Gamma_{\rho}(q^{2}_{2})}
\sqrt{q^{2}_{2}}\Gamma_{\rho}(q^{2}_{2})2A(q^{2}_{2})]\}^{2},
\end{eqnarray}
where \(q^{2}_{1}=(q-k_{1})^{2}\), \(q^{2}_{2}=(q-k_{2})^{2}\),
\(q^{2}_{3}=(q-k_{3})^{2}\), and $k_{1}$, $k_{2}$,and $k_{3}$ are
momentum of the three pions respectively.
There are two decay modes: $\pi^{+}\pi^{-}\pi^{-}$ and
$\pi^{-}\pi^{0}\pi^{0}$ which have equal branching ratio.
The branching ratios are computed to be
\begin{equation}
B(\tau\rightarrow\pi^{+}\pi^{-}\pi^{-}\nu)=
B(\tau\rightarrow\pi^{-}\pi^{0}\pi^{0}\nu)=
6.3\%.
\end{equation}
The comparison with experiments is presented in Table I.
\begin{table}[h]
\begin{center}
\caption {Table I Branching Ratios}
\begin{tabular}{|c|c|c|} \hline
Experiment[25]&$B(2h^{-}h^{+}\nu)\%$&$B(h^{-}2\pi^{0}\nu)\%$ \\ \hline
New W.A.&$9.26\pm0.26$&$9.21\pm0.14$\\ \hline
DELPHI(92-95)&$8.69\pm0.12\pm0.16$&$9.22\pm0.43\pm0.20$\\ \hline
ALEPH(89-93)&$ 9.46\pm0.10\pm0.11$ & $9.32\pm0.13\pm0.10$   \\ \hline
CELLO(90)& & $9.1\pm1.3\pm0.9$ \\  \hline
OPAL(91-94)&$9.83\pm0.10\pm0.24$&  \\ \hline
L3(92)    & &$8.88\pm0.37\pm0.42$\\ \hline
CLEO(93)  &$8.7\pm0.8$    & $8.96\pm0.16\pm0.449$         \\ \hline
CLEO(95)&$9.47\pm0.07\pm0.20$  \\  \hline
CBALL(91)  &    &$5.7\pm0.5\pm1.4$      \\ \hline
ARGUS(93)&$7.3\pm0.1\pm0.5$&      \\ \hline
MAC(87)&  &$8.7\pm0.4\pm0.11$  \\  \hline
BES[16]   & $7.3\pm0.5$($\pi^{+}\pi^{-}\pi^{-}$)  &      \\ \hline
Taula 2.4 &$7.0\pm2.8$&$6.4\pm2.8 $  \\  \hline
This study&6.3$(\pi^{+}\pi^{-}\pi^{-})$&6.3($\pi^{-}\pi^{0}\pi^{0}$)
  \\ \hline
\end{tabular}
\end{center}
\end{table}
The distribution of $d\Gamma(\tau\rightarrow\pi^{+}\pi^{-}\pi^{-}
\nu)/dq$ is shown in
Fig.1. From Fig.(1) the decay width is determined to be
\begin{equation}
\Gamma_{a}=386MeV.
\end{equation}
The data is $\sim400MeV$[28].
The comparison with experiments using the model[19]
is presented in Table II.
The starred results are taken from[19].
\begin{table}[h]
\begin{center}
\caption {Table II Parameters of $a_{1}$ meson}
\begin{tabular}{|c|c|c|} \hline
 Experiment  & $m_{a_{1}}$(GeV)&$\Gamma_{a_{1}}$GeV\\ \hline
ARGUS[27] & $1.211\pm0.007$ & $0.446\pm0.021$         \\ \hline
$DELCO^{*}$  &$1.180\pm0.060$ & $0.430\pm0.190$         \\ \hline
$MARKII^{*}$  &$1.250\pm0.050$ &$0.580\pm0.100$    \\ \hline
$ARGUS^{*}$ & $1.213\pm0.011$ &$0.434\pm0.030$        \\ \hline
This study & 1.20 & 0.386   \\ \hline
\end{tabular}
\end{center}
\end{table}

\begin{figure}
\begin{center}

\setlength{\unitlength}{0.240900pt}
\ifx\plotpoint\undefined\newsavebox{\plotpoint}\fi
\begin{picture}(1500,900)(0,0)
\font\gnuplot=cmr10 at 10pt
\gnuplot
\sbox{\plotpoint}{\rule[-0.500pt]{1.000pt}{1.000pt}}%
\put(220.0,113.0){\rule[-0.500pt]{292.934pt}{1.000pt}}
\put(220.0,113.0){\rule[-0.500pt]{4.818pt}{1.000pt}}
\put(198,113){\makebox(0,0)[r]{0}}
\put(1416.0,113.0){\rule[-0.500pt]{4.818pt}{1.000pt}}
\put(220.0,209.0){\rule[-0.500pt]{4.818pt}{1.000pt}}
\put(198,209){\makebox(0,0)[r]{50}}
\put(1416.0,209.0){\rule[-0.500pt]{4.818pt}{1.000pt}}
\put(220.0,304.0){\rule[-0.500pt]{4.818pt}{1.000pt}}
\put(198,304){\makebox(0,0)[r]{100}}
\put(1416.0,304.0){\rule[-0.500pt]{4.818pt}{1.000pt}}
\put(220.0,400.0){\rule[-0.500pt]{4.818pt}{1.000pt}}
\put(198,400){\makebox(0,0)[r]{150}}
\put(1416.0,400.0){\rule[-0.500pt]{4.818pt}{1.000pt}}
\put(220.0,495.0){\rule[-0.500pt]{4.818pt}{1.000pt}}
\put(198,495){\makebox(0,0)[r]{200}}
\put(1416.0,495.0){\rule[-0.500pt]{4.818pt}{1.000pt}}
\put(220.0,591.0){\rule[-0.500pt]{4.818pt}{1.000pt}}
\put(198,591){\makebox(0,0)[r]{250}}
\put(1416.0,591.0){\rule[-0.500pt]{4.818pt}{1.000pt}}
\put(220.0,686.0){\rule[-0.500pt]{4.818pt}{1.000pt}}
\put(198,686){\makebox(0,0)[r]{300}}
\put(1416.0,686.0){\rule[-0.500pt]{4.818pt}{1.000pt}}
\put(220.0,782.0){\rule[-0.500pt]{4.818pt}{1.000pt}}
\put(198,782){\makebox(0,0)[r]{350}}
\put(1416.0,782.0){\rule[-0.500pt]{4.818pt}{1.000pt}}
\put(220.0,877.0){\rule[-0.500pt]{4.818pt}{1.000pt}}
\put(198,877){\makebox(0,0)[r]{400}}
\put(1416.0,877.0){\rule[-0.500pt]{4.818pt}{1.000pt}}
\put(220.0,113.0){\rule[-0.500pt]{1.000pt}{4.818pt}}
\put(220,68){\makebox(0,0){0.4}}
\put(220.0,857.0){\rule[-0.500pt]{1.000pt}{4.818pt}}
\put(372.0,113.0){\rule[-0.500pt]{1.000pt}{4.818pt}}
\put(372,68){\makebox(0,0){0.6}}
\put(372.0,857.0){\rule[-0.500pt]{1.000pt}{4.818pt}}
\put(524.0,113.0){\rule[-0.500pt]{1.000pt}{4.818pt}}
\put(524,68){\makebox(0,0){0.8}}
\put(524.0,857.0){\rule[-0.500pt]{1.000pt}{4.818pt}}
\put(676.0,113.0){\rule[-0.500pt]{1.000pt}{4.818pt}}
\put(676,68){\makebox(0,0){1}}
\put(676.0,857.0){\rule[-0.500pt]{1.000pt}{4.818pt}}
\put(828.0,113.0){\rule[-0.500pt]{1.000pt}{4.818pt}}
\put(828,68){\makebox(0,0){1.2}}
\put(828.0,857.0){\rule[-0.500pt]{1.000pt}{4.818pt}}
\put(980.0,113.0){\rule[-0.500pt]{1.000pt}{4.818pt}}
\put(980,68){\makebox(0,0){1.4}}
\put(980.0,857.0){\rule[-0.500pt]{1.000pt}{4.818pt}}
\put(1132.0,113.0){\rule[-0.500pt]{1.000pt}{4.818pt}}
\put(1132,68){\makebox(0,0){1.6}}
\put(1132.0,857.0){\rule[-0.500pt]{1.000pt}{4.818pt}}
\put(1284.0,113.0){\rule[-0.500pt]{1.000pt}{4.818pt}}
\put(1284,68){\makebox(0,0){1.8}}
\put(1284.0,857.0){\rule[-0.500pt]{1.000pt}{4.818pt}}
\put(1436.0,113.0){\rule[-0.500pt]{1.000pt}{4.818pt}}
\put(1436,68){\makebox(0,0){2}}
\put(1436.0,857.0){\rule[-0.500pt]{1.000pt}{4.818pt}}
\put(220.0,113.0){\rule[-0.500pt]{292.934pt}{1.000pt}}
\put(1436.0,113.0){\rule[-0.500pt]{1.000pt}{184.048pt}}
\put(220.0,877.0){\rule[-0.500pt]{292.934pt}{1.000pt}}
\put(45,495){\makebox(0,0){${d\Gamma\over d\sqrt{q^{2}}}\times 10^{15}$ }}
\put(828,23){\makebox(0,0){FiG.1     GeV}}
\put(220.0,113.0){\rule[-0.500pt]{1.000pt}{184.048pt}}
\put(372,124){\usebox{\plotpoint}}
\put(372,122.92){\rule{3.373pt}{1.000pt}}
\multiput(372.00,121.92)(7.000,2.000){2}{\rule{1.686pt}{1.000pt}}
\put(386,125.42){\rule{3.373pt}{1.000pt}}
\multiput(386.00,123.92)(7.000,3.000){2}{\rule{1.686pt}{1.000pt}}
\put(400,128.92){\rule{3.132pt}{1.000pt}}
\multiput(400.00,126.92)(6.500,4.000){2}{\rule{1.566pt}{1.000pt}}
\put(413,132.92){\rule{3.373pt}{1.000pt}}
\multiput(413.00,130.92)(7.000,4.000){2}{\rule{1.686pt}{1.000pt}}
\multiput(427.00,138.84)(0.887,0.462){4}{\rule{2.250pt}{0.111pt}}
\multiput(427.00,134.92)(7.330,6.000){2}{\rule{1.125pt}{1.000pt}}
\multiput(439.00,144.84)(0.851,0.475){6}{\rule{2.107pt}{0.114pt}}
\multiput(439.00,140.92)(8.627,7.000){2}{\rule{1.054pt}{1.000pt}}
\multiput(452.00,151.83)(0.544,0.487){12}{\rule{1.450pt}{0.117pt}}
\multiput(452.00,147.92)(8.990,10.000){2}{\rule{0.725pt}{1.000pt}}
\multiput(464.00,161.83)(0.495,0.489){14}{\rule{1.341pt}{0.118pt}}
\multiput(464.00,157.92)(9.217,11.000){2}{\rule{0.670pt}{1.000pt}}
\multiput(477.83,171.00)(0.489,0.591){14}{\rule{0.118pt}{1.523pt}}
\multiput(473.92,171.00)(11.000,10.840){2}{\rule{1.000pt}{0.761pt}}
\multiput(488.83,185.00)(0.489,0.686){14}{\rule{0.118pt}{1.705pt}}
\multiput(484.92,185.00)(11.000,12.462){2}{\rule{1.000pt}{0.852pt}}
\multiput(499.83,201.00)(0.491,0.759){16}{\rule{0.118pt}{1.833pt}}
\multiput(495.92,201.00)(12.000,15.195){2}{\rule{1.000pt}{0.917pt}}
\multiput(511.83,220.00)(0.489,0.878){14}{\rule{0.118pt}{2.068pt}}
\multiput(507.92,220.00)(11.000,15.707){2}{\rule{1.000pt}{1.034pt}}
\multiput(522.83,240.00)(0.487,1.129){12}{\rule{0.117pt}{2.550pt}}
\multiput(518.92,240.00)(10.000,17.707){2}{\rule{1.000pt}{1.275pt}}
\multiput(532.83,263.00)(0.489,1.022){14}{\rule{0.118pt}{2.341pt}}
\multiput(528.92,263.00)(11.000,18.141){2}{\rule{1.000pt}{1.170pt}}
\multiput(543.83,286.00)(0.487,1.182){12}{\rule{0.117pt}{2.650pt}}
\multiput(539.92,286.00)(10.000,18.500){2}{\rule{1.000pt}{1.325pt}}
\multiput(553.83,310.00)(0.487,1.288){12}{\rule{0.117pt}{2.850pt}}
\multiput(549.92,310.00)(10.000,20.085){2}{\rule{1.000pt}{1.425pt}}
\multiput(563.83,336.00)(0.489,1.118){14}{\rule{0.118pt}{2.523pt}}
\multiput(559.92,336.00)(11.000,19.764){2}{\rule{1.000pt}{1.261pt}}
\multiput(574.83,361.00)(0.485,1.441){10}{\rule{0.117pt}{3.139pt}}
\multiput(570.92,361.00)(9.000,19.485){2}{\rule{1.000pt}{1.569pt}}
\multiput(583.83,387.00)(0.487,1.288){12}{\rule{0.117pt}{2.850pt}}
\multiput(579.92,387.00)(10.000,20.085){2}{\rule{1.000pt}{1.425pt}}
\multiput(593.83,413.00)(0.487,1.288){12}{\rule{0.117pt}{2.850pt}}
\multiput(589.92,413.00)(10.000,20.085){2}{\rule{1.000pt}{1.425pt}}
\multiput(603.83,439.00)(0.485,1.441){10}{\rule{0.117pt}{3.139pt}}
\multiput(599.92,439.00)(9.000,19.485){2}{\rule{1.000pt}{1.569pt}}
\multiput(612.83,465.00)(0.487,1.235){12}{\rule{0.117pt}{2.750pt}}
\multiput(608.92,465.00)(10.000,19.292){2}{\rule{1.000pt}{1.375pt}}
\multiput(622.83,490.00)(0.485,1.381){10}{\rule{0.117pt}{3.028pt}}
\multiput(618.92,490.00)(9.000,18.716){2}{\rule{1.000pt}{1.514pt}}
\multiput(631.83,515.00)(0.485,1.321){10}{\rule{0.117pt}{2.917pt}}
\multiput(627.92,515.00)(9.000,17.946){2}{\rule{1.000pt}{1.458pt}}
\multiput(640.83,539.00)(0.485,1.262){10}{\rule{0.117pt}{2.806pt}}
\multiput(636.92,539.00)(9.000,17.177){2}{\rule{1.000pt}{1.403pt}}
\multiput(649.83,562.00)(0.485,1.262){10}{\rule{0.117pt}{2.806pt}}
\multiput(645.92,562.00)(9.000,17.177){2}{\rule{1.000pt}{1.403pt}}
\multiput(658.83,585.00)(0.485,1.142){10}{\rule{0.117pt}{2.583pt}}
\multiput(654.92,585.00)(9.000,15.638){2}{\rule{1.000pt}{1.292pt}}
\multiput(667.83,606.00)(0.481,1.295){8}{\rule{0.116pt}{2.875pt}}
\multiput(663.92,606.00)(8.000,15.033){2}{\rule{1.000pt}{1.438pt}}
\multiput(675.83,627.00)(0.485,1.022){10}{\rule{0.117pt}{2.361pt}}
\multiput(671.92,627.00)(9.000,14.099){2}{\rule{1.000pt}{1.181pt}}
\multiput(684.83,646.00)(0.485,0.962){10}{\rule{0.117pt}{2.250pt}}
\multiput(680.92,646.00)(9.000,13.330){2}{\rule{1.000pt}{1.125pt}}
\multiput(693.83,664.00)(0.481,1.020){8}{\rule{0.116pt}{2.375pt}}
\multiput(689.92,664.00)(8.000,12.071){2}{\rule{1.000pt}{1.188pt}}
\multiput(701.83,681.00)(0.481,0.883){8}{\rule{0.116pt}{2.125pt}}
\multiput(697.92,681.00)(8.000,10.589){2}{\rule{1.000pt}{1.063pt}}
\multiput(709.83,696.00)(0.485,0.723){10}{\rule{0.117pt}{1.806pt}}
\multiput(705.92,696.00)(9.000,10.252){2}{\rule{1.000pt}{0.903pt}}
\multiput(718.83,710.00)(0.481,0.677){8}{\rule{0.116pt}{1.750pt}}
\multiput(714.92,710.00)(8.000,8.368){2}{\rule{1.000pt}{0.875pt}}
\multiput(726.83,722.00)(0.481,0.608){8}{\rule{0.116pt}{1.625pt}}
\multiput(722.92,722.00)(8.000,7.627){2}{\rule{1.000pt}{0.813pt}}
\multiput(734.83,733.00)(0.481,0.470){8}{\rule{0.116pt}{1.375pt}}
\multiput(730.92,733.00)(8.000,6.146){2}{\rule{1.000pt}{0.688pt}}
\multiput(741.00,743.84)(0.444,0.475){6}{\rule{1.393pt}{0.114pt}}
\multiput(741.00,739.92)(5.109,7.000){2}{\rule{0.696pt}{1.000pt}}
\multiput(749.00,750.84)(0.373,0.462){4}{\rule{1.417pt}{0.111pt}}
\multiput(749.00,746.92)(4.060,6.000){2}{\rule{0.708pt}{1.000pt}}
\put(756,754.42){\rule{1.927pt}{1.000pt}}
\multiput(756.00,752.92)(4.000,3.000){2}{\rule{0.964pt}{1.000pt}}
\put(764,756.42){\rule{1.927pt}{1.000pt}}
\multiput(764.00,755.92)(4.000,1.000){2}{\rule{0.964pt}{1.000pt}}
\put(772,755.92){\rule{1.927pt}{1.000pt}}
\multiput(772.00,756.92)(4.000,-2.000){2}{\rule{0.964pt}{1.000pt}}
\put(780,752.92){\rule{1.686pt}{1.000pt}}
\multiput(780.00,754.92)(3.500,-4.000){2}{\rule{0.843pt}{1.000pt}}
\multiput(787.00,750.69)(0.444,-0.475){6}{\rule{1.393pt}{0.114pt}}
\multiput(787.00,750.92)(5.109,-7.000){2}{\rule{0.696pt}{1.000pt}}
\multiput(796.84,739.63)(0.475,-0.525){6}{\rule{0.114pt}{1.536pt}}
\multiput(792.92,742.81)(7.000,-5.813){2}{\rule{1.000pt}{0.768pt}}
\multiput(803.84,728.85)(0.475,-0.769){6}{\rule{0.114pt}{1.964pt}}
\multiput(799.92,732.92)(7.000,-7.923){2}{\rule{1.000pt}{0.982pt}}
\multiput(810.83,716.70)(0.481,-0.814){8}{\rule{0.116pt}{2.000pt}}
\multiput(806.92,720.85)(8.000,-9.849){2}{\rule{1.000pt}{1.000pt}}
\multiput(818.84,699.88)(0.475,-1.176){6}{\rule{0.114pt}{2.679pt}}
\multiput(814.92,705.44)(7.000,-11.440){2}{\rule{1.000pt}{1.339pt}}
\multiput(825.84,681.69)(0.475,-1.339){6}{\rule{0.114pt}{2.964pt}}
\multiput(821.92,687.85)(7.000,-12.847){2}{\rule{1.000pt}{1.482pt}}
\multiput(832.84,661.51)(0.475,-1.502){6}{\rule{0.114pt}{3.250pt}}
\multiput(828.92,668.25)(7.000,-14.254){2}{\rule{1.000pt}{1.625pt}}
\multiput(839.84,639.32)(0.475,-1.665){6}{\rule{0.114pt}{3.536pt}}
\multiput(835.92,646.66)(7.000,-15.661){2}{\rule{1.000pt}{1.768pt}}
\multiput(846.84,614.54)(0.475,-1.909){6}{\rule{0.114pt}{3.964pt}}
\multiput(842.92,622.77)(7.000,-17.772){2}{\rule{1.000pt}{1.982pt}}
\multiput(853.84,588.54)(0.475,-1.909){6}{\rule{0.114pt}{3.964pt}}
\multiput(849.92,596.77)(7.000,-17.772){2}{\rule{1.000pt}{1.982pt}}
\multiput(860.84,560.76)(0.475,-2.153){6}{\rule{0.114pt}{4.393pt}}
\multiput(856.92,569.88)(7.000,-19.882){2}{\rule{1.000pt}{2.196pt}}
\multiput(867.84,531.76)(0.475,-2.153){6}{\rule{0.114pt}{4.393pt}}
\multiput(863.92,540.88)(7.000,-19.882){2}{\rule{1.000pt}{2.196pt}}
\multiput(874.84,502.76)(0.475,-2.153){6}{\rule{0.114pt}{4.393pt}}
\multiput(870.92,511.88)(7.000,-19.882){2}{\rule{1.000pt}{2.196pt}}
\multiput(881.84,472.58)(0.475,-2.316){6}{\rule{0.114pt}{4.679pt}}
\multiput(877.92,482.29)(7.000,-21.289){2}{\rule{1.000pt}{2.339pt}}
\multiput(888.84,443.95)(0.475,-1.990){6}{\rule{0.114pt}{4.107pt}}
\multiput(884.92,452.48)(7.000,-18.475){2}{\rule{1.000pt}{2.054pt}}
\multiput(895.84,413.59)(0.462,-2.530){4}{\rule{0.111pt}{4.917pt}}
\multiput(891.92,423.80)(6.000,-17.795){2}{\rule{1.000pt}{2.458pt}}
\multiput(901.84,389.54)(0.475,-1.909){6}{\rule{0.114pt}{3.964pt}}
\multiput(897.92,397.77)(7.000,-17.772){2}{\rule{1.000pt}{1.982pt}}
\multiput(908.84,362.36)(0.462,-2.119){4}{\rule{0.111pt}{4.250pt}}
\multiput(904.92,371.18)(6.000,-15.179){2}{\rule{1.000pt}{2.125pt}}
\multiput(914.84,341.32)(0.475,-1.665){6}{\rule{0.114pt}{3.536pt}}
\multiput(910.92,348.66)(7.000,-15.661){2}{\rule{1.000pt}{1.768pt}}
\multiput(921.84,317.43)(0.462,-1.811){4}{\rule{0.111pt}{3.750pt}}
\multiput(917.92,325.22)(6.000,-13.217){2}{\rule{1.000pt}{1.875pt}}
\multiput(927.84,299.69)(0.475,-1.339){6}{\rule{0.114pt}{2.964pt}}
\multiput(923.92,305.85)(7.000,-12.847){2}{\rule{1.000pt}{1.482pt}}
\multiput(934.84,279.51)(0.462,-1.503){4}{\rule{0.111pt}{3.250pt}}
\multiput(930.92,286.25)(6.000,-11.254){2}{\rule{1.000pt}{1.625pt}}
\multiput(940.84,265.07)(0.475,-1.013){6}{\rule{0.114pt}{2.393pt}}
\multiput(936.92,270.03)(7.000,-10.034){2}{\rule{1.000pt}{1.196pt}}
\multiput(947.84,249.28)(0.462,-1.092){4}{\rule{0.111pt}{2.583pt}}
\multiput(943.92,254.64)(6.000,-8.638){2}{\rule{1.000pt}{1.292pt}}
\multiput(953.84,236.66)(0.462,-0.887){4}{\rule{0.111pt}{2.250pt}}
\multiput(949.92,241.33)(6.000,-7.330){2}{\rule{1.000pt}{1.125pt}}
\multiput(959.84,226.44)(0.475,-0.688){6}{\rule{0.114pt}{1.821pt}}
\multiput(955.92,230.22)(7.000,-7.220){2}{\rule{1.000pt}{0.911pt}}
\multiput(966.84,215.74)(0.462,-0.579){4}{\rule{0.111pt}{1.750pt}}
\multiput(962.92,219.37)(6.000,-5.368){2}{\rule{1.000pt}{0.875pt}}
\multiput(972.84,206.74)(0.462,-0.579){4}{\rule{0.111pt}{1.750pt}}
\multiput(968.92,210.37)(6.000,-5.368){2}{\rule{1.000pt}{0.875pt}}
\multiput(978.84,199.12)(0.462,-0.373){4}{\rule{0.111pt}{1.417pt}}
\multiput(974.92,202.06)(6.000,-4.060){2}{\rule{1.000pt}{0.708pt}}
\multiput(984.84,192.12)(0.462,-0.373){4}{\rule{0.111pt}{1.417pt}}
\multiput(980.92,195.06)(6.000,-4.060){2}{\rule{1.000pt}{0.708pt}}
\multiput(989.00,188.69)(0.270,-0.462){4}{\rule{1.250pt}{0.111pt}}
\multiput(989.00,188.92)(3.406,-6.000){2}{\rule{0.625pt}{1.000pt}}
\multiput(995.00,182.71)(0.151,-0.424){2}{\rule{1.650pt}{0.102pt}}
\multiput(995.00,182.92)(3.575,-5.000){2}{\rule{0.825pt}{1.000pt}}
\put(1002,175.42){\rule{1.445pt}{1.000pt}}
\multiput(1002.00,177.92)(3.000,-5.000){2}{\rule{0.723pt}{1.000pt}}
\put(1008,170.92){\rule{1.445pt}{1.000pt}}
\multiput(1008.00,172.92)(3.000,-4.000){2}{\rule{0.723pt}{1.000pt}}
\put(1014,166.92){\rule{1.445pt}{1.000pt}}
\multiput(1014.00,168.92)(3.000,-4.000){2}{\rule{0.723pt}{1.000pt}}
\put(1020,163.42){\rule{1.204pt}{1.000pt}}
\multiput(1020.00,164.92)(2.500,-3.000){2}{\rule{0.602pt}{1.000pt}}
\put(1025,160.42){\rule{1.445pt}{1.000pt}}
\multiput(1025.00,161.92)(3.000,-3.000){2}{\rule{0.723pt}{1.000pt}}
\put(1031,157.92){\rule{1.445pt}{1.000pt}}
\multiput(1031.00,158.92)(3.000,-2.000){2}{\rule{0.723pt}{1.000pt}}
\put(1037,155.42){\rule{1.445pt}{1.000pt}}
\multiput(1037.00,156.92)(3.000,-3.000){2}{\rule{0.723pt}{1.000pt}}
\put(1043,152.92){\rule{1.445pt}{1.000pt}}
\multiput(1043.00,153.92)(3.000,-2.000){2}{\rule{0.723pt}{1.000pt}}
\put(1049,151.42){\rule{1.445pt}{1.000pt}}
\multiput(1049.00,151.92)(3.000,-1.000){2}{\rule{0.723pt}{1.000pt}}
\put(1055,149.92){\rule{1.204pt}{1.000pt}}
\multiput(1055.00,150.92)(2.500,-2.000){2}{\rule{0.602pt}{1.000pt}}
\put(1060,148.42){\rule{1.445pt}{1.000pt}}
\multiput(1060.00,148.92)(3.000,-1.000){2}{\rule{0.723pt}{1.000pt}}
\put(1066,146.92){\rule{1.445pt}{1.000pt}}
\multiput(1066.00,147.92)(3.000,-2.000){2}{\rule{0.723pt}{1.000pt}}
\put(1072,145.42){\rule{1.204pt}{1.000pt}}
\multiput(1072.00,145.92)(2.500,-1.000){2}{\rule{0.602pt}{1.000pt}}
\put(1077,144.42){\rule{1.445pt}{1.000pt}}
\multiput(1077.00,144.92)(3.000,-1.000){2}{\rule{0.723pt}{1.000pt}}
\put(1083,143.42){\rule{1.445pt}{1.000pt}}
\multiput(1083.00,143.92)(3.000,-1.000){2}{\rule{0.723pt}{1.000pt}}
\put(1089,142.42){\rule{1.204pt}{1.000pt}}
\multiput(1089.00,142.92)(2.500,-1.000){2}{\rule{0.602pt}{1.000pt}}
\put(1094,141.42){\rule{1.445pt}{1.000pt}}
\multiput(1094.00,141.92)(3.000,-1.000){2}{\rule{0.723pt}{1.000pt}}
\put(1105,140.42){\rule{1.445pt}{1.000pt}}
\multiput(1105.00,140.92)(3.000,-1.000){2}{\rule{0.723pt}{1.000pt}}
\put(1111,139.42){\rule{1.204pt}{1.000pt}}
\multiput(1111.00,139.92)(2.500,-1.000){2}{\rule{0.602pt}{1.000pt}}
\put(1100.0,143.0){\rule[-0.500pt]{1.204pt}{1.000pt}}
\put(1122,138.42){\rule{1.204pt}{1.000pt}}
\multiput(1122.00,138.92)(2.500,-1.000){2}{\rule{0.602pt}{1.000pt}}
\put(1127,137.42){\rule{1.445pt}{1.000pt}}
\multiput(1127.00,137.92)(3.000,-1.000){2}{\rule{0.723pt}{1.000pt}}
\put(1116.0,141.0){\rule[-0.500pt]{1.445pt}{1.000pt}}
\put(1138,136.42){\rule{1.204pt}{1.000pt}}
\multiput(1138.00,136.92)(2.500,-1.000){2}{\rule{0.602pt}{1.000pt}}
\put(1143,135.42){\rule{1.445pt}{1.000pt}}
\multiput(1143.00,135.92)(3.000,-1.000){2}{\rule{0.723pt}{1.000pt}}
\put(1133.0,139.0){\rule[-0.500pt]{1.204pt}{1.000pt}}
\put(1154,134.42){\rule{1.204pt}{1.000pt}}
\multiput(1154.00,134.92)(2.500,-1.000){2}{\rule{0.602pt}{1.000pt}}
\put(1159,133.42){\rule{1.204pt}{1.000pt}}
\multiput(1159.00,133.92)(2.500,-1.000){2}{\rule{0.602pt}{1.000pt}}
\put(1164,132.42){\rule{1.445pt}{1.000pt}}
\multiput(1164.00,132.92)(3.000,-1.000){2}{\rule{0.723pt}{1.000pt}}
\put(1170,131.42){\rule{1.204pt}{1.000pt}}
\multiput(1170.00,131.92)(2.500,-1.000){2}{\rule{0.602pt}{1.000pt}}
\put(1175,130.42){\rule{1.204pt}{1.000pt}}
\multiput(1175.00,130.92)(2.500,-1.000){2}{\rule{0.602pt}{1.000pt}}
\put(1180,129.42){\rule{1.204pt}{1.000pt}}
\multiput(1180.00,129.92)(2.500,-1.000){2}{\rule{0.602pt}{1.000pt}}
\put(1185,127.92){\rule{1.204pt}{1.000pt}}
\multiput(1185.00,128.92)(2.500,-2.000){2}{\rule{0.602pt}{1.000pt}}
\put(1190,126.42){\rule{1.445pt}{1.000pt}}
\multiput(1190.00,126.92)(3.000,-1.000){2}{\rule{0.723pt}{1.000pt}}
\put(1196,125.42){\rule{1.204pt}{1.000pt}}
\multiput(1196.00,125.92)(2.500,-1.000){2}{\rule{0.602pt}{1.000pt}}
\put(1201,124.42){\rule{1.204pt}{1.000pt}}
\multiput(1201.00,124.92)(2.500,-1.000){2}{\rule{0.602pt}{1.000pt}}
\put(1206,122.92){\rule{1.204pt}{1.000pt}}
\multiput(1206.00,123.92)(2.500,-2.000){2}{\rule{0.602pt}{1.000pt}}
\put(1211,121.42){\rule{1.204pt}{1.000pt}}
\multiput(1211.00,121.92)(2.500,-1.000){2}{\rule{0.602pt}{1.000pt}}
\put(1216,119.92){\rule{1.204pt}{1.000pt}}
\multiput(1216.00,120.92)(2.500,-2.000){2}{\rule{0.602pt}{1.000pt}}
\put(1221,118.42){\rule{1.204pt}{1.000pt}}
\multiput(1221.00,118.92)(2.500,-1.000){2}{\rule{0.602pt}{1.000pt}}
\put(1226,117.42){\rule{1.204pt}{1.000pt}}
\multiput(1226.00,117.92)(2.500,-1.000){2}{\rule{0.602pt}{1.000pt}}
\put(1231,115.92){\rule{1.204pt}{1.000pt}}
\multiput(1231.00,116.92)(2.500,-2.000){2}{\rule{0.602pt}{1.000pt}}
\put(1236,114.42){\rule{1.204pt}{1.000pt}}
\multiput(1236.00,114.92)(2.500,-1.000){2}{\rule{0.602pt}{1.000pt}}
\put(1241,113.42){\rule{1.204pt}{1.000pt}}
\multiput(1241.00,113.92)(2.500,-1.000){2}{\rule{0.602pt}{1.000pt}}
\put(1149.0,137.0){\rule[-0.500pt]{1.204pt}{1.000pt}}
\end{picture}

\end{center}
\end{figure}

\section{$a_{1}$ dominance in $\tau\rightarrow f_{1}(1285)\pi\nu$}
$f_{1}$ meson is the chiral partner of $\omega$ meson[11] and
the mass formula of $f_{1}$ meson is derived in Ref.[11]
\begin{equation}
(1-{1\over2\pi^{2}g^{2}})m^{2}_{f_{1}}=6m^{2}+m^{2}_{\omega},
\;\;\;m_{f_{1}}=1.21GeV.
\end{equation}

The vertex of $f_{1}(1285)a_{1}\pi$ is presented in Ref.[11]
(a factor of -4 has been lost),
\begin{equation}
{\cal L}^{f_{1}a_{1}\pi}={1\over \pi^{2}f_{\pi}}{1\over g^{2}}
(1-{1\over2\pi^{2}g^{2}})^{-1}\varepsilon^{\mu\nu\alpha\beta}
f_{\mu}\partial_{\nu}\pi^{i}\partial_{\alpha}a^{i}_{\beta}.
\end{equation}
The narrow width of the decay $f_{1}\rightarrow\rho\pi\pi$ is
revealed from this vertex[11].
Using the substitution(6), the vertex ${\cal L}^{Wf_{1}\pi}$ is
derived. The vertex ${\cal L}^{f_{1}a_{1}\pi}$ has abnormal parity,
hence it belongs to WZW anomaly. Therefore, the WZW anomaly can
be tested in $\tau$ mesonic decay.
Only the axial-vector part of the weak interaction contributes to
the decay $\tau\rightarrow f_{1}\pi\nu$. Using the Lagrangian
${\cal L}^{A}$(10), it is obtained
\begin{equation}
<f_{1}\pi|\bar{\psi}\tau_{+}\gamma_{\mu}\gamma_{5}\psi|0>=
-\frac{1}{\sqrt{4\omega E}}
\frac{1}{\pi^{2}f_{\pi}g^{2}}(1-{1\over2\pi^{2}g^{2}})^{-1}
\frac{g^{2}f_{a}m^{2}_{\rho}-iqf^{-1}_{a}\Gamma_{a}(q^{2})}
{q^{2}-m^{2}_{a}+iq\Gamma_{a}(q^{2})}
\varepsilon^{\mu\nu\alpha\beta}k_{\nu}q_{\alpha}
\epsilon^{*\sigma}_{\beta},
\end{equation}
where k is the momentum of the pion and \(q=p+k\), p is the
momentum of $f_{1}$ meson.

The decay width is derived
\begin{eqnarray}
\lefteqn{\Gamma=\frac{G^{2}}{(2\pi)^{3}}
\frac{cos^{2}\theta_{C}}{128m^{3}_{\tau}}\int dq^{2}
{1\over q^{4}}
(m^{2}_{\tau}-q^{2})^{2}(m^{2}_{\tau}+2q^{2})(q^{2}-m^{2}_{f})^{3}}
\nonumber \\
&&\frac{f^{4}_{a}}{\pi^{4}f^{2}_{\pi}}
\frac{g^{4}f^{2}_{a}m^{4}_{\rho}+q^{2}f^{-2}_{a}\Gamma^{2}_{a}
(q^{2})}
{(q^{2}-m^{2}_{a})^{2}+q^{2}\Gamma^{2}_{a}(q^{2})}.
\end{eqnarray}
$a_{1}$ is dominant in this decay.
The theoretical
prediction of the branching
ratio of this decay is
\begin{equation}
B(\tau\rightarrow f_{1}\pi\nu)=2.91\times10^{-4}.
\end{equation}
The data is $(6.7\pm1.4\pm2.2)\times10^{-4}$[29].
Two factors result in the small branching ratio. Small phase space is
the first factor and the second factor is the anomalous coupling.
The effective theory proposed in Ref.[11] is a theory at low energies,
therefore, derivative expansion is exploited. In this theory the
anomalous couplings are at the fourth order
in derivatives. Comparing
with the couplings at the second order in derivatives, the anomalous
couplings are weaker. This is the reason
why the widths of $\rho$
and $a_{1}$ are broader(the two vertices are at the second order in
derivative expansion) and $\omega$ and $f_{1}\rightarrow\rho\pi\pi$
are narrower.

The distribution of the invariant mass of $f_{1}\pi$ is shown in Fig.2.
The peak of the distribution is resulted by both the effects of the
threshold and the $a_{1}$ resonance.
\begin{figure}
\begin{center}

\setlength{\unitlength}{0.240900pt}
\ifx\plotpoint\undefined\newsavebox{\plotpoint}\fi
\sbox{\plotpoint}{\rule[-0.500pt]{1.000pt}{1.000pt}}%
\begin{picture}(1500,900)(0,0)
\font\gnuplot=cmr10 at 10pt
\gnuplot
\sbox{\plotpoint}{\rule[-0.500pt]{1.000pt}{1.000pt}}%
\put(220.0,113.0){\rule[-0.500pt]{292.934pt}{1.000pt}}
\put(220.0,113.0){\rule[-0.500pt]{4.818pt}{1.000pt}}
\put(198,113){\makebox(0,0)[r]{0}}
\put(1416.0,113.0){\rule[-0.500pt]{4.818pt}{1.000pt}}
\put(220.0,240.0){\rule[-0.500pt]{4.818pt}{1.000pt}}
\put(198,240){\makebox(0,0)[r]{5}}
\put(1416.0,240.0){\rule[-0.500pt]{4.818pt}{1.000pt}}
\put(220.0,368.0){\rule[-0.500pt]{4.818pt}{1.000pt}}
\put(198,368){\makebox(0,0)[r]{10}}
\put(1416.0,368.0){\rule[-0.500pt]{4.818pt}{1.000pt}}
\put(220.0,495.0){\rule[-0.500pt]{4.818pt}{1.000pt}}
\put(198,495){\makebox(0,0)[r]{15}}
\put(1416.0,495.0){\rule[-0.500pt]{4.818pt}{1.000pt}}
\put(220.0,622.0){\rule[-0.500pt]{4.818pt}{1.000pt}}
\put(198,622){\makebox(0,0)[r]{20}}
\put(1416.0,622.0){\rule[-0.500pt]{4.818pt}{1.000pt}}
\put(220.0,750.0){\rule[-0.500pt]{4.818pt}{1.000pt}}
\put(198,750){\makebox(0,0)[r]{25}}
\put(1416.0,750.0){\rule[-0.500pt]{4.818pt}{1.000pt}}
\put(220.0,877.0){\rule[-0.500pt]{4.818pt}{1.000pt}}
\put(198,877){\makebox(0,0)[r]{30}}
\put(1416.0,877.0){\rule[-0.500pt]{4.818pt}{1.000pt}}
\put(220.0,113.0){\rule[-0.500pt]{1.000pt}{4.818pt}}
\put(220,68){\makebox(0,0){1.2}}
\put(220.0,857.0){\rule[-0.500pt]{1.000pt}{4.818pt}}
\put(372.0,113.0){\rule[-0.500pt]{1.000pt}{4.818pt}}
\put(372,68){\makebox(0,0){1.3}}
\put(372.0,857.0){\rule[-0.500pt]{1.000pt}{4.818pt}}
\put(524.0,113.0){\rule[-0.500pt]{1.000pt}{4.818pt}}
\put(524,68){\makebox(0,0){1.4}}
\put(524.0,857.0){\rule[-0.500pt]{1.000pt}{4.818pt}}
\put(676.0,113.0){\rule[-0.500pt]{1.000pt}{4.818pt}}
\put(676,68){\makebox(0,0){1.5}}
\put(676.0,857.0){\rule[-0.500pt]{1.000pt}{4.818pt}}
\put(828.0,113.0){\rule[-0.500pt]{1.000pt}{4.818pt}}
\put(828,68){\makebox(0,0){1.6}}
\put(828.0,857.0){\rule[-0.500pt]{1.000pt}{4.818pt}}
\put(980.0,113.0){\rule[-0.500pt]{1.000pt}{4.818pt}}
\put(980,68){\makebox(0,0){1.7}}
\put(980.0,857.0){\rule[-0.500pt]{1.000pt}{4.818pt}}
\put(1132.0,113.0){\rule[-0.500pt]{1.000pt}{4.818pt}}
\put(1132,68){\makebox(0,0){1.8}}
\put(1132.0,857.0){\rule[-0.500pt]{1.000pt}{4.818pt}}
\put(1284.0,113.0){\rule[-0.500pt]{1.000pt}{4.818pt}}
\put(1284,68){\makebox(0,0){1.9}}
\put(1284.0,857.0){\rule[-0.500pt]{1.000pt}{4.818pt}}
\put(1436.0,113.0){\rule[-0.500pt]{1.000pt}{4.818pt}}
\put(1436,68){\makebox(0,0){2}}
\put(1436.0,857.0){\rule[-0.500pt]{1.000pt}{4.818pt}}
\put(220.0,113.0){\rule[-0.500pt]{292.934pt}{1.000pt}}
\put(1436.0,113.0){\rule[-0.500pt]{1.000pt}{184.048pt}}
\put(220.0,877.0){\rule[-0.500pt]{292.934pt}{1.000pt}}
\put(45,495){\makebox(0,0){${d\Gamma\over d\sqrt{q^{2}}}\times 10^{16}$ }}
\put(828,23){\makebox(0,0){FiG.2        $\sqrt{q^{2}}$        GeV}}
\put(220.0,113.0){\rule[-0.500pt]{1.000pt}{184.048pt}}
\put(561,362){\usebox{\plotpoint}}
\multiput(562.84,362.00)(0.462,1.092){4}{\rule{0.111pt}{2.583pt}}
\multiput(558.92,362.00)(6.000,8.638){2}{\rule{1.000pt}{1.292pt}}
\multiput(568.84,376.00)(0.462,1.298){4}{\rule{0.111pt}{2.917pt}}
\multiput(564.92,376.00)(6.000,9.946){2}{\rule{1.000pt}{1.458pt}}
\multiput(574.84,392.00)(0.462,1.298){4}{\rule{0.111pt}{2.917pt}}
\multiput(570.92,392.00)(6.000,9.946){2}{\rule{1.000pt}{1.458pt}}
\multiput(580.84,408.00)(0.462,1.298){4}{\rule{0.111pt}{2.917pt}}
\multiput(576.92,408.00)(6.000,9.946){2}{\rule{1.000pt}{1.458pt}}
\multiput(586.84,424.00)(0.462,1.400){4}{\rule{0.111pt}{3.083pt}}
\multiput(582.92,424.00)(6.000,10.600){2}{\rule{1.000pt}{1.542pt}}
\multiput(592.84,441.00)(0.462,1.298){4}{\rule{0.111pt}{2.917pt}}
\multiput(588.92,441.00)(6.000,9.946){2}{\rule{1.000pt}{1.458pt}}
\multiput(598.84,457.00)(0.462,1.400){4}{\rule{0.111pt}{3.083pt}}
\multiput(594.92,457.00)(6.000,10.600){2}{\rule{1.000pt}{1.542pt}}
\multiput(604.84,474.00)(0.462,1.298){4}{\rule{0.111pt}{2.917pt}}
\multiput(600.92,474.00)(6.000,9.946){2}{\rule{1.000pt}{1.458pt}}
\multiput(610.84,490.00)(0.462,1.298){4}{\rule{0.111pt}{2.917pt}}
\multiput(606.92,490.00)(6.000,9.946){2}{\rule{1.000pt}{1.458pt}}
\multiput(616.84,506.00)(0.462,1.400){4}{\rule{0.111pt}{3.083pt}}
\multiput(612.92,506.00)(6.000,10.600){2}{\rule{1.000pt}{1.542pt}}
\multiput(622.84,523.00)(0.462,1.298){4}{\rule{0.111pt}{2.917pt}}
\multiput(618.92,523.00)(6.000,9.946){2}{\rule{1.000pt}{1.458pt}}
\multiput(628.84,539.00)(0.462,1.298){4}{\rule{0.111pt}{2.917pt}}
\multiput(624.92,539.00)(6.000,9.946){2}{\rule{1.000pt}{1.458pt}}
\multiput(634.84,555.00)(0.462,1.298){4}{\rule{0.111pt}{2.917pt}}
\multiput(630.92,555.00)(6.000,9.946){2}{\rule{1.000pt}{1.458pt}}
\multiput(640.86,571.00)(0.424,1.679){2}{\rule{0.102pt}{3.450pt}}
\multiput(636.92,571.00)(5.000,8.839){2}{\rule{1.000pt}{1.725pt}}
\multiput(645.84,587.00)(0.462,1.298){4}{\rule{0.111pt}{2.917pt}}
\multiput(641.92,587.00)(6.000,9.946){2}{\rule{1.000pt}{1.458pt}}
\multiput(651.84,603.00)(0.462,1.195){4}{\rule{0.111pt}{2.750pt}}
\multiput(647.92,603.00)(6.000,9.292){2}{\rule{1.000pt}{1.375pt}}
\multiput(657.84,618.00)(0.462,1.298){4}{\rule{0.111pt}{2.917pt}}
\multiput(653.92,618.00)(6.000,9.946){2}{\rule{1.000pt}{1.458pt}}
\multiput(663.86,634.00)(0.424,1.339){2}{\rule{0.102pt}{3.050pt}}
\multiput(659.92,634.00)(5.000,7.670){2}{\rule{1.000pt}{1.525pt}}
\multiput(668.84,648.00)(0.462,1.195){4}{\rule{0.111pt}{2.750pt}}
\multiput(664.92,648.00)(6.000,9.292){2}{\rule{1.000pt}{1.375pt}}
\multiput(674.84,663.00)(0.462,1.092){4}{\rule{0.111pt}{2.583pt}}
\multiput(670.92,663.00)(6.000,8.638){2}{\rule{1.000pt}{1.292pt}}
\multiput(680.84,677.00)(0.462,1.092){4}{\rule{0.111pt}{2.583pt}}
\multiput(676.92,677.00)(6.000,8.638){2}{\rule{1.000pt}{1.292pt}}
\multiput(686.86,691.00)(0.424,1.169){2}{\rule{0.102pt}{2.850pt}}
\multiput(682.92,691.00)(5.000,7.085){2}{\rule{1.000pt}{1.425pt}}
\multiput(691.84,704.00)(0.462,0.989){4}{\rule{0.111pt}{2.417pt}}
\multiput(687.92,704.00)(6.000,7.984){2}{\rule{1.000pt}{1.208pt}}
\multiput(697.84,717.00)(0.462,0.887){4}{\rule{0.111pt}{2.250pt}}
\multiput(693.92,717.00)(6.000,7.330){2}{\rule{1.000pt}{1.125pt}}
\multiput(703.86,729.00)(0.424,1.000){2}{\rule{0.102pt}{2.650pt}}
\multiput(699.92,729.00)(5.000,6.500){2}{\rule{1.000pt}{1.325pt}}
\multiput(708.84,741.00)(0.462,0.784){4}{\rule{0.111pt}{2.083pt}}
\multiput(704.92,741.00)(6.000,6.676){2}{\rule{1.000pt}{1.042pt}}
\multiput(714.84,752.00)(0.462,0.784){4}{\rule{0.111pt}{2.083pt}}
\multiput(710.92,752.00)(6.000,6.676){2}{\rule{1.000pt}{1.042pt}}
\multiput(720.86,763.00)(0.424,0.660){2}{\rule{0.102pt}{2.250pt}}
\multiput(716.92,763.00)(5.000,5.330){2}{\rule{1.000pt}{1.125pt}}
\multiput(725.84,773.00)(0.462,0.579){4}{\rule{0.111pt}{1.750pt}}
\multiput(721.92,773.00)(6.000,5.368){2}{\rule{1.000pt}{0.875pt}}
\multiput(731.84,782.00)(0.462,0.579){4}{\rule{0.111pt}{1.750pt}}
\multiput(727.92,782.00)(6.000,5.368){2}{\rule{1.000pt}{0.875pt}}
\multiput(737.86,791.00)(0.424,0.320){2}{\rule{0.102pt}{1.850pt}}
\multiput(733.92,791.00)(5.000,4.160){2}{\rule{1.000pt}{0.925pt}}
\multiput(742.84,799.00)(0.462,0.476){4}{\rule{0.111pt}{1.583pt}}
\multiput(738.92,799.00)(6.000,4.714){2}{\rule{1.000pt}{0.792pt}}
\multiput(748.86,807.00)(0.424,0.151){2}{\rule{0.102pt}{1.650pt}}
\multiput(744.92,807.00)(5.000,3.575){2}{\rule{1.000pt}{0.825pt}}
\multiput(752.00,815.84)(0.270,0.462){4}{\rule{1.250pt}{0.111pt}}
\multiput(752.00,811.92)(3.406,6.000){2}{\rule{0.625pt}{1.000pt}}
\put(758,820.42){\rule{1.204pt}{1.000pt}}
\multiput(758.00,817.92)(2.500,5.000){2}{\rule{0.602pt}{1.000pt}}
\put(763,825.42){\rule{1.445pt}{1.000pt}}
\multiput(763.00,822.92)(3.000,5.000){2}{\rule{0.723pt}{1.000pt}}
\put(769,829.92){\rule{1.204pt}{1.000pt}}
\multiput(769.00,827.92)(2.500,4.000){2}{\rule{0.602pt}{1.000pt}}
\put(774,833.42){\rule{1.445pt}{1.000pt}}
\multiput(774.00,831.92)(3.000,3.000){2}{\rule{0.723pt}{1.000pt}}
\put(780,836.42){\rule{1.204pt}{1.000pt}}
\multiput(780.00,834.92)(2.500,3.000){2}{\rule{0.602pt}{1.000pt}}
\put(785,838.42){\rule{1.445pt}{1.000pt}}
\multiput(785.00,837.92)(3.000,1.000){2}{\rule{0.723pt}{1.000pt}}
\put(791,839.42){\rule{1.204pt}{1.000pt}}
\multiput(791.00,838.92)(2.500,1.000){2}{\rule{0.602pt}{1.000pt}}
\put(802,839.42){\rule{1.204pt}{1.000pt}}
\multiput(802.00,839.92)(2.500,-1.000){2}{\rule{0.602pt}{1.000pt}}
\put(807,838.42){\rule{1.445pt}{1.000pt}}
\multiput(807.00,838.92)(3.000,-1.000){2}{\rule{0.723pt}{1.000pt}}
\put(813,836.42){\rule{1.204pt}{1.000pt}}
\multiput(813.00,837.92)(2.500,-3.000){2}{\rule{0.602pt}{1.000pt}}
\put(818,833.42){\rule{1.204pt}{1.000pt}}
\multiput(818.00,834.92)(2.500,-3.000){2}{\rule{0.602pt}{1.000pt}}
\put(823,829.92){\rule{1.445pt}{1.000pt}}
\multiput(823.00,831.92)(3.000,-4.000){2}{\rule{0.723pt}{1.000pt}}
\put(829,825.42){\rule{1.204pt}{1.000pt}}
\multiput(829.00,827.92)(2.500,-5.000){2}{\rule{0.602pt}{1.000pt}}
\put(834,820.42){\rule{1.204pt}{1.000pt}}
\multiput(834.00,822.92)(2.500,-5.000){2}{\rule{0.602pt}{1.000pt}}
\multiput(839.00,817.69)(0.270,-0.462){4}{\rule{1.250pt}{0.111pt}}
\multiput(839.00,817.92)(3.406,-6.000){2}{\rule{0.625pt}{1.000pt}}
\multiput(846.86,807.15)(0.424,-0.151){2}{\rule{0.102pt}{1.650pt}}
\multiput(842.92,810.58)(5.000,-3.575){2}{\rule{1.000pt}{0.825pt}}
\multiput(851.86,799.32)(0.424,-0.320){2}{\rule{0.102pt}{1.850pt}}
\multiput(847.92,803.16)(5.000,-4.160){2}{\rule{1.000pt}{0.925pt}}
\multiput(856.84,791.74)(0.462,-0.579){4}{\rule{0.111pt}{1.750pt}}
\multiput(852.92,795.37)(6.000,-5.368){2}{\rule{1.000pt}{0.875pt}}
\multiput(862.86,781.49)(0.424,-0.490){2}{\rule{0.102pt}{2.050pt}}
\multiput(858.92,785.75)(5.000,-4.745){2}{\rule{1.000pt}{1.025pt}}
\multiput(867.86,770.83)(0.424,-0.830){2}{\rule{0.102pt}{2.450pt}}
\multiput(863.92,775.91)(5.000,-5.915){2}{\rule{1.000pt}{1.225pt}}
\multiput(872.84,762.04)(0.462,-0.681){4}{\rule{0.111pt}{1.917pt}}
\multiput(868.92,766.02)(6.000,-6.022){2}{\rule{1.000pt}{0.958pt}}
\multiput(878.86,749.00)(0.424,-1.000){2}{\rule{0.102pt}{2.650pt}}
\multiput(874.92,754.50)(5.000,-6.500){2}{\rule{1.000pt}{1.325pt}}
\multiput(883.86,737.00)(0.424,-1.000){2}{\rule{0.102pt}{2.650pt}}
\multiput(879.92,742.50)(5.000,-6.500){2}{\rule{1.000pt}{1.325pt}}
\multiput(888.86,724.17)(0.424,-1.169){2}{\rule{0.102pt}{2.850pt}}
\multiput(884.92,730.08)(5.000,-7.085){2}{\rule{1.000pt}{1.425pt}}
\multiput(893.84,712.28)(0.462,-1.092){4}{\rule{0.111pt}{2.583pt}}
\multiput(889.92,717.64)(6.000,-8.638){2}{\rule{1.000pt}{1.292pt}}
\multiput(899.86,696.34)(0.424,-1.339){2}{\rule{0.102pt}{3.050pt}}
\multiput(895.92,702.67)(5.000,-7.670){2}{\rule{1.000pt}{1.525pt}}
\multiput(904.86,682.34)(0.424,-1.339){2}{\rule{0.102pt}{3.050pt}}
\multiput(900.92,688.67)(5.000,-7.670){2}{\rule{1.000pt}{1.525pt}}
\multiput(909.86,666.68)(0.424,-1.679){2}{\rule{0.102pt}{3.450pt}}
\multiput(905.92,673.84)(5.000,-8.839){2}{\rule{1.000pt}{1.725pt}}
\multiput(914.86,650.68)(0.424,-1.679){2}{\rule{0.102pt}{3.450pt}}
\multiput(910.92,657.84)(5.000,-8.839){2}{\rule{1.000pt}{1.725pt}}
\multiput(919.84,636.89)(0.462,-1.298){4}{\rule{0.111pt}{2.917pt}}
\multiput(915.92,642.95)(6.000,-9.946){2}{\rule{1.000pt}{1.458pt}}
\multiput(925.86,617.85)(0.424,-1.848){2}{\rule{0.102pt}{3.650pt}}
\multiput(921.92,625.42)(5.000,-9.424){2}{\rule{1.000pt}{1.825pt}}
\multiput(930.86,600.85)(0.424,-1.848){2}{\rule{0.102pt}{3.650pt}}
\multiput(926.92,608.42)(5.000,-9.424){2}{\rule{1.000pt}{1.825pt}}
\multiput(935.86,583.02)(0.424,-2.018){2}{\rule{0.102pt}{3.850pt}}
\multiput(931.92,591.01)(5.000,-10.009){2}{\rule{1.000pt}{1.925pt}}
\multiput(940.86,565.02)(0.424,-2.018){2}{\rule{0.102pt}{3.850pt}}
\multiput(936.92,573.01)(5.000,-10.009){2}{\rule{1.000pt}{1.925pt}}
\multiput(945.86,547.02)(0.424,-2.018){2}{\rule{0.102pt}{3.850pt}}
\multiput(941.92,555.01)(5.000,-10.009){2}{\rule{1.000pt}{1.925pt}}
\multiput(950.86,528.19)(0.424,-2.188){2}{\rule{0.102pt}{4.050pt}}
\multiput(946.92,536.59)(5.000,-10.594){2}{\rule{1.000pt}{2.025pt}}
\multiput(955.84,512.51)(0.462,-1.503){4}{\rule{0.111pt}{3.250pt}}
\multiput(951.92,519.25)(6.000,-11.254){2}{\rule{1.000pt}{1.625pt}}
\multiput(961.86,491.19)(0.424,-2.188){2}{\rule{0.102pt}{4.050pt}}
\multiput(957.92,499.59)(5.000,-10.594){2}{\rule{1.000pt}{2.025pt}}
\multiput(966.86,471.36)(0.424,-2.358){2}{\rule{0.102pt}{4.250pt}}
\multiput(962.92,480.18)(5.000,-11.179){2}{\rule{1.000pt}{2.125pt}}
\multiput(971.86,452.19)(0.424,-2.188){2}{\rule{0.102pt}{4.050pt}}
\multiput(967.92,460.59)(5.000,-10.594){2}{\rule{1.000pt}{2.025pt}}
\multiput(976.86,433.19)(0.424,-2.188){2}{\rule{0.102pt}{4.050pt}}
\multiput(972.92,441.59)(5.000,-10.594){2}{\rule{1.000pt}{2.025pt}}
\multiput(981.86,413.36)(0.424,-2.358){2}{\rule{0.102pt}{4.250pt}}
\multiput(977.92,422.18)(5.000,-11.179){2}{\rule{1.000pt}{2.125pt}}
\multiput(986.86,394.19)(0.424,-2.188){2}{\rule{0.102pt}{4.050pt}}
\multiput(982.92,402.59)(5.000,-10.594){2}{\rule{1.000pt}{2.025pt}}
\multiput(991.86,375.19)(0.424,-2.188){2}{\rule{0.102pt}{4.050pt}}
\multiput(987.92,383.59)(5.000,-10.594){2}{\rule{1.000pt}{2.025pt}}
\multiput(996.86,356.19)(0.424,-2.188){2}{\rule{0.102pt}{4.050pt}}
\multiput(992.92,364.59)(5.000,-10.594){2}{\rule{1.000pt}{2.025pt}}
\multiput(1001.86,337.19)(0.424,-2.188){2}{\rule{0.102pt}{4.050pt}}
\multiput(997.92,345.59)(5.000,-10.594){2}{\rule{1.000pt}{2.025pt}}
\multiput(1006.86,319.02)(0.424,-2.018){2}{\rule{0.102pt}{3.850pt}}
\multiput(1002.92,327.01)(5.000,-10.009){2}{\rule{1.000pt}{1.925pt}}
\multiput(1011.86,300.19)(0.424,-2.188){2}{\rule{0.102pt}{4.050pt}}
\multiput(1007.92,308.59)(5.000,-10.594){2}{\rule{1.000pt}{2.025pt}}
\multiput(1016.86,282.85)(0.424,-1.848){2}{\rule{0.102pt}{3.650pt}}
\multiput(1012.92,290.42)(5.000,-9.424){2}{\rule{1.000pt}{1.825pt}}
\multiput(1021.86,265.02)(0.424,-2.018){2}{\rule{0.102pt}{3.850pt}}
\multiput(1017.92,273.01)(5.000,-10.009){2}{\rule{1.000pt}{1.925pt}}
\multiput(1026.86,248.68)(0.424,-1.679){2}{\rule{0.102pt}{3.450pt}}
\multiput(1022.92,255.84)(5.000,-8.839){2}{\rule{1.000pt}{1.725pt}}
\multiput(1031.86,232.68)(0.424,-1.679){2}{\rule{0.102pt}{3.450pt}}
\multiput(1027.92,239.84)(5.000,-8.839){2}{\rule{1.000pt}{1.725pt}}
\multiput(1036.86,216.68)(0.424,-1.679){2}{\rule{0.102pt}{3.450pt}}
\multiput(1032.92,223.84)(5.000,-8.839){2}{\rule{1.000pt}{1.725pt}}
\multiput(1041.86,202.34)(0.424,-1.339){2}{\rule{0.102pt}{3.050pt}}
\multiput(1037.92,208.67)(5.000,-7.670){2}{\rule{1.000pt}{1.525pt}}
\multiput(1046.86,188.34)(0.424,-1.339){2}{\rule{0.102pt}{3.050pt}}
\multiput(1042.92,194.67)(5.000,-7.670){2}{\rule{1.000pt}{1.525pt}}
\multiput(1051.86,175.17)(0.424,-1.169){2}{\rule{0.102pt}{2.850pt}}
\multiput(1047.92,181.08)(5.000,-7.085){2}{\rule{1.000pt}{1.425pt}}
\multiput(1056.86,163.00)(0.424,-1.000){2}{\rule{0.102pt}{2.650pt}}
\multiput(1052.92,168.50)(5.000,-6.500){2}{\rule{1.000pt}{1.325pt}}
\put(1059.92,151){\rule{1.000pt}{2.650pt}}
\multiput(1057.92,156.50)(4.000,-5.500){2}{\rule{1.000pt}{1.325pt}}
\multiput(1065.86,142.49)(0.424,-0.490){2}{\rule{0.102pt}{2.050pt}}
\multiput(1061.92,146.75)(5.000,-4.745){2}{\rule{1.000pt}{1.025pt}}
\multiput(1070.86,133.49)(0.424,-0.490){2}{\rule{0.102pt}{2.050pt}}
\multiput(1066.92,137.75)(5.000,-4.745){2}{\rule{1.000pt}{1.025pt}}
\multiput(1075.86,126.15)(0.424,-0.151){2}{\rule{0.102pt}{1.650pt}}
\multiput(1071.92,129.58)(5.000,-3.575){2}{\rule{1.000pt}{0.825pt}}
\put(1079,121.42){\rule{1.204pt}{1.000pt}}
\multiput(1079.00,123.92)(2.500,-5.000){2}{\rule{0.602pt}{1.000pt}}
\put(1084,116.42){\rule{1.204pt}{1.000pt}}
\multiput(1084.00,118.92)(2.500,-5.000){2}{\rule{0.602pt}{1.000pt}}
\put(1089,112.92){\rule{1.204pt}{1.000pt}}
\multiput(1089.00,113.92)(2.500,-2.000){2}{\rule{0.602pt}{1.000pt}}
\put(796.0,842.0){\rule[-0.500pt]{1.445pt}{1.000pt}}
\end{picture}

\end{center}
\end{figure}

The experimental measurement of this decay is a test of the
Wess-Zumino-Witten anomaly and the mechanism proposed in this paper.

\section{$\tau\rightarrow K^{*}(892)K\nu$}
The processes $\tau\rightarrow KK\pi\nu$ have been studied by many
authors. The earlist study is done by using a chiral Lagrangian
[3]. In Ref.[17] a chiral Lagrangian and three resonances
$\rho$, $\rho'$ and $\rho''$ are used. In Ref.[4] the $\rho$
meson field is introduced to a chiral Lagrangian. In Refs.[5]
a comprehensive resonace model including vector and
axial-vector resonances has been exploited.
In this paper the process $\tau\rightarrow K^{*}K\nu$ is studied
in terms of the same formulas(5,10) used to study $\tau\rightarrow
3\pi\nu$ and $f_{1}\pi\nu$.

Both the vector and the axial-vector currents of the weak interactions
contribute to the decay $\tau^{-}\rightarrow(K^{*}K)^{-}\nu$. The
vector part comes from the anomaly
and the vertex is presented in Ref.[11]
\begin{equation}
{\cal L}^{K^{*}K\rho}=-\frac{N_{C}}{\pi^{2}g^{2}f_{\pi}}\varepsilon
^{\mu\nu\alpha\beta}d_{abi}K^{a}_{\mu}\partial_{\nu}\rho^{i}_{
\alpha}\partial_{\beta}K^{b},
\end{equation}
${\cal L}^{K^{*}KW}$ can be derived by using the substitution(4).
There are three vertices in the axial-vector part: $\pi K^{*}K$,
$a_{1}K^{*}K$, and $WK^{*}K$. As mentioned above, the later can
be derived by using the substitution(6) in the vertex $a_{1}K^{*}K$.
The vertex ${\cal L}^{\pi K^{*}K}$ is given in Ref.[11]
\begin{eqnarray}
\lefteqn{{\cal L}^{\pi K^{*}K}=if_{K^{*}K\pi}(q^{2})
\{-{1\over\sqrt{2}}
K^{0}_{\mu}(\pi^{+}\partial^{\mu}K^{-}-K^{-}\partial^{\mu}\pi^{+})
-{1\over\sqrt{2}}K^{+}_{\mu}(\pi^{-}\partial^{\mu}\bar{K}^{0}
-\bar{K}^{0}\partial^{\mu}\pi^{-})}\nonumber \\
&&+{1\over2}K^{0}_{\mu}(\pi^{0}\partial^{\mu}\bar{K}^{0}-\bar{K}^{0}
\partial^{\mu}\pi^{0})-{1\over2}K^{+}_{\mu}(\pi^{0}\partial^{\mu}
K^{-}-K^{-}\partial^{\mu}\pi^{0})\}+h.c. ,
\end{eqnarray}
where $q^{2}$ is the momentum squared of $K^{*}$.
In the limit of \(m_{q}=0\), $f_{K^{*}K\pi}(q^{2})$ is the same as
$f_{\rho\pi\pi}(q^{2})$(37).
The vertex $a_{1}K^{*}K$ is derived from Ref.[11]
\begin{equation}
{\cal L}^{a_{1}K^{*}K}=f_{abi}K^{a}_{\mu}K^{b}a^{i}_{\nu}
\{A(q^{2})_{K^{*}}g_{\mu\nu}+Bk_{\mu}k_{\nu}\}
-f_{abi}DK^{a}_{\mu}\partial^{\mu}K^{b}\partial^{\nu}a^{i}_{\nu},
\end{equation}
where $A(q^{2})_{K^{*}}$ is defined in Eq.(76).
In the limit of \(m_{q}=0\), A, B, and D are the same as (27,29,30).
The vector matrix element is obtained from ${\cal L}^{V}$(3) and the
vertex(47)
\begin{equation}
<K^{-}K^{*0}|\bar{\psi}\tau_{+}\gamma_{\mu}\psi|0>=-{1\over
\sqrt{4\omega E}}\frac{N_{C}}{\sqrt{2}\pi^{2}gf_{\pi}}
\frac{m^{2}_{\rho}-i\sqrt{q^{2}}\Gamma_{\rho}(q^{2})}
{q^{2}-m^{2}_{\rho}+i\sqrt{q^{2}}\Gamma_{\rho}(q^{2})}
\varepsilon^{\mu\nu\alpha\beta}\epsilon^{*\sigma}_{\nu}q_{\alpha}
k_{\beta},
\end{equation}
where k is the momentum of kaon and \(q=p+k\), p is the momentum of
$K^{*}$.
Because of $q^{2}>4m^{2}_{K}$, the decay mode of $\rho\rightarrow
K\bar{K}$ is open and we have
\begin{eqnarray}
\lefteqn{\Gamma(\rho\rightarrow\pi\pi)=\frac{f^{2}_{\rho\pi\pi}
(q^{2})}{48\pi}{q^{2}\over m_{\rho}}(1-{4m^{2}_{\pi}\over q^{2}})^
{{3\over2}},}\nonumber \\
&&\Gamma(\rho\rightarrow K\bar{K})=\frac{f^{2}_{\rho\pi\pi}
(q^{2})}{96\pi}{q^{2}\over m_{\rho}}(1-{4m^{2}_{K}\over q^{2}})^
{{3\over2}},\nonumber \\
&&\Gamma_{\rho}(q^{2})=\Gamma(\rho\rightarrow\pi\pi)+
\Gamma(\rho\rightarrow K\bar{K}).
\end{eqnarray}

The axial-vector matrix element is derived by using ${\cal L}
^{K^{*}K\pi}$(48), ${\cal L}^{a_{1}K^{*}K}$(49),
${\cal L}^{WK^{*}K}$, and ${\cal L}^{A}$(10)
\begin{eqnarray}
\lefteqn{<K^{-}K^{*0}|\bar{\psi}\tau_{-}\gamma_{\mu}\gamma_{5}\psi|0>
=-{i\over
\sqrt{4\omega E}}
{1\over\sqrt{2}}
\frac{g^{2}f_{a}m^{2}_{\rho}-i\sqrt{q^{2}}f^{-1}_{a}\Gamma_{a}(q^{2})}
{q^{2}-m^{2}_{a}+im_{a}\Gamma_{a}(q^{2})}}\nonumber \\
&&({q_{\mu}q_{\nu}\over q^{2}}-g_{\mu\nu})(A(q^{2})_{K^{*}}g_{\nu
\lambda}+Bk_{\nu}k_{\lambda}-Dk^{\nu}p^{\lambda})\epsilon^{*\lambda}
_{\sigma}.
\end{eqnarray}
The decay width is derived
\begin{eqnarray}
\lefteqn{{d\Gamma\over dq^{2}}(\tau^{-}\rightarrow K^{*0}K^{-}\nu)=
\frac{G^{2}cos^{2}\theta_{C}}{64m^{3}_{\tau}q^{4}}{1\over(2\pi)^{3}}
[(q^{2}+m^{2}_{K^{*}}-m^{2}_{K})^{2}
-4q^{2}m^{2}_{K^{*}}]^{{1\over2}}
(m^{2}_{\tau}-q^{2})^{2}
}\nonumber \\
&&(m^{2}_{\tau}+2q^{2})
\{\frac{3}{\pi^{4}g^{2}f^{2}_{\pi}}\frac{m^{4}_{\rho}+q^{2}
\Gamma^{2}_{\rho}(q^{2})
}{(q^{2}-m^{2}_{\rho})^{2}+q^{2}\Gamma^{2}_{\rho}(q^{2})}
[(p\cdot q)^{2}-q^{2}m^{2}_{K^{*}}]\nonumber \\
&&+{1\over 2}\frac{g^{4}f^{2}_{a}m^{4}_{\rho}+f^{-2}_{a}
q^{2}\Gamma^{2}_{a}(q^{2})}
{(q^{2}-m^{2}_{a}
)^{2}+q^{2}\Gamma^{2}_{a}(q^{2})}[A^{2}(q^{2})_{K^{*}}-{1\over3}
\frac{(q\cdot k)^{2}}{q^{2}}(2A(q^{2}_{K^{*}}B-m^{2}_{K^{*}}D^{2})]\},
\end{eqnarray}
The distribution of ${d\Gamma\over dq}$ is shown in Fig.3.
There is a peak located at 1.51GeV
which is caused by both the threshold and resonance effects.
The branching ratio is computed to be
\[B(\tau\rightarrow K^{*0}K^{-}\nu)=0.392\%,\]
The calculation shows that the vector current is the
dominant contributor
and the contribution of the axial-vector current is only $7.5\%$ of the
decay rate.
The data are: CLEO[30] $0.32\pm0.08\pm0.12\%$; ARGUS[31] $0.20\pm0.05
\pm0.04\%$.
The branching ratio of $\tau^{-}\rightarrow K^{*-}K^{0}\nu$ is the same
as $\tau^{-}\rightarrow K^{*0}K^{-}\nu$.

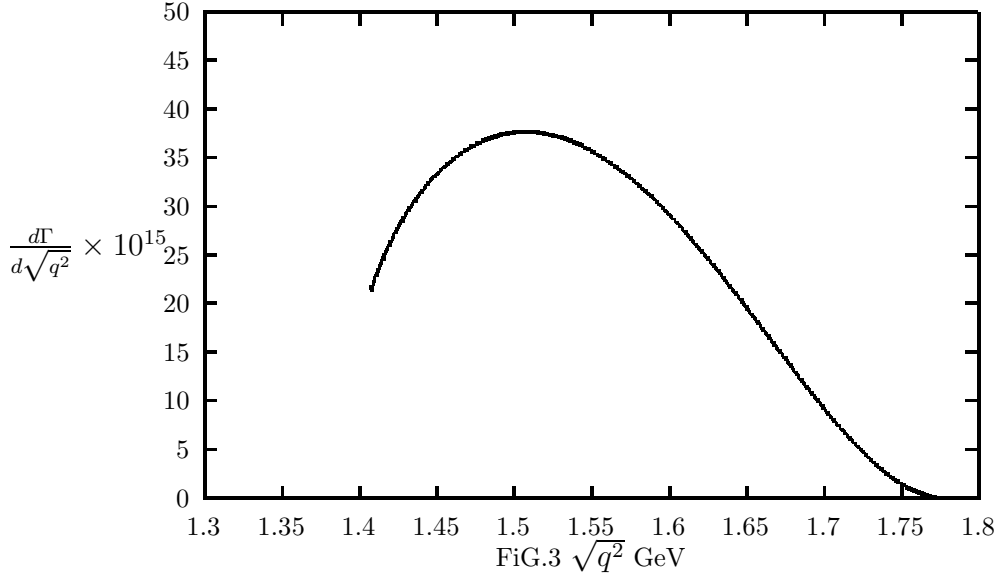
\begin{figure}
\begin{center}

\setlength{\unitlength}{0.240900pt}
\ifx\plotpoint\undefined\newsavebox{\plotpoint}\fi
\begin{picture}(1500,900)(0,0)
\font\gnuplot=cmr10 at 10pt
\gnuplot
\sbox{\plotpoint}{\rule[-0.500pt]{1.000pt}{1.000pt}}%
\put(220.0,113.0){\rule[-0.500pt]{292.934pt}{1.000pt}}
\put(220.0,113.0){\rule[-0.500pt]{4.818pt}{1.000pt}}
\put(198,113){\makebox(0,0)[r]{0}}
\put(1416.0,113.0){\rule[-0.500pt]{4.818pt}{1.000pt}}
\put(220.0,189.0){\rule[-0.500pt]{4.818pt}{1.000pt}}
\put(198,189){\makebox(0,0)[r]{5}}
\put(1416.0,189.0){\rule[-0.500pt]{4.818pt}{1.000pt}}
\put(220.0,266.0){\rule[-0.500pt]{4.818pt}{1.000pt}}
\put(198,266){\makebox(0,0)[r]{10}}
\put(1416.0,266.0){\rule[-0.500pt]{4.818pt}{1.000pt}}
\put(220.0,342.0){\rule[-0.500pt]{4.818pt}{1.000pt}}
\put(198,342){\makebox(0,0)[r]{15}}
\put(1416.0,342.0){\rule[-0.500pt]{4.818pt}{1.000pt}}
\put(220.0,419.0){\rule[-0.500pt]{4.818pt}{1.000pt}}
\put(198,419){\makebox(0,0)[r]{20}}
\put(1416.0,419.0){\rule[-0.500pt]{4.818pt}{1.000pt}}
\put(220.0,495.0){\rule[-0.500pt]{4.818pt}{1.000pt}}
\put(198,495){\makebox(0,0)[r]{25}}
\put(1416.0,495.0){\rule[-0.500pt]{4.818pt}{1.000pt}}
\put(220.0,571.0){\rule[-0.500pt]{4.818pt}{1.000pt}}
\put(198,571){\makebox(0,0)[r]{30}}
\put(1416.0,571.0){\rule[-0.500pt]{4.818pt}{1.000pt}}
\put(220.0,648.0){\rule[-0.500pt]{4.818pt}{1.000pt}}
\put(198,648){\makebox(0,0)[r]{35}}
\put(1416.0,648.0){\rule[-0.500pt]{4.818pt}{1.000pt}}
\put(220.0,724.0){\rule[-0.500pt]{4.818pt}{1.000pt}}
\put(198,724){\makebox(0,0)[r]{40}}
\put(1416.0,724.0){\rule[-0.500pt]{4.818pt}{1.000pt}}
\put(220.0,801.0){\rule[-0.500pt]{4.818pt}{1.000pt}}
\put(198,801){\makebox(0,0)[r]{45}}
\put(1416.0,801.0){\rule[-0.500pt]{4.818pt}{1.000pt}}
\put(220.0,877.0){\rule[-0.500pt]{4.818pt}{1.000pt}}
\put(198,877){\makebox(0,0)[r]{50}}
\put(1416.0,877.0){\rule[-0.500pt]{4.818pt}{1.000pt}}
\put(220.0,113.0){\rule[-0.500pt]{1.000pt}{4.818pt}}
\put(220,68){\makebox(0,0){1.3}}
\put(220.0,857.0){\rule[-0.500pt]{1.000pt}{4.818pt}}
\put(342.0,113.0){\rule[-0.500pt]{1.000pt}{4.818pt}}
\put(342,68){\makebox(0,0){1.35}}
\put(342.0,857.0){\rule[-0.500pt]{1.000pt}{4.818pt}}
\put(463.0,113.0){\rule[-0.500pt]{1.000pt}{4.818pt}}
\put(463,68){\makebox(0,0){1.4}}
\put(463.0,857.0){\rule[-0.500pt]{1.000pt}{4.818pt}}
\put(585.0,113.0){\rule[-0.500pt]{1.000pt}{4.818pt}}
\put(585,68){\makebox(0,0){1.45}}
\put(585.0,857.0){\rule[-0.500pt]{1.000pt}{4.818pt}}
\put(706.0,113.0){\rule[-0.500pt]{1.000pt}{4.818pt}}
\put(706,68){\makebox(0,0){1.5}}
\put(706.0,857.0){\rule[-0.500pt]{1.000pt}{4.818pt}}
\put(828.0,113.0){\rule[-0.500pt]{1.000pt}{4.818pt}}
\put(828,68){\makebox(0,0){1.55}}
\put(828.0,857.0){\rule[-0.500pt]{1.000pt}{4.818pt}}
\put(950.0,113.0){\rule[-0.500pt]{1.000pt}{4.818pt}}
\put(950,68){\makebox(0,0){1.6}}
\put(950.0,857.0){\rule[-0.500pt]{1.000pt}{4.818pt}}
\put(1071.0,113.0){\rule[-0.500pt]{1.000pt}{4.818pt}}
\put(1071,68){\makebox(0,0){1.65}}
\put(1071.0,857.0){\rule[-0.500pt]{1.000pt}{4.818pt}}
\put(1193.0,113.0){\rule[-0.500pt]{1.000pt}{4.818pt}}
\put(1193,68){\makebox(0,0){1.7}}
\put(1193.0,857.0){\rule[-0.500pt]{1.000pt}{4.818pt}}
\put(1314.0,113.0){\rule[-0.500pt]{1.000pt}{4.818pt}}
\put(1314,68){\makebox(0,0){1.75}}
\put(1314.0,857.0){\rule[-0.500pt]{1.000pt}{4.818pt}}
\put(1436.0,113.0){\rule[-0.500pt]{1.000pt}{4.818pt}}
\put(1436,68){\makebox(0,0){1.8}}
\put(1436.0,857.0){\rule[-0.500pt]{1.000pt}{4.818pt}}
\put(220.0,113.0){\rule[-0.500pt]{292.934pt}{1.000pt}}
\put(1436.0,113.0){\rule[-0.500pt]{1.000pt}{184.048pt}}
\put(220.0,877.0){\rule[-0.500pt]{292.934pt}{1.000pt}}
\put(45,495){\makebox(0,0){${d\Gamma\over d\sqrt{q^{2}}}\times 10^{15}$ }}
\put(828,23){\makebox(0,0){FiG.3        $\sqrt{q^{2}}$        GeV}}
\put(220.0,113.0){\rule[-0.500pt]{1.000pt}{184.048pt}}
\put(483,440){\usebox{\plotpoint}}
\multiput(484.83,440.00)(0.487,1.341){12}{\rule{0.117pt}{2.950pt}}
\multiput(480.92,440.00)(10.000,20.877){2}{\rule{1.000pt}{1.475pt}}
\multiput(494.83,467.00)(0.487,1.235){12}{\rule{0.117pt}{2.750pt}}
\multiput(490.92,467.00)(10.000,19.292){2}{\rule{1.000pt}{1.375pt}}
\multiput(504.83,492.00)(0.487,1.022){12}{\rule{0.117pt}{2.350pt}}
\multiput(500.92,492.00)(10.000,16.122){2}{\rule{1.000pt}{1.175pt}}
\multiput(514.83,513.00)(0.487,0.969){12}{\rule{0.117pt}{2.250pt}}
\multiput(510.92,513.00)(10.000,15.330){2}{\rule{1.000pt}{1.125pt}}
\multiput(524.83,533.00)(0.487,0.863){12}{\rule{0.117pt}{2.050pt}}
\multiput(520.92,533.00)(10.000,13.745){2}{\rule{1.000pt}{1.025pt}}
\multiput(534.83,551.00)(0.487,0.756){12}{\rule{0.117pt}{1.850pt}}
\multiput(530.92,551.00)(10.000,12.160){2}{\rule{1.000pt}{0.925pt}}
\multiput(544.83,567.00)(0.487,0.703){12}{\rule{0.117pt}{1.750pt}}
\multiput(540.92,567.00)(10.000,11.368){2}{\rule{1.000pt}{0.875pt}}
\multiput(554.83,582.00)(0.487,0.597){12}{\rule{0.117pt}{1.550pt}}
\multiput(550.92,582.00)(10.000,9.783){2}{\rule{1.000pt}{0.775pt}}
\multiput(564.83,595.00)(0.487,0.597){12}{\rule{0.117pt}{1.550pt}}
\multiput(560.92,595.00)(10.000,9.783){2}{\rule{1.000pt}{0.775pt}}
\multiput(574.83,608.00)(0.487,0.491){12}{\rule{0.117pt}{1.350pt}}
\multiput(570.92,608.00)(10.000,8.198){2}{\rule{1.000pt}{0.675pt}}
\multiput(583.00,620.83)(0.437,0.487){12}{\rule{1.250pt}{0.117pt}}
\multiput(583.00,616.92)(7.406,10.000){2}{\rule{0.625pt}{1.000pt}}
\multiput(593.00,630.83)(0.483,0.485){10}{\rule{1.361pt}{0.117pt}}
\multiput(593.00,626.92)(7.175,9.000){2}{\rule{0.681pt}{1.000pt}}
\multiput(603.00,639.83)(0.539,0.481){8}{\rule{1.500pt}{0.116pt}}
\multiput(603.00,635.92)(6.887,8.000){2}{\rule{0.750pt}{1.000pt}}
\multiput(613.00,647.83)(0.539,0.481){8}{\rule{1.500pt}{0.116pt}}
\multiput(613.00,643.92)(6.887,8.000){2}{\rule{0.750pt}{1.000pt}}
\multiput(623.00,655.84)(0.579,0.462){4}{\rule{1.750pt}{0.111pt}}
\multiput(623.00,651.92)(5.368,6.000){2}{\rule{0.875pt}{1.000pt}}
\multiput(632.00,661.84)(0.681,0.462){4}{\rule{1.917pt}{0.111pt}}
\multiput(632.00,657.92)(6.022,6.000){2}{\rule{0.958pt}{1.000pt}}
\multiput(642.00,667.86)(0.660,0.424){2}{\rule{2.250pt}{0.102pt}}
\multiput(642.00,663.92)(5.330,5.000){2}{\rule{1.125pt}{1.000pt}}
\multiput(652.00,672.86)(0.490,0.424){2}{\rule{2.050pt}{0.102pt}}
\multiput(652.00,668.92)(4.745,5.000){2}{\rule{1.025pt}{1.000pt}}
\put(661,675.42){\rule{2.409pt}{1.000pt}}
\multiput(661.00,673.92)(5.000,3.000){2}{\rule{1.204pt}{1.000pt}}
\put(671,678.42){\rule{2.409pt}{1.000pt}}
\multiput(671.00,676.92)(5.000,3.000){2}{\rule{1.204pt}{1.000pt}}
\put(681,681.42){\rule{2.168pt}{1.000pt}}
\multiput(681.00,679.92)(4.500,3.000){2}{\rule{1.084pt}{1.000pt}}
\put(690,683.42){\rule{2.409pt}{1.000pt}}
\multiput(690.00,682.92)(5.000,1.000){2}{\rule{1.204pt}{1.000pt}}
\put(700,684.92){\rule{2.409pt}{1.000pt}}
\multiput(700.00,683.92)(5.000,2.000){2}{\rule{1.204pt}{1.000pt}}
\put(738,685.42){\rule{2.409pt}{1.000pt}}
\multiput(738.00,685.92)(5.000,-1.000){2}{\rule{1.204pt}{1.000pt}}
\put(748,683.92){\rule{2.168pt}{1.000pt}}
\multiput(748.00,684.92)(4.500,-2.000){2}{\rule{1.084pt}{1.000pt}}
\put(757,681.92){\rule{2.168pt}{1.000pt}}
\multiput(757.00,682.92)(4.500,-2.000){2}{\rule{1.084pt}{1.000pt}}
\put(766,679.92){\rule{2.409pt}{1.000pt}}
\multiput(766.00,680.92)(5.000,-2.000){2}{\rule{1.204pt}{1.000pt}}
\put(776,677.42){\rule{2.168pt}{1.000pt}}
\multiput(776.00,678.92)(4.500,-3.000){2}{\rule{1.084pt}{1.000pt}}
\put(785,674.42){\rule{2.168pt}{1.000pt}}
\multiput(785.00,675.92)(4.500,-3.000){2}{\rule{1.084pt}{1.000pt}}
\put(794,670.92){\rule{2.409pt}{1.000pt}}
\multiput(794.00,672.92)(5.000,-4.000){2}{\rule{1.204pt}{1.000pt}}
\put(804,666.92){\rule{2.168pt}{1.000pt}}
\multiput(804.00,668.92)(4.500,-4.000){2}{\rule{1.084pt}{1.000pt}}
\multiput(813.00,664.71)(0.490,-0.424){2}{\rule{2.050pt}{0.102pt}}
\multiput(813.00,664.92)(4.745,-5.000){2}{\rule{1.025pt}{1.000pt}}
\multiput(822.00,659.71)(0.660,-0.424){2}{\rule{2.250pt}{0.102pt}}
\multiput(822.00,659.92)(5.330,-5.000){2}{\rule{1.125pt}{1.000pt}}
\multiput(832.00,654.69)(0.579,-0.462){4}{\rule{1.750pt}{0.111pt}}
\multiput(832.00,654.92)(5.368,-6.000){2}{\rule{0.875pt}{1.000pt}}
\multiput(841.00,648.71)(0.490,-0.424){2}{\rule{2.050pt}{0.102pt}}
\multiput(841.00,648.92)(4.745,-5.000){2}{\rule{1.025pt}{1.000pt}}
\multiput(850.00,643.69)(0.525,-0.475){6}{\rule{1.536pt}{0.114pt}}
\multiput(850.00,643.92)(5.813,-7.000){2}{\rule{0.768pt}{1.000pt}}
\multiput(859.00,636.69)(0.579,-0.462){4}{\rule{1.750pt}{0.111pt}}
\multiput(859.00,636.92)(5.368,-6.000){2}{\rule{0.875pt}{1.000pt}}
\multiput(868.00,630.69)(0.525,-0.475){6}{\rule{1.536pt}{0.114pt}}
\multiput(868.00,630.92)(5.813,-7.000){2}{\rule{0.768pt}{1.000pt}}
\multiput(877.00,623.69)(0.606,-0.475){6}{\rule{1.679pt}{0.114pt}}
\multiput(877.00,623.92)(6.516,-7.000){2}{\rule{0.839pt}{1.000pt}}
\multiput(887.00,616.68)(0.470,-0.481){8}{\rule{1.375pt}{0.116pt}}
\multiput(887.00,616.92)(6.146,-8.000){2}{\rule{0.688pt}{1.000pt}}
\multiput(896.00,608.68)(0.470,-0.481){8}{\rule{1.375pt}{0.116pt}}
\multiput(896.00,608.92)(6.146,-8.000){2}{\rule{0.688pt}{1.000pt}}
\multiput(905.00,600.68)(0.470,-0.481){8}{\rule{1.375pt}{0.116pt}}
\multiput(905.00,600.92)(6.146,-8.000){2}{\rule{0.688pt}{1.000pt}}
\multiput(914.00,592.68)(0.470,-0.481){8}{\rule{1.375pt}{0.116pt}}
\multiput(914.00,592.92)(6.146,-8.000){2}{\rule{0.688pt}{1.000pt}}
\multiput(923.00,584.68)(0.423,-0.485){10}{\rule{1.250pt}{0.117pt}}
\multiput(923.00,584.92)(6.406,-9.000){2}{\rule{0.625pt}{1.000pt}}
\multiput(932.00,575.68)(0.423,-0.485){10}{\rule{1.250pt}{0.117pt}}
\multiput(932.00,575.92)(6.406,-9.000){2}{\rule{0.625pt}{1.000pt}}
\multiput(941.00,566.68)(0.423,-0.485){10}{\rule{1.250pt}{0.117pt}}
\multiput(941.00,566.92)(6.406,-9.000){2}{\rule{0.625pt}{1.000pt}}
\multiput(951.83,554.35)(0.485,-0.483){10}{\rule{0.117pt}{1.361pt}}
\multiput(947.92,557.17)(9.000,-7.175){2}{\rule{1.000pt}{0.681pt}}
\multiput(959.00,547.68)(0.423,-0.485){10}{\rule{1.250pt}{0.117pt}}
\multiput(959.00,547.92)(6.406,-9.000){2}{\rule{0.625pt}{1.000pt}}
\multiput(969.83,535.35)(0.485,-0.483){10}{\rule{0.117pt}{1.361pt}}
\multiput(965.92,538.17)(9.000,-7.175){2}{\rule{1.000pt}{0.681pt}}
\multiput(978.83,524.77)(0.481,-0.539){8}{\rule{0.116pt}{1.500pt}}
\multiput(974.92,527.89)(8.000,-6.887){2}{\rule{1.000pt}{0.750pt}}
\multiput(986.83,514.89)(0.485,-0.543){10}{\rule{0.117pt}{1.472pt}}
\multiput(982.92,517.94)(9.000,-7.944){2}{\rule{1.000pt}{0.736pt}}
\multiput(995.83,504.35)(0.485,-0.483){10}{\rule{0.117pt}{1.361pt}}
\multiput(991.92,507.17)(9.000,-7.175){2}{\rule{1.000pt}{0.681pt}}
\multiput(1004.83,493.89)(0.485,-0.543){10}{\rule{0.117pt}{1.472pt}}
\multiput(1000.92,496.94)(9.000,-7.944){2}{\rule{1.000pt}{0.736pt}}
\multiput(1013.83,483.35)(0.485,-0.483){10}{\rule{0.117pt}{1.361pt}}
\multiput(1009.92,486.17)(9.000,-7.175){2}{\rule{1.000pt}{0.681pt}}
\multiput(1022.83,472.89)(0.485,-0.543){10}{\rule{0.117pt}{1.472pt}}
\multiput(1018.92,475.94)(9.000,-7.944){2}{\rule{1.000pt}{0.736pt}}
\multiput(1031.83,461.25)(0.481,-0.608){8}{\rule{0.116pt}{1.625pt}}
\multiput(1027.92,464.63)(8.000,-7.627){2}{\rule{1.000pt}{0.813pt}}
\multiput(1039.83,450.43)(0.485,-0.603){10}{\rule{0.117pt}{1.583pt}}
\multiput(1035.92,453.71)(9.000,-8.714){2}{\rule{1.000pt}{0.792pt}}
\multiput(1048.83,438.89)(0.485,-0.543){10}{\rule{0.117pt}{1.472pt}}
\multiput(1044.92,441.94)(9.000,-7.944){2}{\rule{1.000pt}{0.736pt}}
\multiput(1057.83,427.89)(0.485,-0.543){10}{\rule{0.117pt}{1.472pt}}
\multiput(1053.92,430.94)(9.000,-7.944){2}{\rule{1.000pt}{0.736pt}}
\multiput(1066.83,416.25)(0.481,-0.608){8}{\rule{0.116pt}{1.625pt}}
\multiput(1062.92,419.63)(8.000,-7.627){2}{\rule{1.000pt}{0.813pt}}
\multiput(1074.83,405.43)(0.485,-0.603){10}{\rule{0.117pt}{1.583pt}}
\multiput(1070.92,408.71)(9.000,-8.714){2}{\rule{1.000pt}{0.792pt}}
\multiput(1083.83,393.89)(0.485,-0.543){10}{\rule{0.117pt}{1.472pt}}
\multiput(1079.92,396.94)(9.000,-7.944){2}{\rule{1.000pt}{0.736pt}}
\multiput(1092.83,381.74)(0.481,-0.677){8}{\rule{0.116pt}{1.750pt}}
\multiput(1088.92,385.37)(8.000,-8.368){2}{\rule{1.000pt}{0.875pt}}
\multiput(1100.83,370.89)(0.485,-0.543){10}{\rule{0.117pt}{1.472pt}}
\multiput(1096.92,373.94)(9.000,-7.944){2}{\rule{1.000pt}{0.736pt}}
\multiput(1109.83,359.25)(0.481,-0.608){8}{\rule{0.116pt}{1.625pt}}
\multiput(1105.92,362.63)(8.000,-7.627){2}{\rule{1.000pt}{0.813pt}}
\multiput(1117.83,348.43)(0.485,-0.603){10}{\rule{0.117pt}{1.583pt}}
\multiput(1113.92,351.71)(9.000,-8.714){2}{\rule{1.000pt}{0.792pt}}
\multiput(1126.83,336.89)(0.485,-0.543){10}{\rule{0.117pt}{1.472pt}}
\multiput(1122.92,339.94)(9.000,-7.944){2}{\rule{1.000pt}{0.736pt}}
\multiput(1135.83,325.25)(0.481,-0.608){8}{\rule{0.116pt}{1.625pt}}
\multiput(1131.92,328.63)(8.000,-7.627){2}{\rule{1.000pt}{0.813pt}}
\multiput(1143.83,314.43)(0.485,-0.603){10}{\rule{0.117pt}{1.583pt}}
\multiput(1139.92,317.71)(9.000,-8.714){2}{\rule{1.000pt}{0.792pt}}
\multiput(1152.83,302.25)(0.481,-0.608){8}{\rule{0.116pt}{1.625pt}}
\multiput(1148.92,305.63)(8.000,-7.627){2}{\rule{1.000pt}{0.813pt}}
\multiput(1160.83,291.89)(0.485,-0.543){10}{\rule{0.117pt}{1.472pt}}
\multiput(1156.92,294.94)(9.000,-7.944){2}{\rule{1.000pt}{0.736pt}}
\multiput(1169.83,280.77)(0.481,-0.539){8}{\rule{0.116pt}{1.500pt}}
\multiput(1165.92,283.89)(8.000,-6.887){2}{\rule{1.000pt}{0.750pt}}
\multiput(1177.83,270.89)(0.485,-0.543){10}{\rule{0.117pt}{1.472pt}}
\multiput(1173.92,273.94)(9.000,-7.944){2}{\rule{1.000pt}{0.736pt}}
\multiput(1186.83,259.77)(0.481,-0.539){8}{\rule{0.116pt}{1.500pt}}
\multiput(1182.92,262.89)(8.000,-6.887){2}{\rule{1.000pt}{0.750pt}}
\multiput(1194.83,249.25)(0.481,-0.608){8}{\rule{0.116pt}{1.625pt}}
\multiput(1190.92,252.63)(8.000,-7.627){2}{\rule{1.000pt}{0.813pt}}
\multiput(1202.83,239.35)(0.485,-0.483){10}{\rule{0.117pt}{1.361pt}}
\multiput(1198.92,242.17)(9.000,-7.175){2}{\rule{1.000pt}{0.681pt}}
\multiput(1211.83,228.77)(0.481,-0.539){8}{\rule{0.116pt}{1.500pt}}
\multiput(1207.92,231.89)(8.000,-6.887){2}{\rule{1.000pt}{0.750pt}}
\multiput(1218.00,222.68)(0.423,-0.485){10}{\rule{1.250pt}{0.117pt}}
\multiput(1218.00,222.92)(6.406,-9.000){2}{\rule{0.625pt}{1.000pt}}
\multiput(1228.83,210.29)(0.481,-0.470){8}{\rule{0.116pt}{1.375pt}}
\multiput(1224.92,213.15)(8.000,-6.146){2}{\rule{1.000pt}{0.688pt}}
\multiput(1236.83,201.29)(0.481,-0.470){8}{\rule{0.116pt}{1.375pt}}
\multiput(1232.92,204.15)(8.000,-6.146){2}{\rule{1.000pt}{0.688pt}}
\multiput(1243.00,195.68)(0.423,-0.485){10}{\rule{1.250pt}{0.117pt}}
\multiput(1243.00,195.92)(6.406,-9.000){2}{\rule{0.625pt}{1.000pt}}
\multiput(1252.00,186.68)(0.402,-0.481){8}{\rule{1.250pt}{0.116pt}}
\multiput(1252.00,186.92)(5.406,-8.000){2}{\rule{0.625pt}{1.000pt}}
\multiput(1260.00,178.68)(0.402,-0.481){8}{\rule{1.250pt}{0.116pt}}
\multiput(1260.00,178.92)(5.406,-8.000){2}{\rule{0.625pt}{1.000pt}}
\multiput(1268.00,170.68)(0.402,-0.481){8}{\rule{1.250pt}{0.116pt}}
\multiput(1268.00,170.92)(5.406,-8.000){2}{\rule{0.625pt}{1.000pt}}
\multiput(1276.00,162.69)(0.525,-0.475){6}{\rule{1.536pt}{0.114pt}}
\multiput(1276.00,162.92)(5.813,-7.000){2}{\rule{0.768pt}{1.000pt}}
\multiput(1285.00,155.69)(0.444,-0.475){6}{\rule{1.393pt}{0.114pt}}
\multiput(1285.00,155.92)(5.109,-7.000){2}{\rule{0.696pt}{1.000pt}}
\multiput(1293.00,148.69)(0.476,-0.462){4}{\rule{1.583pt}{0.111pt}}
\multiput(1293.00,148.92)(4.714,-6.000){2}{\rule{0.792pt}{1.000pt}}
\multiput(1301.00,142.69)(0.476,-0.462){4}{\rule{1.583pt}{0.111pt}}
\multiput(1301.00,142.92)(4.714,-6.000){2}{\rule{0.792pt}{1.000pt}}
\multiput(1309.00,136.71)(0.490,-0.424){2}{\rule{2.050pt}{0.102pt}}
\multiput(1309.00,136.92)(4.745,-5.000){2}{\rule{1.025pt}{1.000pt}}
\multiput(1318.00,131.71)(0.320,-0.424){2}{\rule{1.850pt}{0.102pt}}
\multiput(1318.00,131.92)(4.160,-5.000){2}{\rule{0.925pt}{1.000pt}}
\put(1326,124.92){\rule{1.927pt}{1.000pt}}
\multiput(1326.00,126.92)(4.000,-4.000){2}{\rule{0.964pt}{1.000pt}}
\put(1334,121.42){\rule{1.927pt}{1.000pt}}
\multiput(1334.00,122.92)(4.000,-3.000){2}{\rule{0.964pt}{1.000pt}}
\put(1342,118.42){\rule{1.927pt}{1.000pt}}
\multiput(1342.00,119.92)(4.000,-3.000){2}{\rule{0.964pt}{1.000pt}}
\put(1350,115.42){\rule{1.927pt}{1.000pt}}
\multiput(1350.00,116.92)(4.000,-3.000){2}{\rule{0.964pt}{1.000pt}}
\put(1358,112.92){\rule{1.927pt}{1.000pt}}
\multiput(1358.00,113.92)(4.000,-2.000){2}{\rule{0.964pt}{1.000pt}}
\put(1366,111.42){\rule{1.927pt}{1.000pt}}
\multiput(1366.00,111.92)(4.000,-1.000){2}{\rule{0.964pt}{1.000pt}}
\put(710.0,688.0){\rule[-0.500pt]{6.745pt}{1.000pt}}
\end{picture}

\end{center}
\end{figure}

\section{CVC and $e^{+}e^{-}\rightarrow\pi\pi$}
As discussed in Ref.[16] that CVC works
very well in both meson productions in $e^{+}e^{-}$ annihilation
and $\tau$ decays.
As a test of the theory explored in this paper,
the same theory[11] is used to study $e^{+}e^{-}
\rightarrow\pi^{+}\pi^{-}$.

The expression of the VMD[11] is
\begin{equation}
{e\over 2}g\{-{1\over2}F^{\mu\nu}(\partial_{\mu}\rho^{0}_{\nu}-
\partial_{\nu}\rho^{0}_{\mu})+A^{\mu}j_{\nu}\}.
\end{equation}
Using the substitution
\[\rho_{\mu}\rightarrow{e\over2}gA_{\mu},\]
the current $j_{\mu}$ is obtained from ${\cal L}^{\rho\pi\pi}$(31).
There are two diagrams: the photon is coupled to $\pi\pi$ directly
and the photon is via the $\rho$ meson coupled to $\pi\pi$.
The matrix element is derived
\begin{eqnarray}
\lefteqn{<\pi^{+}\pi^{-}|\bar{\psi}\tau_{3}\gamma_{\mu}\psi|0>=
\frac{1}{\sqrt
{4\omega_{1}\omega_{2}}}
gf_{\rho\pi\pi}(q^{2})
\{1-\frac{q^{2}}
{q^{2}-m^{2}_{\rho}+iq\Gamma_{\rho}(q^{2})}\}(k_{1}-k_{2})_{\mu}}
\nonumber \\
&&=\frac{1}{\sqrt
{4\omega_{1}\omega_{2}}}gf_{\rho\pi\pi}(q^{2})\frac{-m^{2}_{\rho}+iq
\Gamma_{\rho}(q^{2})}{q^{2}-m^{2}_{\rho}+iq\Gamma_{\rho}(q^{2})}(k_{1}
-k_{2})_{\mu},
\end{eqnarray}
where $k_{1}$ and $k_{2}$ are momentum of $\pi^{+}$ and $\pi^{-}$
respectively, \(q^{2}=(k_{1}+k_{2})^{2}\). The cross section
is found to be
\begin{equation}
\sigma=\frac{\pi\alpha^{2}}{12}{1\over q^{2}}(1-{4m^{2}_{\pi}\over
q^{2}})^{{3\over2}}g^{2}f^{2}_{\rho\pi\pi}(q^{2})
\frac{m^{4}_{\rho}+q^{2}\Gamma
^{2}_{\rho}(q^{2})
}{(q^{2}-m^{2}_{\rho})^{2}+q^{2}\Gamma^{2}_{\rho}(q^{2})}.
\end{equation}
The numerical
results are shown in Fig.(4). Theoretical results are in good agreement
with data[32].
Systematic study of meson production in $e^{+}e^{-}$ collisions
will be presented somewhere else.

The pion form factor is found from Eq.(56)
\[|F(q^{2})|^{2}=({g\over2}f_{\rho\pi\pi}(q^{2}))^{2}
\frac{m^{4}_{\rho}+q^{2}\Gamma^{2}_{\rho}(q^{2})}
{(q^{2}-m^{2}_{\rho})^{2}+q^{2}\Gamma^{2}_{\rho}(q^{2})}.\]

The new point in this study is that the coupling
$f_{\rho\pi\pi}$(37) is a function of $q^{2}$.
As a matter of fact, $f_{\rho\pi\pi}(q^{2})$ is the form factor of
the vertex $\rho\pi\pi$. The chiral theory of mesons presented
in Ref.[11] is a theory at low energies(the energy scale $\Lambda$
is determined to be 1.6GeV[11]) and covariant derivative
expansion is exploited. In Eq.(37) part of
the $q^{2}$ dependence of $f_{\rho\pi\pi}$, $\frac{q^{2}}{2\pi^{2}
f^{2}_{\pi}}(1-{2c\over g})^{2}$ comes from the
the fourth order in derivatives and $\frac{q^{2}}{2\pi^{2}
f^{2}_{\pi}}(-4\pi^{2}c^{2})$ comes from
\[-{1\over 8}Tr\rho^{\mu\nu}\rho_{\mu\nu},\]
where $\rho_{\mu\nu}$ is the strength of the nonabelian $\rho$
field[11]. The radius of the charged pion is derived from the
the pion form factor in the space-like region of $q^{2}$
\[<r^{2}>_{\pi}={6\over m^{2}_{\rho}}+{3\over \pi^{2}f^{2}_{\pi}}
\{(1-{2c\over g})^{2}-4\pi^{2}c^{2}\}.\]
In Ref.[11] \(g=0.35\) is chosen and the last two terms are cancelled
out. In this paper we choose \(g=0.39\) to have better fitts and
\[<r^{2}>_{\pi}=(0.393+0.0549)fm^{2}=0.447 fm^{2}.\]
The first number is from the $\rho$ pole and the second comes from
the form factor $f_{\rho\pi\pi}$ and it is $12.2\%$ of the total
value.\\
The data[33] is $(0.44\pm0.01)fm^{2}$.\\
It is necessary to point out that the new expression of the pion
form factor is still resulted in the VMD and
the $\rho\pi\pi$ coupling constant is substituted by
the form factor of $\rho\pi\pi$. On the other hand, the $\rho\pi\pi$
form factor increases the value of the pion form factor
at higher $q^{2}$.

In the chiral limit, the form factors of the vertices $\rho
K\bar{K}$, $\rho K^{*}K$, $K^{*}K\pi$, and $K^{*}K\eta$ are the same.
These form factors result physical effects in corresponding $tau$
decays.

\section{$\tau\rightarrow\omega\pi\nu$}
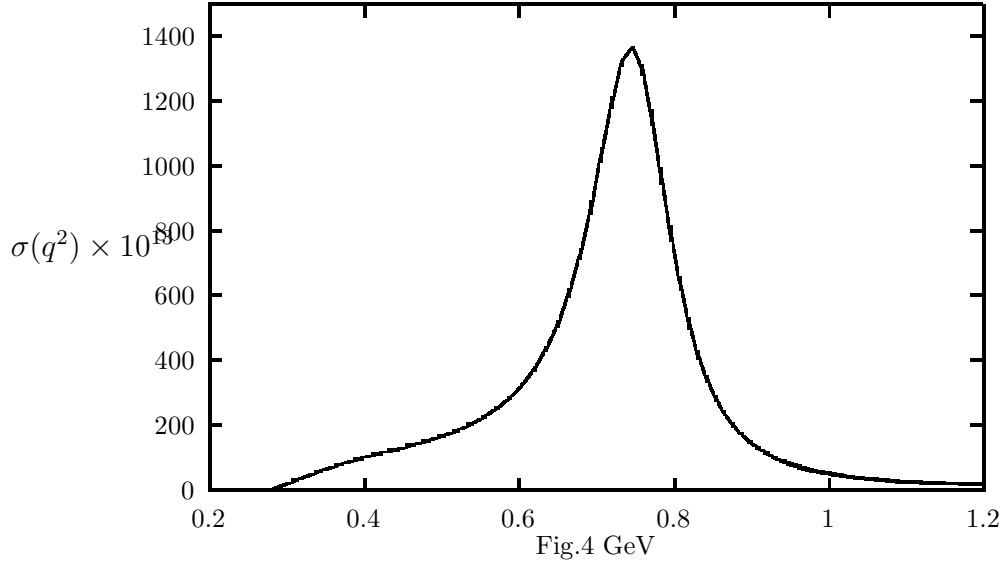
\begin{figure}
\begin{center}

\setlength{\unitlength}{0.240900pt}
\ifx\plotpoint\undefined\newsavebox{\plotpoint}\fi
\begin{picture}(1500,900)(0,0)
\font\gnuplot=cmr10 at 10pt
\gnuplot
\sbox{\plotpoint}{\rule[-0.500pt]{1.000pt}{1.000pt}}%
\put(220.0,113.0){\rule[-0.500pt]{292.934pt}{1.000pt}}
\put(220.0,113.0){\rule[-0.500pt]{4.818pt}{1.000pt}}
\put(198,113){\makebox(0,0)[r]{0}}
\put(1416.0,113.0){\rule[-0.500pt]{4.818pt}{1.000pt}}
\put(220.0,215.0){\rule[-0.500pt]{4.818pt}{1.000pt}}
\put(198,215){\makebox(0,0)[r]{200}}
\put(1416.0,215.0){\rule[-0.500pt]{4.818pt}{1.000pt}}
\put(220.0,317.0){\rule[-0.500pt]{4.818pt}{1.000pt}}
\put(198,317){\makebox(0,0)[r]{400}}
\put(1416.0,317.0){\rule[-0.500pt]{4.818pt}{1.000pt}}
\put(220.0,419.0){\rule[-0.500pt]{4.818pt}{1.000pt}}
\put(198,419){\makebox(0,0)[r]{600}}
\put(1416.0,419.0){\rule[-0.500pt]{4.818pt}{1.000pt}}
\put(220.0,520.0){\rule[-0.500pt]{4.818pt}{1.000pt}}
\put(198,520){\makebox(0,0)[r]{800}}
\put(1416.0,520.0){\rule[-0.500pt]{4.818pt}{1.000pt}}
\put(220.0,622.0){\rule[-0.500pt]{4.818pt}{1.000pt}}
\put(198,622){\makebox(0,0)[r]{1000}}
\put(1416.0,622.0){\rule[-0.500pt]{4.818pt}{1.000pt}}
\put(220.0,724.0){\rule[-0.500pt]{4.818pt}{1.000pt}}
\put(198,724){\makebox(0,0)[r]{1200}}
\put(1416.0,724.0){\rule[-0.500pt]{4.818pt}{1.000pt}}
\put(220.0,826.0){\rule[-0.500pt]{4.818pt}{1.000pt}}
\put(198,826){\makebox(0,0)[r]{1400}}
\put(1416.0,826.0){\rule[-0.500pt]{4.818pt}{1.000pt}}
\put(220.0,113.0){\rule[-0.500pt]{1.000pt}{4.818pt}}
\put(220,68){\makebox(0,0){0.2}}
\put(220.0,857.0){\rule[-0.500pt]{1.000pt}{4.818pt}}
\put(463.0,113.0){\rule[-0.500pt]{1.000pt}{4.818pt}}
\put(463,68){\makebox(0,0){0.4}}
\put(463.0,857.0){\rule[-0.500pt]{1.000pt}{4.818pt}}
\put(706.0,113.0){\rule[-0.500pt]{1.000pt}{4.818pt}}
\put(706,68){\makebox(0,0){0.6}}
\put(706.0,857.0){\rule[-0.500pt]{1.000pt}{4.818pt}}
\put(950.0,113.0){\rule[-0.500pt]{1.000pt}{4.818pt}}
\put(950,68){\makebox(0,0){0.8}}
\put(950.0,857.0){\rule[-0.500pt]{1.000pt}{4.818pt}}
\put(1193.0,113.0){\rule[-0.500pt]{1.000pt}{4.818pt}}
\put(1193,68){\makebox(0,0){1}}
\put(1193.0,857.0){\rule[-0.500pt]{1.000pt}{4.818pt}}
\put(1436.0,113.0){\rule[-0.500pt]{1.000pt}{4.818pt}}
\put(1436,68){\makebox(0,0){1.2}}
\put(1436.0,857.0){\rule[-0.500pt]{1.000pt}{4.818pt}}
\put(220.0,113.0){\rule[-0.500pt]{292.934pt}{1.000pt}}
\put(1436.0,113.0){\rule[-0.500pt]{1.000pt}{184.048pt}}
\put(220.0,877.0){\rule[-0.500pt]{292.934pt}{1.000pt}}
\put(45,495){\makebox(0,0){$\sigma(q^{2})\times10^{13}$ }}
\put(828,23){\makebox(0,0){Fig.4   GeV}}
\put(220.0,113.0){\rule[-0.500pt]{1.000pt}{184.048pt}}
\put(316,113){\usebox{\plotpoint}}
\multiput(316.00,114.83)(1.328,0.493){22}{\rule{2.917pt}{0.119pt}}
\multiput(316.00,110.92)(33.946,15.000){2}{\rule{1.458pt}{1.000pt}}
\multiput(356.00,129.83)(1.278,0.492){20}{\rule{2.821pt}{0.119pt}}
\multiput(356.00,125.92)(30.144,14.000){2}{\rule{1.411pt}{1.000pt}}
\multiput(392.00,143.83)(1.501,0.489){14}{\rule{3.250pt}{0.118pt}}
\multiput(392.00,139.92)(26.254,11.000){2}{\rule{1.625pt}{1.000pt}}
\multiput(425.00,154.83)(1.681,0.485){10}{\rule{3.583pt}{0.117pt}}
\multiput(425.00,150.92)(22.563,9.000){2}{\rule{1.792pt}{1.000pt}}
\multiput(455.00,163.83)(1.845,0.481){8}{\rule{3.875pt}{0.116pt}}
\multiput(455.00,159.92)(20.957,8.000){2}{\rule{1.938pt}{1.000pt}}
\multiput(484.00,171.84)(2.427,0.462){4}{\rule{4.750pt}{0.111pt}}
\multiput(484.00,167.92)(17.141,6.000){2}{\rule{2.375pt}{1.000pt}}
\multiput(511.00,177.84)(1.909,0.475){6}{\rule{3.964pt}{0.114pt}}
\multiput(511.00,173.92)(17.772,7.000){2}{\rule{1.982pt}{1.000pt}}
\multiput(537.00,184.84)(1.827,0.475){6}{\rule{3.821pt}{0.114pt}}
\multiput(537.00,180.92)(17.068,7.000){2}{\rule{1.911pt}{1.000pt}}
\multiput(562.00,191.83)(1.501,0.481){8}{\rule{3.250pt}{0.116pt}}
\multiput(562.00,187.92)(17.254,8.000){2}{\rule{1.625pt}{1.000pt}}
\multiput(586.00,199.83)(1.433,0.481){8}{\rule{3.125pt}{0.116pt}}
\multiput(586.00,195.92)(16.514,8.000){2}{\rule{1.563pt}{1.000pt}}
\multiput(609.00,207.83)(1.075,0.487){12}{\rule{2.450pt}{0.117pt}}
\multiput(609.00,203.92)(16.915,10.000){2}{\rule{1.225pt}{1.000pt}}
\multiput(631.00,217.83)(0.847,0.491){16}{\rule{2.000pt}{0.118pt}}
\multiput(631.00,213.92)(16.849,12.000){2}{\rule{1.000pt}{1.000pt}}
\multiput(652.00,229.83)(0.723,0.492){20}{\rule{1.750pt}{0.119pt}}
\multiput(652.00,225.92)(17.368,14.000){2}{\rule{0.875pt}{1.000pt}}
\multiput(673.00,243.83)(0.599,0.494){24}{\rule{1.500pt}{0.119pt}}
\multiput(673.00,239.92)(16.887,16.000){2}{\rule{0.750pt}{1.000pt}}
\multiput(693.00,259.83)(0.478,0.495){32}{\rule{1.250pt}{0.119pt}}
\multiput(693.00,255.92)(17.406,20.000){2}{\rule{0.625pt}{1.000pt}}
\multiput(714.83,278.00)(0.495,0.638){30}{\rule{0.119pt}{1.566pt}}
\multiput(710.92,278.00)(19.000,21.750){2}{\rule{1.000pt}{0.783pt}}
\multiput(733.83,303.00)(0.495,0.845){28}{\rule{0.119pt}{1.972pt}}
\multiput(729.92,303.00)(18.000,26.907){2}{\rule{1.000pt}{0.986pt}}
\multiput(751.83,334.00)(0.495,1.015){30}{\rule{0.119pt}{2.303pt}}
\multiput(747.92,334.00)(19.000,34.221){2}{\rule{1.000pt}{1.151pt}}
\multiput(770.83,373.00)(0.494,1.440){26}{\rule{0.119pt}{3.132pt}}
\multiput(766.92,373.00)(17.000,42.499){2}{\rule{1.000pt}{1.566pt}}
\multiput(787.83,422.00)(0.495,1.700){28}{\rule{0.119pt}{3.639pt}}
\multiput(783.92,422.00)(18.000,53.447){2}{\rule{1.000pt}{1.819pt}}
\multiput(805.83,483.00)(0.494,2.165){26}{\rule{0.119pt}{4.544pt}}
\multiput(801.92,483.00)(17.000,63.568){2}{\rule{1.000pt}{2.272pt}}
\multiput(822.83,556.00)(0.494,2.660){24}{\rule{0.119pt}{5.500pt}}
\multiput(818.92,556.00)(16.000,72.584){2}{\rule{1.000pt}{2.750pt}}
\multiput(838.83,640.00)(0.494,2.468){26}{\rule{0.119pt}{5.132pt}}
\multiput(834.92,640.00)(17.000,72.348){2}{\rule{1.000pt}{2.566pt}}
\multiput(855.83,723.00)(0.494,2.048){24}{\rule{0.119pt}{4.312pt}}
\multiput(851.92,723.00)(16.000,56.049){2}{\rule{1.000pt}{2.156pt}}
\multiput(871.83,788.00)(0.494,0.631){24}{\rule{0.119pt}{1.562pt}}
\multiput(867.92,788.00)(16.000,17.757){2}{\rule{1.000pt}{0.781pt}}
\multiput(887.83,798.83)(0.493,-1.087){22}{\rule{0.119pt}{2.450pt}}
\multiput(883.92,803.91)(15.000,-27.915){2}{\rule{1.000pt}{1.225pt}}
\multiput(902.83,753.93)(0.493,-2.568){22}{\rule{0.119pt}{5.317pt}}
\multiput(898.92,764.96)(15.000,-64.965){2}{\rule{1.000pt}{2.658pt}}
\multiput(917.83,672.95)(0.493,-3.188){22}{\rule{0.119pt}{6.517pt}}
\multiput(913.92,686.47)(15.000,-80.474){2}{\rule{1.000pt}{3.258pt}}
\multiput(932.83,579.78)(0.493,-3.085){22}{\rule{0.119pt}{6.317pt}}
\multiput(928.92,592.89)(15.000,-77.889){2}{\rule{1.000pt}{3.158pt}}
\multiput(947.83,492.38)(0.493,-2.637){22}{\rule{0.119pt}{5.450pt}}
\multiput(943.92,503.69)(15.000,-66.688){2}{\rule{1.000pt}{2.725pt}}
\multiput(962.83,417.28)(0.492,-2.278){20}{\rule{0.119pt}{4.750pt}}
\multiput(958.92,427.14)(14.000,-53.141){2}{\rule{1.000pt}{2.375pt}}
\multiput(976.83,358.43)(0.492,-1.760){20}{\rule{0.119pt}{3.750pt}}
\multiput(972.92,366.22)(14.000,-41.217){2}{\rule{1.000pt}{1.875pt}}
\multiput(990.83,312.69)(0.492,-1.352){20}{\rule{0.119pt}{2.964pt}}
\multiput(986.92,318.85)(14.000,-31.847){2}{\rule{1.000pt}{1.482pt}}
\multiput(1004.83,277.07)(0.492,-1.056){20}{\rule{0.119pt}{2.393pt}}
\multiput(1000.92,282.03)(14.000,-25.034){2}{\rule{1.000pt}{1.196pt}}
\multiput(1018.83,248.30)(0.492,-0.900){18}{\rule{0.118pt}{2.096pt}}
\multiput(1014.92,252.65)(13.000,-19.649){2}{\rule{1.000pt}{1.048pt}}
\multiput(1031.83,226.63)(0.492,-0.612){20}{\rule{0.119pt}{1.536pt}}
\multiput(1027.92,229.81)(14.000,-14.813){2}{\rule{1.000pt}{0.768pt}}
\multiput(1045.83,209.17)(0.492,-0.540){18}{\rule{0.118pt}{1.404pt}}
\multiput(1041.92,212.09)(13.000,-12.086){2}{\rule{1.000pt}{0.702pt}}
\multiput(1057.00,197.68)(0.498,-0.491){16}{\rule{1.333pt}{0.118pt}}
\multiput(1057.00,197.92)(10.233,-12.000){2}{\rule{0.667pt}{1.000pt}}
\multiput(1070.00,185.68)(0.663,-0.485){10}{\rule{1.694pt}{0.117pt}}
\multiput(1070.00,185.92)(9.483,-9.000){2}{\rule{0.847pt}{1.000pt}}
\multiput(1083.00,176.68)(0.745,-0.481){8}{\rule{1.875pt}{0.116pt}}
\multiput(1083.00,176.92)(9.108,-8.000){2}{\rule{0.937pt}{1.000pt}}
\multiput(1096.00,168.69)(0.769,-0.475){6}{\rule{1.964pt}{0.114pt}}
\multiput(1096.00,168.92)(7.923,-7.000){2}{\rule{0.982pt}{1.000pt}}
\multiput(1108.00,161.69)(0.989,-0.462){4}{\rule{2.417pt}{0.111pt}}
\multiput(1108.00,161.92)(7.984,-6.000){2}{\rule{1.208pt}{1.000pt}}
\put(1121,153.92){\rule{2.891pt}{1.000pt}}
\multiput(1121.00,155.92)(6.000,-4.000){2}{\rule{1.445pt}{1.000pt}}
\put(1133,149.92){\rule{2.891pt}{1.000pt}}
\multiput(1133.00,151.92)(6.000,-4.000){2}{\rule{1.445pt}{1.000pt}}
\put(1145,145.92){\rule{3.132pt}{1.000pt}}
\multiput(1145.00,147.92)(6.500,-4.000){2}{\rule{1.566pt}{1.000pt}}
\put(1158,142.42){\rule{2.891pt}{1.000pt}}
\multiput(1158.00,143.92)(6.000,-3.000){2}{\rule{1.445pt}{1.000pt}}
\put(1170,139.92){\rule{2.650pt}{1.000pt}}
\multiput(1170.00,140.92)(5.500,-2.000){2}{\rule{1.325pt}{1.000pt}}
\put(1181,137.92){\rule{2.891pt}{1.000pt}}
\multiput(1181.00,138.92)(6.000,-2.000){2}{\rule{1.445pt}{1.000pt}}
\put(1193,135.92){\rule{2.891pt}{1.000pt}}
\multiput(1193.00,136.92)(6.000,-2.000){2}{\rule{1.445pt}{1.000pt}}
\put(1205,133.92){\rule{2.650pt}{1.000pt}}
\multiput(1205.00,134.92)(5.500,-2.000){2}{\rule{1.325pt}{1.000pt}}
\put(1216,131.92){\rule{2.891pt}{1.000pt}}
\multiput(1216.00,132.92)(6.000,-2.000){2}{\rule{1.445pt}{1.000pt}}
\put(1228,130.42){\rule{2.650pt}{1.000pt}}
\multiput(1228.00,130.92)(5.500,-1.000){2}{\rule{1.325pt}{1.000pt}}
\put(1239,129.42){\rule{2.650pt}{1.000pt}}
\multiput(1239.00,129.92)(5.500,-1.000){2}{\rule{1.325pt}{1.000pt}}
\put(1250,128.42){\rule{2.650pt}{1.000pt}}
\multiput(1250.00,128.92)(5.500,-1.000){2}{\rule{1.325pt}{1.000pt}}
\put(1261,127.42){\rule{2.650pt}{1.000pt}}
\multiput(1261.00,127.92)(5.500,-1.000){2}{\rule{1.325pt}{1.000pt}}
\put(1272,126.42){\rule{2.650pt}{1.000pt}}
\multiput(1272.00,126.92)(5.500,-1.000){2}{\rule{1.325pt}{1.000pt}}
\put(1283,125.42){\rule{2.650pt}{1.000pt}}
\multiput(1283.00,125.92)(5.500,-1.000){2}{\rule{1.325pt}{1.000pt}}
\put(1294,124.42){\rule{2.650pt}{1.000pt}}
\multiput(1294.00,124.92)(5.500,-1.000){2}{\rule{1.325pt}{1.000pt}}
\put(1315,123.42){\rule{2.650pt}{1.000pt}}
\multiput(1315.00,123.92)(5.500,-1.000){2}{\rule{1.325pt}{1.000pt}}
\put(1305.0,126.0){\rule[-0.500pt]{2.409pt}{1.000pt}}
\put(1337,122.42){\rule{2.409pt}{1.000pt}}
\multiput(1337.00,122.92)(5.000,-1.000){2}{\rule{1.204pt}{1.000pt}}
\put(1326.0,125.0){\rule[-0.500pt]{2.650pt}{1.000pt}}
\put(1368,121.42){\rule{2.409pt}{1.000pt}}
\multiput(1368.00,121.92)(5.000,-1.000){2}{\rule{1.204pt}{1.000pt}}
\put(1347.0,124.0){\rule[-0.500pt]{5.059pt}{1.000pt}}
\put(1398,120.42){\rule{2.409pt}{1.000pt}}
\multiput(1398.00,120.92)(5.000,-1.000){2}{\rule{1.204pt}{1.000pt}}
\put(1378.0,123.0){\rule[-0.500pt]{4.818pt}{1.000pt}}
\put(1408.0,122.0){\rule[-0.500pt]{6.745pt}{1.000pt}}
\end{picture}

\end{center}
\end{figure}
This process has been studied in Ref.[8] by using the abnormal
vertex $\rho\omega\pi$. The effects of excited $\rho$ mesons have
been taken into account.
In this paper we
only take the contribution of the $\rho$ meson.
The $\omega\rho\pi$ vertex
is presented in Ref.[11]
\[{\cal L}^{\omega\rho\pi}=-{N_{C}\over \pi^{2}g^{2}f_{\pi}}
\varepsilon^{\mu\nu\alpha\beta}\partial_{\mu}\omega_{\nu}\rho^{i}
_{\alpha}\partial_{\beta}\pi^{i}.\]
The vertices of $\pi^{0}\gamma\gamma$, $\omega\pi\gamma$,
$\rho\pi\gamma$, and $\omega3\pi$ are via the VMD derived from this
vertex and theoretical results agree with data well.
The coupling constant of this $\omega\rho\pi$ vertex is different
with the one presented in Ref.[8].
The decay width of $\tau\rightarrow\omega\pi\nu$ is
derived
\begin{eqnarray}
\lefteqn{\Gamma=\frac{G^{2}}{128m^{3}_{\tau}}\frac{cos^{2}\theta_{C}}
{(2\pi)^{3}}\int dq^{2}{1\over q^{4}}(m^{2}_{\tau}-q^{2})^{2}
(m^{2}_{\tau}+2q^{2})(q^{2}-m^{2}_{\omega})^{3}}\nonumber \\
&&\frac{3}{\pi^{4}g^{2}f^{2}_{\pi}}\frac{m^{4}_{\rho}+q^{2}\Gamma^{2}
_{\rho}(q^{2})}{(q^{2}-m^{2}_{\rho})^{2}+q^{2}\Gamma^{2}_{\rho}(q^{2})}.
\end{eqnarray}
The numerical result of the branching ratio is
\[B=1.2\%,\]
and the experiment is $1.6\pm0.5\%$[28].
This result is the same as the one obtained in Ref.[8] when
only $\rho$ meson is taken.
The distribution of the invariant mass of $\omega$ and $\pi$ is shown in
Fig.(5).

\begin{figure}
\begin{center}
\setlength{\unitlength}{0.240900pt}
\ifx\plotpoint\undefined\newsavebox{\plotpoint}\fi
\begin{picture}(1500,900)(0,0)
\font\gnuplot=cmr10 at 10pt
\gnuplot
\sbox{\plotpoint}{\rule[-0.500pt]{1.000pt}{1.000pt}}%
\put(220.0,113.0){\rule[-0.500pt]{292.934pt}{1.000pt}}
\put(220.0,113.0){\rule[-0.500pt]{4.818pt}{1.000pt}}
\put(198,113){\makebox(0,0)[r]{0}}
\put(1416.0,113.0){\rule[-0.500pt]{4.818pt}{1.000pt}}
\put(220.0,189.0){\rule[-0.500pt]{4.818pt}{1.000pt}}
\put(198,189){\makebox(0,0)[r]{5}}
\put(1416.0,189.0){\rule[-0.500pt]{4.818pt}{1.000pt}}
\put(220.0,266.0){\rule[-0.500pt]{4.818pt}{1.000pt}}
\put(198,266){\makebox(0,0)[r]{10}}
\put(1416.0,266.0){\rule[-0.500pt]{4.818pt}{1.000pt}}
\put(220.0,342.0){\rule[-0.500pt]{4.818pt}{1.000pt}}
\put(198,342){\makebox(0,0)[r]{15}}
\put(1416.0,342.0){\rule[-0.500pt]{4.818pt}{1.000pt}}
\put(220.0,419.0){\rule[-0.500pt]{4.818pt}{1.000pt}}
\put(198,419){\makebox(0,0)[r]{20}}
\put(1416.0,419.0){\rule[-0.500pt]{4.818pt}{1.000pt}}
\put(220.0,495.0){\rule[-0.500pt]{4.818pt}{1.000pt}}
\put(198,495){\makebox(0,0)[r]{25}}
\put(1416.0,495.0){\rule[-0.500pt]{4.818pt}{1.000pt}}
\put(220.0,571.0){\rule[-0.500pt]{4.818pt}{1.000pt}}
\put(198,571){\makebox(0,0)[r]{30}}
\put(1416.0,571.0){\rule[-0.500pt]{4.818pt}{1.000pt}}
\put(220.0,648.0){\rule[-0.500pt]{4.818pt}{1.000pt}}
\put(198,648){\makebox(0,0)[r]{35}}
\put(1416.0,648.0){\rule[-0.500pt]{4.818pt}{1.000pt}}
\put(220.0,724.0){\rule[-0.500pt]{4.818pt}{1.000pt}}
\put(198,724){\makebox(0,0)[r]{40}}
\put(1416.0,724.0){\rule[-0.500pt]{4.818pt}{1.000pt}}
\put(220.0,801.0){\rule[-0.500pt]{4.818pt}{1.000pt}}
\put(198,801){\makebox(0,0)[r]{45}}
\put(1416.0,801.0){\rule[-0.500pt]{4.818pt}{1.000pt}}
\put(220.0,877.0){\rule[-0.500pt]{4.818pt}{1.000pt}}
\put(198,877){\makebox(0,0)[r]{50}}
\put(1416.0,877.0){\rule[-0.500pt]{4.818pt}{1.000pt}}
\put(220.0,113.0){\rule[-0.500pt]{1.000pt}{4.818pt}}
\put(220,68){\makebox(0,0){0.8}}
\put(220.0,857.0){\rule[-0.500pt]{1.000pt}{4.818pt}}
\put(463.0,113.0){\rule[-0.500pt]{1.000pt}{4.818pt}}
\put(463,68){\makebox(0,0){1}}
\put(463.0,857.0){\rule[-0.500pt]{1.000pt}{4.818pt}}
\put(706.0,113.0){\rule[-0.500pt]{1.000pt}{4.818pt}}
\put(706,68){\makebox(0,0){1.2}}
\put(706.0,857.0){\rule[-0.500pt]{1.000pt}{4.818pt}}
\put(950.0,113.0){\rule[-0.500pt]{1.000pt}{4.818pt}}
\put(950,68){\makebox(0,0){1.4}}
\put(950.0,857.0){\rule[-0.500pt]{1.000pt}{4.818pt}}
\put(1193.0,113.0){\rule[-0.500pt]{1.000pt}{4.818pt}}
\put(1193,68){\makebox(0,0){1.6}}
\put(1193.0,857.0){\rule[-0.500pt]{1.000pt}{4.818pt}}
\put(1436.0,113.0){\rule[-0.500pt]{1.000pt}{4.818pt}}
\put(1436,68){\makebox(0,0){1.8}}
\put(1436.0,857.0){\rule[-0.500pt]{1.000pt}{4.818pt}}
\put(220.0,113.0){\rule[-0.500pt]{292.934pt}{1.000pt}}
\put(1436.0,113.0){\rule[-0.500pt]{1.000pt}{184.048pt}}
\put(220.0,877.0){\rule[-0.500pt]{292.934pt}{1.000pt}}
\put(45,495){\makebox(0,0){${d\Gamma\over d\sqrt{q^{2}}}\times 10^{15}$ }}
\put(828,23){\makebox(0,0){FiG.5   $\sqrt{q^{2}}$        GeV}}
\put(220.0,113.0){\rule[-0.500pt]{1.000pt}{184.048pt}}
\put(408,369){\usebox{\plotpoint}}
\multiput(409.83,369.00)(0.493,0.502){22}{\rule{0.119pt}{1.317pt}}
\multiput(405.92,369.00)(15.000,13.267){2}{\rule{1.000pt}{0.658pt}}
\multiput(424.83,385.00)(0.492,0.538){20}{\rule{0.119pt}{1.393pt}}
\multiput(420.92,385.00)(14.000,13.109){2}{\rule{1.000pt}{0.696pt}}
\multiput(438.83,401.00)(0.492,0.501){20}{\rule{0.119pt}{1.321pt}}
\multiput(434.92,401.00)(14.000,12.257){2}{\rule{1.000pt}{0.661pt}}
\multiput(452.83,416.00)(0.492,0.500){18}{\rule{0.118pt}{1.327pt}}
\multiput(448.92,416.00)(13.000,11.246){2}{\rule{1.000pt}{0.663pt}}
\multiput(464.00,431.83)(0.500,0.492){18}{\rule{1.327pt}{0.118pt}}
\multiput(464.00,427.92)(11.246,13.000){2}{\rule{0.663pt}{1.000pt}}
\multiput(478.00,444.83)(0.498,0.491){16}{\rule{1.333pt}{0.118pt}}
\multiput(478.00,440.92)(10.233,12.000){2}{\rule{0.667pt}{1.000pt}}
\multiput(491.00,456.83)(0.541,0.491){16}{\rule{1.417pt}{0.118pt}}
\multiput(491.00,452.92)(11.060,12.000){2}{\rule{0.708pt}{1.000pt}}
\multiput(505.00,468.83)(0.498,0.491){16}{\rule{1.333pt}{0.118pt}}
\multiput(505.00,464.92)(10.233,12.000){2}{\rule{0.667pt}{1.000pt}}
\multiput(518.00,480.83)(0.543,0.489){14}{\rule{1.432pt}{0.118pt}}
\multiput(518.00,476.92)(10.028,11.000){2}{\rule{0.716pt}{1.000pt}}
\multiput(531.00,491.83)(0.543,0.489){14}{\rule{1.432pt}{0.118pt}}
\multiput(531.00,487.92)(10.028,11.000){2}{\rule{0.716pt}{1.000pt}}
\multiput(544.00,502.83)(0.495,0.489){14}{\rule{1.341pt}{0.118pt}}
\multiput(544.00,498.92)(9.217,11.000){2}{\rule{0.670pt}{1.000pt}}
\multiput(556.00,513.83)(0.543,0.489){14}{\rule{1.432pt}{0.118pt}}
\multiput(556.00,509.92)(10.028,11.000){2}{\rule{0.716pt}{1.000pt}}
\multiput(569.00,524.83)(0.543,0.489){14}{\rule{1.432pt}{0.118pt}}
\multiput(569.00,520.92)(10.028,11.000){2}{\rule{0.716pt}{1.000pt}}
\multiput(582.00,535.83)(0.544,0.487){12}{\rule{1.450pt}{0.117pt}}
\multiput(582.00,531.92)(8.990,10.000){2}{\rule{0.725pt}{1.000pt}}
\multiput(594.00,545.83)(0.495,0.489){14}{\rule{1.341pt}{0.118pt}}
\multiput(594.00,541.92)(9.217,11.000){2}{\rule{0.670pt}{1.000pt}}
\multiput(606.00,556.83)(0.543,0.489){14}{\rule{1.432pt}{0.118pt}}
\multiput(606.00,552.92)(10.028,11.000){2}{\rule{0.716pt}{1.000pt}}
\multiput(619.00,567.83)(0.495,0.489){14}{\rule{1.341pt}{0.118pt}}
\multiput(619.00,563.92)(9.217,11.000){2}{\rule{0.670pt}{1.000pt}}
\multiput(631.00,578.83)(0.544,0.487){12}{\rule{1.450pt}{0.117pt}}
\multiput(631.00,574.92)(8.990,10.000){2}{\rule{0.725pt}{1.000pt}}
\multiput(643.00,588.83)(0.447,0.489){14}{\rule{1.250pt}{0.118pt}}
\multiput(643.00,584.92)(8.406,11.000){2}{\rule{0.625pt}{1.000pt}}
\multiput(654.00,599.83)(0.495,0.489){14}{\rule{1.341pt}{0.118pt}}
\multiput(654.00,595.92)(9.217,11.000){2}{\rule{0.670pt}{1.000pt}}
\multiput(666.00,610.83)(0.495,0.489){14}{\rule{1.341pt}{0.118pt}}
\multiput(666.00,606.92)(9.217,11.000){2}{\rule{0.670pt}{1.000pt}}
\multiput(679.83,620.00)(0.489,0.495){14}{\rule{0.118pt}{1.341pt}}
\multiput(675.92,620.00)(11.000,9.217){2}{\rule{1.000pt}{0.670pt}}
\multiput(689.00,633.83)(0.495,0.489){14}{\rule{1.341pt}{0.118pt}}
\multiput(689.00,629.92)(9.217,11.000){2}{\rule{0.670pt}{1.000pt}}
\multiput(701.00,644.83)(0.447,0.489){14}{\rule{1.250pt}{0.118pt}}
\multiput(701.00,640.92)(8.406,11.000){2}{\rule{0.625pt}{1.000pt}}
\multiput(712.00,655.83)(0.495,0.489){14}{\rule{1.341pt}{0.118pt}}
\multiput(712.00,651.92)(9.217,11.000){2}{\rule{0.670pt}{1.000pt}}
\multiput(725.83,665.00)(0.489,0.495){14}{\rule{0.118pt}{1.341pt}}
\multiput(721.92,665.00)(11.000,9.217){2}{\rule{1.000pt}{0.670pt}}
\multiput(735.00,678.83)(0.447,0.489){14}{\rule{1.250pt}{0.118pt}}
\multiput(735.00,674.92)(8.406,11.000){2}{\rule{0.625pt}{1.000pt}}
\multiput(746.00,689.83)(0.447,0.489){14}{\rule{1.250pt}{0.118pt}}
\multiput(746.00,685.92)(8.406,11.000){2}{\rule{0.625pt}{1.000pt}}
\multiput(757.00,700.83)(0.447,0.489){14}{\rule{1.250pt}{0.118pt}}
\multiput(757.00,696.92)(8.406,11.000){2}{\rule{0.625pt}{1.000pt}}
\multiput(768.00,711.83)(0.447,0.489){14}{\rule{1.250pt}{0.118pt}}
\multiput(768.00,707.92)(8.406,11.000){2}{\rule{0.625pt}{1.000pt}}
\multiput(779.00,722.83)(0.447,0.489){14}{\rule{1.250pt}{0.118pt}}
\multiput(779.00,718.92)(8.406,11.000){2}{\rule{0.625pt}{1.000pt}}
\multiput(791.83,732.00)(0.487,0.491){12}{\rule{0.117pt}{1.350pt}}
\multiput(787.92,732.00)(10.000,8.198){2}{\rule{1.000pt}{0.675pt}}
\multiput(800.00,744.83)(0.491,0.487){12}{\rule{1.350pt}{0.117pt}}
\multiput(800.00,740.92)(8.198,10.000){2}{\rule{0.675pt}{1.000pt}}
\multiput(811.00,754.83)(0.491,0.487){12}{\rule{1.350pt}{0.117pt}}
\multiput(811.00,750.92)(8.198,10.000){2}{\rule{0.675pt}{1.000pt}}
\multiput(822.00,764.83)(0.437,0.487){12}{\rule{1.250pt}{0.117pt}}
\multiput(822.00,760.92)(7.406,10.000){2}{\rule{0.625pt}{1.000pt}}
\multiput(832.00,774.83)(0.491,0.487){12}{\rule{1.350pt}{0.117pt}}
\multiput(832.00,770.92)(8.198,10.000){2}{\rule{0.675pt}{1.000pt}}
\multiput(843.00,784.83)(0.483,0.485){10}{\rule{1.361pt}{0.117pt}}
\multiput(843.00,780.92)(7.175,9.000){2}{\rule{0.681pt}{1.000pt}}
\multiput(853.00,793.83)(0.483,0.485){10}{\rule{1.361pt}{0.117pt}}
\multiput(853.00,789.92)(7.175,9.000){2}{\rule{0.681pt}{1.000pt}}
\multiput(863.00,802.83)(0.543,0.485){10}{\rule{1.472pt}{0.117pt}}
\multiput(863.00,798.92)(7.944,9.000){2}{\rule{0.736pt}{1.000pt}}
\multiput(874.00,811.84)(0.606,0.475){6}{\rule{1.679pt}{0.114pt}}
\multiput(874.00,807.92)(6.516,7.000){2}{\rule{0.839pt}{1.000pt}}
\multiput(884.00,818.83)(0.539,0.481){8}{\rule{1.500pt}{0.116pt}}
\multiput(884.00,814.92)(6.887,8.000){2}{\rule{0.750pt}{1.000pt}}
\multiput(894.00,826.84)(0.606,0.475){6}{\rule{1.679pt}{0.114pt}}
\multiput(894.00,822.92)(6.516,7.000){2}{\rule{0.839pt}{1.000pt}}
\multiput(904.00,833.84)(0.681,0.462){4}{\rule{1.917pt}{0.111pt}}
\multiput(904.00,829.92)(6.022,6.000){2}{\rule{0.958pt}{1.000pt}}
\multiput(914.00,839.86)(0.660,0.424){2}{\rule{2.250pt}{0.102pt}}
\multiput(914.00,835.92)(5.330,5.000){2}{\rule{1.125pt}{1.000pt}}
\multiput(924.00,844.86)(0.660,0.424){2}{\rule{2.250pt}{0.102pt}}
\multiput(924.00,840.92)(5.330,5.000){2}{\rule{1.125pt}{1.000pt}}
\put(934,847.92){\rule{2.409pt}{1.000pt}}
\multiput(934.00,845.92)(5.000,4.000){2}{\rule{1.204pt}{1.000pt}}
\put(944,851.92){\rule{2.409pt}{1.000pt}}
\multiput(944.00,849.92)(5.000,4.000){2}{\rule{1.204pt}{1.000pt}}
\put(954,854.92){\rule{2.168pt}{1.000pt}}
\multiput(954.00,853.92)(4.500,2.000){2}{\rule{1.084pt}{1.000pt}}
\put(963,856.92){\rule{2.409pt}{1.000pt}}
\multiput(963.00,855.92)(5.000,2.000){2}{\rule{1.204pt}{1.000pt}}
\put(973,858.42){\rule{2.409pt}{1.000pt}}
\multiput(973.00,857.92)(5.000,1.000){2}{\rule{1.204pt}{1.000pt}}
\put(992,858.42){\rule{2.409pt}{1.000pt}}
\multiput(992.00,858.92)(5.000,-1.000){2}{\rule{1.204pt}{1.000pt}}
\put(1002,856.92){\rule{2.168pt}{1.000pt}}
\multiput(1002.00,857.92)(4.500,-2.000){2}{\rule{1.084pt}{1.000pt}}
\put(1011,854.42){\rule{2.168pt}{1.000pt}}
\multiput(1011.00,855.92)(4.500,-3.000){2}{\rule{1.084pt}{1.000pt}}
\put(1020,850.92){\rule{2.409pt}{1.000pt}}
\multiput(1020.00,852.92)(5.000,-4.000){2}{\rule{1.204pt}{1.000pt}}
\put(1030,846.92){\rule{2.168pt}{1.000pt}}
\multiput(1030.00,848.92)(4.500,-4.000){2}{\rule{1.084pt}{1.000pt}}
\multiput(1039.00,844.69)(0.579,-0.462){4}{\rule{1.750pt}{0.111pt}}
\multiput(1039.00,844.92)(5.368,-6.000){2}{\rule{0.875pt}{1.000pt}}
\multiput(1048.00,838.69)(0.606,-0.475){6}{\rule{1.679pt}{0.114pt}}
\multiput(1048.00,838.92)(6.516,-7.000){2}{\rule{0.839pt}{1.000pt}}
\multiput(1058.00,831.68)(0.470,-0.481){8}{\rule{1.375pt}{0.116pt}}
\multiput(1058.00,831.92)(6.146,-8.000){2}{\rule{0.688pt}{1.000pt}}
\multiput(1067.00,823.68)(0.423,-0.485){10}{\rule{1.250pt}{0.117pt}}
\multiput(1067.00,823.92)(6.406,-9.000){2}{\rule{0.625pt}{1.000pt}}
\multiput(1077.83,811.35)(0.485,-0.483){10}{\rule{0.117pt}{1.361pt}}
\multiput(1073.92,814.17)(9.000,-7.175){2}{\rule{1.000pt}{0.681pt}}
\multiput(1086.83,800.89)(0.485,-0.543){10}{\rule{0.117pt}{1.472pt}}
\multiput(1082.92,803.94)(9.000,-7.944){2}{\rule{1.000pt}{0.736pt}}
\multiput(1095.83,789.43)(0.485,-0.603){10}{\rule{0.117pt}{1.583pt}}
\multiput(1091.92,792.71)(9.000,-8.714){2}{\rule{1.000pt}{0.792pt}}
\multiput(1104.83,776.97)(0.485,-0.663){10}{\rule{0.117pt}{1.694pt}}
\multiput(1100.92,780.48)(9.000,-9.483){2}{\rule{1.000pt}{0.847pt}}
\multiput(1113.83,763.97)(0.485,-0.663){10}{\rule{0.117pt}{1.694pt}}
\multiput(1109.92,767.48)(9.000,-9.483){2}{\rule{1.000pt}{0.847pt}}
\multiput(1122.83,751.43)(0.485,-0.603){10}{\rule{0.117pt}{1.583pt}}
\multiput(1118.92,754.71)(9.000,-8.714){2}{\rule{1.000pt}{0.792pt}}
\multiput(1131.83,739.77)(0.481,-0.539){8}{\rule{0.116pt}{1.500pt}}
\multiput(1127.92,742.89)(8.000,-6.887){2}{\rule{1.000pt}{0.750pt}}
\multiput(1139.83,730.35)(0.485,-0.483){10}{\rule{0.117pt}{1.361pt}}
\multiput(1135.92,733.17)(9.000,-7.175){2}{\rule{1.000pt}{0.681pt}}
\multiput(1148.83,719.89)(0.485,-0.543){10}{\rule{0.117pt}{1.472pt}}
\multiput(1144.92,722.94)(9.000,-7.944){2}{\rule{1.000pt}{0.736pt}}
\multiput(1157.83,708.43)(0.485,-0.603){10}{\rule{0.117pt}{1.583pt}}
\multiput(1153.92,711.71)(9.000,-8.714){2}{\rule{1.000pt}{0.792pt}}
\multiput(1166.83,694.70)(0.481,-0.814){8}{\rule{0.116pt}{2.000pt}}
\multiput(1162.92,698.85)(8.000,-9.849){2}{\rule{1.000pt}{1.000pt}}
\multiput(1174.83,681.50)(0.485,-0.723){10}{\rule{0.117pt}{1.806pt}}
\multiput(1170.92,685.25)(9.000,-10.252){2}{\rule{1.000pt}{0.903pt}}
\multiput(1183.83,666.12)(0.485,-0.902){10}{\rule{0.117pt}{2.139pt}}
\multiput(1179.92,670.56)(9.000,-12.561){2}{\rule{1.000pt}{1.069pt}}
\multiput(1192.83,648.14)(0.481,-1.020){8}{\rule{0.116pt}{2.375pt}}
\multiput(1188.92,653.07)(8.000,-12.071){2}{\rule{1.000pt}{1.188pt}}
\multiput(1200.83,631.20)(0.485,-1.022){10}{\rule{0.117pt}{2.361pt}}
\multiput(1196.92,636.10)(9.000,-14.099){2}{\rule{1.000pt}{1.181pt}}
\multiput(1209.83,610.07)(0.481,-1.295){8}{\rule{0.116pt}{2.875pt}}
\multiput(1205.92,616.03)(8.000,-15.033){2}{\rule{1.000pt}{1.438pt}}
\multiput(1217.83,588.55)(0.481,-1.364){8}{\rule{0.116pt}{3.000pt}}
\multiput(1213.92,594.77)(8.000,-15.773){2}{\rule{1.000pt}{1.500pt}}
\multiput(1225.83,567.35)(0.485,-1.262){10}{\rule{0.117pt}{2.806pt}}
\multiput(1221.92,573.18)(9.000,-17.177){2}{\rule{1.000pt}{1.403pt}}
\multiput(1234.83,542.51)(0.481,-1.501){8}{\rule{0.116pt}{3.250pt}}
\multiput(1230.92,549.25)(8.000,-17.254){2}{\rule{1.000pt}{1.625pt}}
\multiput(1242.83,519.43)(0.485,-1.381){10}{\rule{0.117pt}{3.028pt}}
\multiput(1238.92,525.72)(9.000,-18.716){2}{\rule{1.000pt}{1.514pt}}
\multiput(1251.83,492.99)(0.481,-1.570){8}{\rule{0.116pt}{3.375pt}}
\multiput(1247.92,500.00)(8.000,-17.995){2}{\rule{1.000pt}{1.688pt}}
\multiput(1259.83,467.47)(0.481,-1.639){8}{\rule{0.116pt}{3.500pt}}
\multiput(1255.92,474.74)(8.000,-18.736){2}{\rule{1.000pt}{1.750pt}}
\multiput(1267.83,440.95)(0.481,-1.707){8}{\rule{0.116pt}{3.625pt}}
\multiput(1263.92,448.48)(8.000,-19.476){2}{\rule{1.000pt}{1.813pt}}
\multiput(1275.83,413.95)(0.481,-1.707){8}{\rule{0.116pt}{3.625pt}}
\multiput(1271.92,421.48)(8.000,-19.476){2}{\rule{1.000pt}{1.813pt}}
\multiput(1283.83,388.51)(0.485,-1.501){10}{\rule{0.117pt}{3.250pt}}
\multiput(1279.92,395.25)(9.000,-20.254){2}{\rule{1.000pt}{1.625pt}}
\multiput(1292.83,360.47)(0.481,-1.639){8}{\rule{0.116pt}{3.500pt}}
\multiput(1288.92,367.74)(8.000,-18.736){2}{\rule{1.000pt}{1.750pt}}
\multiput(1300.83,334.47)(0.481,-1.639){8}{\rule{0.116pt}{3.500pt}}
\multiput(1296.92,341.74)(8.000,-18.736){2}{\rule{1.000pt}{1.750pt}}
\multiput(1308.83,308.47)(0.481,-1.639){8}{\rule{0.116pt}{3.500pt}}
\multiput(1304.92,315.74)(8.000,-18.736){2}{\rule{1.000pt}{1.750pt}}
\multiput(1316.83,282.99)(0.481,-1.570){8}{\rule{0.116pt}{3.375pt}}
\multiput(1312.92,290.00)(8.000,-17.995){2}{\rule{1.000pt}{1.688pt}}
\multiput(1324.83,258.51)(0.481,-1.501){8}{\rule{0.116pt}{3.250pt}}
\multiput(1320.92,265.25)(8.000,-17.254){2}{\rule{1.000pt}{1.625pt}}
\multiput(1332.83,235.55)(0.481,-1.364){8}{\rule{0.116pt}{3.000pt}}
\multiput(1328.92,241.77)(8.000,-15.773){2}{\rule{1.000pt}{1.500pt}}
\multiput(1340.83,213.55)(0.481,-1.364){8}{\rule{0.116pt}{3.000pt}}
\multiput(1336.92,219.77)(8.000,-15.773){2}{\rule{1.000pt}{1.500pt}}
\multiput(1348.83,193.10)(0.481,-1.158){8}{\rule{0.116pt}{2.625pt}}
\multiput(1344.92,198.55)(8.000,-13.552){2}{\rule{1.000pt}{1.313pt}}
\multiput(1356.83,174.62)(0.481,-1.089){8}{\rule{0.116pt}{2.500pt}}
\multiput(1352.92,179.81)(8.000,-12.811){2}{\rule{1.000pt}{1.250pt}}
\multiput(1364.84,157.07)(0.475,-1.013){6}{\rule{0.114pt}{2.393pt}}
\multiput(1360.92,162.03)(7.000,-10.034){2}{\rule{1.000pt}{1.196pt}}
\multiput(1371.83,143.70)(0.481,-0.814){8}{\rule{0.116pt}{2.000pt}}
\multiput(1367.92,147.85)(8.000,-9.849){2}{\rule{1.000pt}{1.000pt}}
\multiput(1379.83,131.77)(0.481,-0.539){8}{\rule{0.116pt}{1.500pt}}
\multiput(1375.92,134.89)(8.000,-6.887){2}{\rule{1.000pt}{0.750pt}}
\multiput(1386.00,125.68)(0.402,-0.481){8}{\rule{1.250pt}{0.116pt}}
\multiput(1386.00,125.92)(5.406,-8.000){2}{\rule{0.625pt}{1.000pt}}
\multiput(1394.00,117.71)(0.320,-0.424){2}{\rule{1.850pt}{0.102pt}}
\multiput(1394.00,117.92)(4.160,-5.000){2}{\rule{0.925pt}{1.000pt}}
\put(983.0,861.0){\rule[-0.500pt]{2.168pt}{1.000pt}}
\end{picture}

\end{center}
\end{figure}

\section{$\tau\rightarrow K\bar{K}\nu$}
This process has been studied by several groups[34].
The $a_{1}$ field doesn't couple to $K\bar{K}$. The vertex
$a_{1}K\bar{K}$ in which the tensor $\varepsilon^{\mu\nu\alpha\beta}$
must be involved cannot be constructed. Therefore,
only the vector current contributes to this decay. The vertex
${\cal L}^{vK\bar{K}}$ has been used to calculate the electric form
factors of charged kaon and neutral kaon and theoretical predictions
are in good agreements with data[11]. We use
the same vertex(only the isovector
part) to calculate the decay rate of $\tau\rightarrow K\bar{K}
\nu$. This is a test of CVC. The related vertex is presented in
Ref.[11]
\begin{equation}
{\cal L}^{\rho K\bar{K}}=
{i\over\sqrt{2}}f_{\rho KK}\{\rho^{-}_{\mu}(K^{+}\partial^{\mu}\bar
{K}^{0}-\bar{K}^{0}\partial^{\mu}K^{+})-\rho^{+}_{\mu}(K^{-}\partial
^{\mu}K^{0}-K^{0}\partial^{\mu}K^{-})\},
\end{equation}
where $f_{\rho KK}$ is the same as the $f_{\rho\pi\pi}$(37) in the limit
of \(m_{q}=0\). By using VMD(3) and the vertex(58),
the matrix element is derived
\begin{equation}
<K^{-}K^{0}|\bar{\psi}\tau_{+}\gamma_{\mu}\psi|0>=
\frac{1}{\sqrt{4\omega_{1}\omega_{2}}}{1\over\sqrt{2}}(k_{1}-
k_{2})_{\mu}gf_{\rho\pi\pi}(q^{2})\frac{-m^{2}_{\rho}+iq\Gamma
_{\rho}(q^{2})}{q^{2}-m^{2}_{\rho}+iq\Gamma_{\rho}(q^{2})},
\end{equation}
where $k_{1}$ and $k_{2}$ are momentum of two kaons respectively and
\(q=k_{1}+k_{2}\).
Using this matrix element, the decay width is obtained
\begin{equation}
\frac{d\Gamma}{dq^{2}}(\tau\rightarrow K^{0}\bar{K}^{-}\nu)=
\frac{G^{2}}{(2\pi)
^{3}}\frac{cos^{2}\theta_{C}}
{384m^{3}_{\tau}}(m^{2}_{\tau}-q^{2})^{2}
(m^{2}_{\tau}+2q^{2})(1-{4m^{2}_{K}\over q^{2}})^{{3\over2}}
g^{2}f^{2}_{\rho\pi\pi}(q^{2})\frac{m^{4}_{\rho}+q^{2}\Gamma
^{2}_{\rho}(q^{2})}{(q^{2}-m^{2}_{\rho})^{2}+q^{2}
\Gamma^{2}_{\rho}(q^{2})}.
\end{equation}
The branching
ratio is computed to be
\[B=0.27\%\]
and the data are TPC/2$\gamma$[35]:$< 0.26$, ALEPH[36]: $0.26\pm0.09
\pm0.02$, CLEO[37]: $0.151\pm0.021\pm0.022$.
The distribution of the decay rate is shown in Fig.6.

\begin{figure}
\begin{center}

\setlength{\unitlength}{0.240900pt}
\ifx\plotpoint\undefined\newsavebox{\plotpoint}\fi
\begin{picture}(1500,900)(0,0)
\font\gnuplot=cmr10 at 10pt
\gnuplot
\sbox{\plotpoint}{\rule[-0.500pt]{1.000pt}{1.000pt}}%
\put(220.0,113.0){\rule[-0.500pt]{292.934pt}{1.000pt}}
\put(220.0,113.0){\rule[-0.500pt]{4.818pt}{1.000pt}}
\put(198,113){\makebox(0,0)[r]{0}}
\put(1416.0,113.0){\rule[-0.500pt]{4.818pt}{1.000pt}}
\put(220.0,266.0){\rule[-0.500pt]{4.818pt}{1.000pt}}
\put(198,266){\makebox(0,0)[r]{2}}
\put(1416.0,266.0){\rule[-0.500pt]{4.818pt}{1.000pt}}
\put(220.0,419.0){\rule[-0.500pt]{4.818pt}{1.000pt}}
\put(198,419){\makebox(0,0)[r]{4}}
\put(1416.0,419.0){\rule[-0.500pt]{4.818pt}{1.000pt}}
\put(220.0,571.0){\rule[-0.500pt]{4.818pt}{1.000pt}}
\put(198,571){\makebox(0,0)[r]{6}}
\put(1416.0,571.0){\rule[-0.500pt]{4.818pt}{1.000pt}}
\put(220.0,724.0){\rule[-0.500pt]{4.818pt}{1.000pt}}
\put(198,724){\makebox(0,0)[r]{8}}
\put(1416.0,724.0){\rule[-0.500pt]{4.818pt}{1.000pt}}
\put(220.0,877.0){\rule[-0.500pt]{4.818pt}{1.000pt}}
\put(198,877){\makebox(0,0)[r]{10}}
\put(1416.0,877.0){\rule[-0.500pt]{4.818pt}{1.000pt}}
\put(220.0,113.0){\rule[-0.500pt]{1.000pt}{4.818pt}}
\put(220,68){\makebox(0,0){0.8}}
\put(220.0,857.0){\rule[-0.500pt]{1.000pt}{4.818pt}}
\put(463.0,113.0){\rule[-0.500pt]{1.000pt}{4.818pt}}
\put(463,68){\makebox(0,0){1}}
\put(463.0,857.0){\rule[-0.500pt]{1.000pt}{4.818pt}}
\put(706.0,113.0){\rule[-0.500pt]{1.000pt}{4.818pt}}
\put(706,68){\makebox(0,0){1.2}}
\put(706.0,857.0){\rule[-0.500pt]{1.000pt}{4.818pt}}
\put(950.0,113.0){\rule[-0.500pt]{1.000pt}{4.818pt}}
\put(950,68){\makebox(0,0){1.4}}
\put(950.0,857.0){\rule[-0.500pt]{1.000pt}{4.818pt}}
\put(1193.0,113.0){\rule[-0.500pt]{1.000pt}{4.818pt}}
\put(1193,68){\makebox(0,0){1.6}}
\put(1193.0,857.0){\rule[-0.500pt]{1.000pt}{4.818pt}}
\put(1436.0,113.0){\rule[-0.500pt]{1.000pt}{4.818pt}}
\put(1436,68){\makebox(0,0){1.8}}
\put(1436.0,857.0){\rule[-0.500pt]{1.000pt}{4.818pt}}
\put(220.0,113.0){\rule[-0.500pt]{292.934pt}{1.000pt}}
\put(1436.0,113.0){\rule[-0.500pt]{1.000pt}{184.048pt}}
\put(220.0,877.0){\rule[-0.500pt]{292.934pt}{1.000pt}}
\put(45,495){\makebox(0,0){${d\Gamma\over d\sqrt{q^{2}}}\times 10^{15}$ }}
\put(828,23){\makebox(0,0){FiG.6       $\sqrt{q^{2}}$        GeV}}
\put(220.0,113.0){\rule[-0.500pt]{1.000pt}{184.048pt}}
\put(451,113){\usebox{\plotpoint}}
\multiput(452.83,113.00)(0.492,1.180){18}{\rule{0.118pt}{2.635pt}}
\multiput(448.92,113.00)(13.000,25.532){2}{\rule{1.000pt}{1.317pt}}
\multiput(465.83,144.00)(0.492,1.686){20}{\rule{0.119pt}{3.607pt}}
\multiput(461.92,144.00)(14.000,39.513){2}{\rule{1.000pt}{1.804pt}}
\multiput(479.83,191.00)(0.492,2.021){18}{\rule{0.118pt}{4.250pt}}
\multiput(475.92,191.00)(13.000,43.179){2}{\rule{1.000pt}{2.125pt}}
\multiput(492.83,243.00)(0.491,2.198){16}{\rule{0.118pt}{4.583pt}}
\multiput(488.92,243.00)(12.000,42.487){2}{\rule{1.000pt}{2.292pt}}
\multiput(504.83,295.00)(0.492,1.941){18}{\rule{0.118pt}{4.096pt}}
\multiput(500.92,295.00)(13.000,41.498){2}{\rule{1.000pt}{2.048pt}}
\multiput(517.83,345.00)(0.492,1.781){18}{\rule{0.118pt}{3.788pt}}
\multiput(513.92,345.00)(13.000,38.137){2}{\rule{1.000pt}{1.894pt}}
\multiput(530.83,391.00)(0.491,1.806){16}{\rule{0.118pt}{3.833pt}}
\multiput(526.92,391.00)(12.000,35.044){2}{\rule{1.000pt}{1.917pt}}
\multiput(542.83,434.00)(0.492,1.460){18}{\rule{0.118pt}{3.173pt}}
\multiput(538.92,434.00)(13.000,31.414){2}{\rule{1.000pt}{1.587pt}}
\multiput(555.83,472.00)(0.491,1.457){16}{\rule{0.118pt}{3.167pt}}
\multiput(551.92,472.00)(12.000,28.427){2}{\rule{1.000pt}{1.583pt}}
\multiput(567.83,507.00)(0.491,1.283){16}{\rule{0.118pt}{2.833pt}}
\multiput(563.92,507.00)(12.000,25.119){2}{\rule{1.000pt}{1.417pt}}
\multiput(579.83,538.00)(0.491,1.108){16}{\rule{0.118pt}{2.500pt}}
\multiput(575.92,538.00)(12.000,21.811){2}{\rule{1.000pt}{1.250pt}}
\multiput(591.83,565.00)(0.491,0.977){16}{\rule{0.118pt}{2.250pt}}
\multiput(587.92,565.00)(12.000,19.330){2}{\rule{1.000pt}{1.125pt}}
\multiput(603.83,589.00)(0.491,0.803){16}{\rule{0.118pt}{1.917pt}}
\multiput(599.92,589.00)(12.000,16.022){2}{\rule{1.000pt}{0.958pt}}
\multiput(615.83,609.00)(0.491,0.716){16}{\rule{0.118pt}{1.750pt}}
\multiput(611.92,609.00)(12.000,14.368){2}{\rule{1.000pt}{0.875pt}}
\multiput(627.83,627.00)(0.491,0.585){16}{\rule{0.118pt}{1.500pt}}
\multiput(623.92,627.00)(12.000,11.887){2}{\rule{1.000pt}{0.750pt}}
\multiput(639.83,642.00)(0.489,0.495){14}{\rule{0.118pt}{1.341pt}}
\multiput(635.92,642.00)(11.000,9.217){2}{\rule{1.000pt}{0.670pt}}
\multiput(649.00,655.83)(0.495,0.489){14}{\rule{1.341pt}{0.118pt}}
\multiput(649.00,651.92)(9.217,11.000){2}{\rule{0.670pt}{1.000pt}}
\multiput(661.00,666.83)(0.608,0.481){8}{\rule{1.625pt}{0.116pt}}
\multiput(661.00,662.92)(7.627,8.000){2}{\rule{0.813pt}{1.000pt}}
\multiput(672.00,674.84)(0.784,0.462){4}{\rule{2.083pt}{0.111pt}}
\multiput(672.00,670.92)(6.676,6.000){2}{\rule{1.042pt}{1.000pt}}
\multiput(683.00,680.86)(0.830,0.424){2}{\rule{2.450pt}{0.102pt}}
\multiput(683.00,676.92)(5.915,5.000){2}{\rule{1.225pt}{1.000pt}}
\put(694,683.42){\rule{2.650pt}{1.000pt}}
\multiput(694.00,681.92)(5.500,3.000){2}{\rule{1.325pt}{1.000pt}}
\put(705,685.92){\rule{2.650pt}{1.000pt}}
\multiput(705.00,684.92)(5.500,2.000){2}{\rule{1.325pt}{1.000pt}}
\put(716,687.42){\rule{2.650pt}{1.000pt}}
\multiput(716.00,686.92)(5.500,1.000){2}{\rule{1.325pt}{1.000pt}}
\put(727,687.42){\rule{2.650pt}{1.000pt}}
\multiput(727.00,687.92)(5.500,-1.000){2}{\rule{1.325pt}{1.000pt}}
\put(738,686.42){\rule{2.650pt}{1.000pt}}
\multiput(738.00,686.92)(5.500,-1.000){2}{\rule{1.325pt}{1.000pt}}
\put(749,684.42){\rule{2.650pt}{1.000pt}}
\multiput(749.00,685.92)(5.500,-3.000){2}{\rule{1.325pt}{1.000pt}}
\put(760,681.42){\rule{2.409pt}{1.000pt}}
\multiput(760.00,682.92)(5.000,-3.000){2}{\rule{1.204pt}{1.000pt}}
\put(770,677.92){\rule{2.650pt}{1.000pt}}
\multiput(770.00,679.92)(5.500,-4.000){2}{\rule{1.325pt}{1.000pt}}
\multiput(781.00,675.71)(0.660,-0.424){2}{\rule{2.250pt}{0.102pt}}
\multiput(781.00,675.92)(5.330,-5.000){2}{\rule{1.125pt}{1.000pt}}
\multiput(791.00,670.69)(0.784,-0.462){4}{\rule{2.083pt}{0.111pt}}
\multiput(791.00,670.92)(6.676,-6.000){2}{\rule{1.042pt}{1.000pt}}
\multiput(802.00,664.69)(0.681,-0.462){4}{\rule{1.917pt}{0.111pt}}
\multiput(802.00,664.92)(6.022,-6.000){2}{\rule{0.958pt}{1.000pt}}
\multiput(812.00,658.69)(0.606,-0.475){6}{\rule{1.679pt}{0.114pt}}
\multiput(812.00,658.92)(6.516,-7.000){2}{\rule{0.839pt}{1.000pt}}
\multiput(822.00,651.69)(0.688,-0.475){6}{\rule{1.821pt}{0.114pt}}
\multiput(822.00,651.92)(7.220,-7.000){2}{\rule{0.911pt}{1.000pt}}
\multiput(833.00,644.69)(0.606,-0.475){6}{\rule{1.679pt}{0.114pt}}
\multiput(833.00,644.92)(6.516,-7.000){2}{\rule{0.839pt}{1.000pt}}
\multiput(843.00,637.68)(0.539,-0.481){8}{\rule{1.500pt}{0.116pt}}
\multiput(843.00,637.92)(6.887,-8.000){2}{\rule{0.750pt}{1.000pt}}
\multiput(853.00,629.68)(0.539,-0.481){8}{\rule{1.500pt}{0.116pt}}
\multiput(853.00,629.92)(6.887,-8.000){2}{\rule{0.750pt}{1.000pt}}
\multiput(863.00,621.68)(0.483,-0.485){10}{\rule{1.361pt}{0.117pt}}
\multiput(863.00,621.92)(7.175,-9.000){2}{\rule{0.681pt}{1.000pt}}
\multiput(873.00,612.68)(0.483,-0.485){10}{\rule{1.361pt}{0.117pt}}
\multiput(873.00,612.92)(7.175,-9.000){2}{\rule{0.681pt}{1.000pt}}
\multiput(883.00,603.68)(0.423,-0.485){10}{\rule{1.250pt}{0.117pt}}
\multiput(883.00,603.92)(6.406,-9.000){2}{\rule{0.625pt}{1.000pt}}
\multiput(892.00,594.68)(0.483,-0.485){10}{\rule{1.361pt}{0.117pt}}
\multiput(892.00,594.92)(7.175,-9.000){2}{\rule{0.681pt}{1.000pt}}
\multiput(902.00,585.68)(0.437,-0.487){12}{\rule{1.250pt}{0.117pt}}
\multiput(902.00,585.92)(7.406,-10.000){2}{\rule{0.625pt}{1.000pt}}
\multiput(912.00,575.68)(0.437,-0.487){12}{\rule{1.250pt}{0.117pt}}
\multiput(912.00,575.92)(7.406,-10.000){2}{\rule{0.625pt}{1.000pt}}
\multiput(922.00,565.68)(0.423,-0.485){10}{\rule{1.250pt}{0.117pt}}
\multiput(922.00,565.92)(6.406,-9.000){2}{\rule{0.625pt}{1.000pt}}
\multiput(931.00,556.68)(0.437,-0.487){12}{\rule{1.250pt}{0.117pt}}
\multiput(931.00,556.92)(7.406,-10.000){2}{\rule{0.625pt}{1.000pt}}
\multiput(942.83,542.89)(0.485,-0.543){10}{\rule{0.117pt}{1.472pt}}
\multiput(938.92,545.94)(9.000,-7.944){2}{\rule{1.000pt}{0.736pt}}
\multiput(950.00,535.68)(0.437,-0.487){12}{\rule{1.250pt}{0.117pt}}
\multiput(950.00,535.92)(7.406,-10.000){2}{\rule{0.625pt}{1.000pt}}
\multiput(961.83,522.35)(0.485,-0.483){10}{\rule{0.117pt}{1.361pt}}
\multiput(957.92,525.17)(9.000,-7.175){2}{\rule{1.000pt}{0.681pt}}
\multiput(970.83,511.89)(0.485,-0.543){10}{\rule{0.117pt}{1.472pt}}
\multiput(966.92,514.94)(9.000,-7.944){2}{\rule{1.000pt}{0.736pt}}
\multiput(978.00,504.68)(0.437,-0.487){12}{\rule{1.250pt}{0.117pt}}
\multiput(978.00,504.92)(7.406,-10.000){2}{\rule{0.625pt}{1.000pt}}
\multiput(989.83,490.89)(0.485,-0.543){10}{\rule{0.117pt}{1.472pt}}
\multiput(985.92,493.94)(9.000,-7.944){2}{\rule{1.000pt}{0.736pt}}
\multiput(998.83,480.35)(0.485,-0.483){10}{\rule{0.117pt}{1.361pt}}
\multiput(994.92,483.17)(9.000,-7.175){2}{\rule{1.000pt}{0.681pt}}
\multiput(1007.83,469.89)(0.485,-0.543){10}{\rule{0.117pt}{1.472pt}}
\multiput(1003.92,472.94)(9.000,-7.944){2}{\rule{1.000pt}{0.736pt}}
\multiput(1016.83,458.89)(0.485,-0.543){10}{\rule{0.117pt}{1.472pt}}
\multiput(1012.92,461.94)(9.000,-7.944){2}{\rule{1.000pt}{0.736pt}}
\multiput(1025.83,448.35)(0.485,-0.483){10}{\rule{0.117pt}{1.361pt}}
\multiput(1021.92,451.17)(9.000,-7.175){2}{\rule{1.000pt}{0.681pt}}
\multiput(1034.83,437.89)(0.485,-0.543){10}{\rule{0.117pt}{1.472pt}}
\multiput(1030.92,440.94)(9.000,-7.944){2}{\rule{1.000pt}{0.736pt}}
\multiput(1043.83,427.35)(0.485,-0.483){10}{\rule{0.117pt}{1.361pt}}
\multiput(1039.92,430.17)(9.000,-7.175){2}{\rule{1.000pt}{0.681pt}}
\multiput(1052.83,416.89)(0.485,-0.543){10}{\rule{0.117pt}{1.472pt}}
\multiput(1048.92,419.94)(9.000,-7.944){2}{\rule{1.000pt}{0.736pt}}
\multiput(1061.83,405.89)(0.485,-0.543){10}{\rule{0.117pt}{1.472pt}}
\multiput(1057.92,408.94)(9.000,-7.944){2}{\rule{1.000pt}{0.736pt}}
\multiput(1070.83,395.35)(0.485,-0.483){10}{\rule{0.117pt}{1.361pt}}
\multiput(1066.92,398.17)(9.000,-7.175){2}{\rule{1.000pt}{0.681pt}}
\multiput(1079.83,385.35)(0.485,-0.483){10}{\rule{0.117pt}{1.361pt}}
\multiput(1075.92,388.17)(9.000,-7.175){2}{\rule{1.000pt}{0.681pt}}
\multiput(1088.83,374.89)(0.485,-0.543){10}{\rule{0.117pt}{1.472pt}}
\multiput(1084.92,377.94)(9.000,-7.944){2}{\rule{1.000pt}{0.736pt}}
\multiput(1097.83,363.77)(0.481,-0.539){8}{\rule{0.116pt}{1.500pt}}
\multiput(1093.92,366.89)(8.000,-6.887){2}{\rule{1.000pt}{0.750pt}}
\multiput(1105.83,354.35)(0.485,-0.483){10}{\rule{0.117pt}{1.361pt}}
\multiput(1101.92,357.17)(9.000,-7.175){2}{\rule{1.000pt}{0.681pt}}
\multiput(1114.83,343.89)(0.485,-0.543){10}{\rule{0.117pt}{1.472pt}}
\multiput(1110.92,346.94)(9.000,-7.944){2}{\rule{1.000pt}{0.736pt}}
\multiput(1123.83,332.77)(0.481,-0.539){8}{\rule{0.116pt}{1.500pt}}
\multiput(1119.92,335.89)(8.000,-6.887){2}{\rule{1.000pt}{0.750pt}}
\multiput(1130.00,326.68)(0.423,-0.485){10}{\rule{1.250pt}{0.117pt}}
\multiput(1130.00,326.92)(6.406,-9.000){2}{\rule{0.625pt}{1.000pt}}
\multiput(1140.83,313.77)(0.481,-0.539){8}{\rule{0.116pt}{1.500pt}}
\multiput(1136.92,316.89)(8.000,-6.887){2}{\rule{1.000pt}{0.750pt}}
\multiput(1148.83,304.35)(0.485,-0.483){10}{\rule{0.117pt}{1.361pt}}
\multiput(1144.92,307.17)(9.000,-7.175){2}{\rule{1.000pt}{0.681pt}}
\multiput(1157.83,293.77)(0.481,-0.539){8}{\rule{0.116pt}{1.500pt}}
\multiput(1153.92,296.89)(8.000,-6.887){2}{\rule{1.000pt}{0.750pt}}
\multiput(1165.83,284.29)(0.481,-0.470){8}{\rule{0.116pt}{1.375pt}}
\multiput(1161.92,287.15)(8.000,-6.146){2}{\rule{1.000pt}{0.688pt}}
\multiput(1172.00,278.68)(0.423,-0.485){10}{\rule{1.250pt}{0.117pt}}
\multiput(1172.00,278.92)(6.406,-9.000){2}{\rule{0.625pt}{1.000pt}}
\multiput(1182.83,266.29)(0.481,-0.470){8}{\rule{0.116pt}{1.375pt}}
\multiput(1178.92,269.15)(8.000,-6.146){2}{\rule{1.000pt}{0.688pt}}
\multiput(1190.83,257.29)(0.481,-0.470){8}{\rule{0.116pt}{1.375pt}}
\multiput(1186.92,260.15)(8.000,-6.146){2}{\rule{1.000pt}{0.688pt}}
\multiput(1197.00,251.68)(0.423,-0.485){10}{\rule{1.250pt}{0.117pt}}
\multiput(1197.00,251.92)(6.406,-9.000){2}{\rule{0.625pt}{1.000pt}}
\multiput(1207.83,239.29)(0.481,-0.470){8}{\rule{0.116pt}{1.375pt}}
\multiput(1203.92,242.15)(8.000,-6.146){2}{\rule{1.000pt}{0.688pt}}
\multiput(1214.00,233.68)(0.402,-0.481){8}{\rule{1.250pt}{0.116pt}}
\multiput(1214.00,233.92)(5.406,-8.000){2}{\rule{0.625pt}{1.000pt}}
\multiput(1222.00,225.68)(0.402,-0.481){8}{\rule{1.250pt}{0.116pt}}
\multiput(1222.00,225.92)(5.406,-8.000){2}{\rule{0.625pt}{1.000pt}}
\multiput(1230.00,217.68)(0.402,-0.481){8}{\rule{1.250pt}{0.116pt}}
\multiput(1230.00,217.92)(5.406,-8.000){2}{\rule{0.625pt}{1.000pt}}
\multiput(1238.00,209.68)(0.402,-0.481){8}{\rule{1.250pt}{0.116pt}}
\multiput(1238.00,209.92)(5.406,-8.000){2}{\rule{0.625pt}{1.000pt}}
\multiput(1246.00,201.68)(0.402,-0.481){8}{\rule{1.250pt}{0.116pt}}
\multiput(1246.00,201.92)(5.406,-8.000){2}{\rule{0.625pt}{1.000pt}}
\multiput(1254.00,193.69)(0.525,-0.475){6}{\rule{1.536pt}{0.114pt}}
\multiput(1254.00,193.92)(5.813,-7.000){2}{\rule{0.768pt}{1.000pt}}
\multiput(1263.00,186.69)(0.362,-0.475){6}{\rule{1.250pt}{0.114pt}}
\multiput(1263.00,186.92)(4.406,-7.000){2}{\rule{0.625pt}{1.000pt}}
\multiput(1270.00,179.69)(0.444,-0.475){6}{\rule{1.393pt}{0.114pt}}
\multiput(1270.00,179.92)(5.109,-7.000){2}{\rule{0.696pt}{1.000pt}}
\multiput(1278.00,172.69)(0.476,-0.462){4}{\rule{1.583pt}{0.111pt}}
\multiput(1278.00,172.92)(4.714,-6.000){2}{\rule{0.792pt}{1.000pt}}
\multiput(1286.00,166.69)(0.444,-0.475){6}{\rule{1.393pt}{0.114pt}}
\multiput(1286.00,166.92)(5.109,-7.000){2}{\rule{0.696pt}{1.000pt}}
\multiput(1294.00,159.71)(0.320,-0.424){2}{\rule{1.850pt}{0.102pt}}
\multiput(1294.00,159.92)(4.160,-5.000){2}{\rule{0.925pt}{1.000pt}}
\multiput(1302.00,154.69)(0.476,-0.462){4}{\rule{1.583pt}{0.111pt}}
\multiput(1302.00,154.92)(4.714,-6.000){2}{\rule{0.792pt}{1.000pt}}
\multiput(1310.00,148.71)(0.320,-0.424){2}{\rule{1.850pt}{0.102pt}}
\multiput(1310.00,148.92)(4.160,-5.000){2}{\rule{0.925pt}{1.000pt}}
\multiput(1318.00,143.71)(0.320,-0.424){2}{\rule{1.850pt}{0.102pt}}
\multiput(1318.00,143.92)(4.160,-5.000){2}{\rule{0.925pt}{1.000pt}}
\multiput(1326.00,138.71)(0.151,-0.424){2}{\rule{1.650pt}{0.102pt}}
\multiput(1326.00,138.92)(3.575,-5.000){2}{\rule{0.825pt}{1.000pt}}
\put(1333,131.92){\rule{1.927pt}{1.000pt}}
\multiput(1333.00,133.92)(4.000,-4.000){2}{\rule{0.964pt}{1.000pt}}
\put(1341,127.92){\rule{1.927pt}{1.000pt}}
\multiput(1341.00,129.92)(4.000,-4.000){2}{\rule{0.964pt}{1.000pt}}
\put(1349,124.42){\rule{1.686pt}{1.000pt}}
\multiput(1349.00,125.92)(3.500,-3.000){2}{\rule{0.843pt}{1.000pt}}
\put(1356,121.42){\rule{1.927pt}{1.000pt}}
\multiput(1356.00,122.92)(4.000,-3.000){2}{\rule{0.964pt}{1.000pt}}
\put(1364,118.42){\rule{1.927pt}{1.000pt}}
\multiput(1364.00,119.92)(4.000,-3.000){2}{\rule{0.964pt}{1.000pt}}
\put(1372,115.92){\rule{1.686pt}{1.000pt}}
\multiput(1372.00,116.92)(3.500,-2.000){2}{\rule{0.843pt}{1.000pt}}
\put(1379,113.92){\rule{1.927pt}{1.000pt}}
\multiput(1379.00,114.92)(4.000,-2.000){2}{\rule{0.964pt}{1.000pt}}
\put(1387,112.42){\rule{1.686pt}{1.000pt}}
\multiput(1387.00,112.92)(3.500,-1.000){2}{\rule{0.843pt}{1.000pt}}
\put(1394,111.42){\rule{1.927pt}{1.000pt}}
\multiput(1394.00,111.92)(4.000,-1.000){2}{\rule{0.964pt}{1.000pt}}
\end{picture}

\end{center}
\end{figure}
Due to the effects of the threshold and the $\rho$ resonance
there is a peak in
the distribution and it is positioned at 1.17GeV.

\section{The form factors of $\pi^{-}\rightarrow e\gamma\nu$}
As a test of ${\cal L}^{V,A}$(3,10), we study
the vector and the axial-vector form
factors of $\pi^{-}\rightarrow e\gamma\nu$.
In $\pi^{-}\rightarrow\gamma e\nu$ there are inner bremsstrahlung
from the lepton and meson and structure-dependent term[28]. The form
factors of the structure-dependent term have been studied[28].

The the structure-dependent form factors are defined as[28]
\begin{eqnarray}
\lefteqn{M^{V}_{SD}+M^{A}_{SD}=e{G\over\sqrt{2}}
cos\theta_{C}m_{\pi}
\epsilon^{*}_{\nu}\bar{\nu}\gamma_{\mu}(1-\gamma_{5})e}\nonumber
\\
&&\{F^{V}\varepsilon^{\mu\nu\alpha\beta}p_{\pi\alpha}k_{\beta}
+iF^{A}[g^{\mu\nu}k\cdot p_{\pi}-k^{\mu}p^{\nu}_{\pi}]+
iRtg^{\mu\nu}\},
\end{eqnarray}
where k and $p_{\pi}$ are momentum of photon and pion respectively,
\(t=k^{2}\).
It is known that $F^{V}$ is via CVC determined by the amplitude of
$\pi^{0}\rightarrow2\gamma$. In this theory the amplitude of
$\pi^{0}\rightarrow2\gamma$ obtained by triangle anomaly is obtained
by combining the vertex ${\cal L}^{\omega\rho\pi}$ and VMD.
Using $L^{V}$(3), ${\cal L}^{\omega\rho\pi}$, and VMD, it is
obtained
\begin{equation}
F^{V}=\frac{m_{\pi}}{2\sqrt{2}\pi^{2}f_{\pi}}=0.0268.
\end{equation}
The experiments[28] are
\(0.014\pm 0.009\), \(0.023^{+0.015}_{-0.013}\).

The form factors $F^{A}$ and R are determined by calculating
the matrix element $<\gamma|\bar{\psi}\tau_{-}\gamma_{\mu}
\gamma_{5}\psi|\pi^{+}>$ which is via VMD found from
$<\rho^{0}|\bar{\psi}\tau_{-}\gamma_{\mu}\gamma_{5}\psi|\pi^{+}>$
(35). However, for pion weak radiative decay besides the
axial-vector current conservation(in the limit \(m_{q}=0\)) the
electric current is conserved too.
In order to satisfy the
electric current conservation, the divergence of
$\rho$ field which is ignored
in the vertex ${\cal L}^{a_{1}\rho\pi}$(26)
must be kept and the term derived from the effective Lagrangian of
Ref.[11] is
\begin{equation}
-D\epsilon_{ijk}a^{i}_{\mu}\partial^{\mu}\pi^{j}
\partial^{\nu}\rho^{k}_{\nu},.
\end{equation}
where D is given in Eq.(30)
Adding the term(63) to the matrix element(35) we obtain
\begin{eqnarray}
<\gamma|\bar{\psi}\tau_{-}\gamma_{\mu}\gamma_{5}\psi|
\pi^{+}>=\frac{ie}{\sqrt{4\omega_{\pi}\omega_{\gamma}}}
({q_{\mu}q_{\nu}\over q^{2}}-g_{\mu\nu})\{Ag_{\lambda\nu}+Bp_{\pi
\nu}p_{\pi\lambda}+Dk_{\nu}k_{\lambda}\}
\epsilon^{*\lambda}_{\sigma}
\frac{1}{2}g^{3}f_{a}\frac{1}{q^{2}-m^{2}_{a}}\frac{m^{2}_{\rho}}
{m^{2}_{\rho}-k^{2}},
\end{eqnarray}
where \(q=p_{\pi}-k\). Using
the expressions of A(27), B(29), and D(30), it is proved that
\begin{equation}
A+k\cdot p_{\pi}B+Dk^{2}=0.
\end{equation}
Eq.(65) guarantees the electric current conservation. Ignoring
$q^{2}$ and $k^{2}$, the two form factors are found
\begin{eqnarray}
\lefteqn{F^{A}=\frac{1}{2\sqrt{2}\pi^{2}}\frac{m_{\pi}}{f_{\pi}}
\frac{m^{2}_{\rho}}{m^{2}_{a}}(1-{2c\over g})(1-{1\over2\pi^{2}
g^{2}})^{-1}=0.0102,}\\
&&R=\frac{g^{2}}{2\sqrt{2}}\frac{m_{\pi}}{f_{\pi}}\frac{m^{2}_{\rho}}
{m^{2}_{a}}\{{2c\over g}+{1\over\pi^{2}g^{2}}(1-{2c\over g})\}
(1-{1\over2\pi^{2}g^{2}})^{-1}.
\end{eqnarray}
The experimental values[28] of $F^{A}$ are $0.0106\pm0.006$, $0.0135\pm
0.0016$, $0.011\pm0.003$.

\section{Effective Lagrangian of \(\Delta s=1\) weak interactions}
It is natural to generalize the expressions of ${\cal L}^{V,A}$(3,10)
to the case of three flavors. The vector part of the weak
interaction,
instead $\rho$ meson in Eq.(3) the $K^{*}(892)$ meson takes part in
\begin{eqnarray}
\lefteqn{{\cal L}^{Vs}={g_{W}\over4}sin\theta_{C}
{1\over f_{K^{*}}}\{-{1\over 2}
(\partial_{\mu}W^{+}_{\nu}-\partial_{\nu}
W^{+}_{\mu})
(\partial^{\mu}K^{-\nu}-\partial^{\nu}K^{-\mu})}\nonumber \\
&&+(\partial_{\mu}W^{-}_{\nu}-\partial_{\nu}
W^{-}_{\mu})
(\partial^{\mu}K^{+\nu}-\partial^{\nu}K^{+\mu})
+W^{+}_{\mu}j^{-\mu}+W^{-}_{\mu}j^{+\mu}\},
\end{eqnarray}
Where $j^{\pm}_{\mu}$ are obtained by substituting $K^{\pm}_{\mu}
\rightarrow{g_{W}\over4}{1\over f_{K^{*}}}sin\theta_{C}W^{\pm}_{\mu}$
into the vertex
in which $K_{\mu}$ fields are involved.
In the chiral limit, $f_{K^{*}}$ is determined to be $g^{-1}$[11].
This Lagrangian has been used to calculate the form factors of
$K_{l3}$[11] and the results are in good agreement with data.

For axial-vector part ${\cal L}^{A}$ there are two $1^{+}$ K-mesons:
$K_{1}(1400)$ and $K_{1}(1275)$.
In Ref.[11] the chiral partner of
$K^{*}(892)$ meson, the $K_{1}$ meson,
is coupled to
\begin{equation}
\bar{\psi}\lambda^{a}\gamma_{\mu}\gamma_{5}\psi.
\end{equation}
The mass of this $K_{1}$ meson is derived as
\begin{equation}
(1-{1\over2\pi^{2}g^{2}})m^{2}_{K_{1}}=6m^{2}+m^{2}_{K^{*}},\;\;\;
m_{K_{1}}=1.32GeV.
\end{equation}
Theoretical value of $m_{K_{1}}$ is lower than the mass of
$K_{1}(1400)$ and greater than $K_{1}(1270)$'s mass.
The widths of three decay modes
($K_{1}\rightarrow K^{*}\pi$, $K\rho$,
$K\omega$) are calculated[11]. It is found that $K^{*}\pi$ channel
is dominant, however, $B(K\rho)$ is about $11\%$. The data[28]
show that the branching ratio of
$K_{1}(1400)$ decaying into $K\rho$ is very small. Therefore,
the meson coupled to the quark axial-vector current is not a pure
$K_{1}(1400)$ state, instead, it is a mixture of the two $K_{1}$
mesons. This state is coupled to the quark axial-vector current(69)
and it is $K_{a}$.
\begin{eqnarray}
K_{a}=cos\theta K_{1}(1400)+sin\theta K_{1}(1270),\nonumber \\
K_{b}=-sin\theta K_{1}(1400)+cos\theta K_{1}(1270).
\end{eqnarray}
In this theory $K_{a}$ is coupled to the
quark axial-vector current and the amplitudes of
$\tau\rightarrow K_{a}\nu$ is from the tree diagrams and at $O(N_{C})$
[11].
The production of $K_{b}$ in $\tau$
decay is through loop diagrams of mesons which is at $O(1)$
in large $N_{c}$ expansion[11]. This theory predicts a small
branching ratio for $K_{b}$ production in $\tau$ decays.

In the limit of \(m_{q}=0\), the currents
$\bar{\psi}\lambda_{a}\gamma_{\mu}\psi$ and
$\bar{\psi}\lambda_{a}\gamma_{\mu}\gamma_{5}\psi$
form an algebra of $SU(3)_{L}\times SU(3)_{R}$.
In this theory $K_{a}$ is taken as the chiral partner of
$K^{*}$ meson. The axial-vector
part of the weak interaction ${\cal L}^{A}$(10) is generalized
to the case of \(\Delta s= 1\)
\begin{eqnarray}
\lefteqn{{\cal L}^{As}=-
{g_{W}\over 4}{1\over f_{a}}
sin\theta_{C}\{-{1\over 2}(\partial_{\mu}W^{\pm}
_{\nu}-\partial_{\nu}W^{\pm}_{\mu})
(\partial^{\mu}K^{\mp
\nu}_{a}-\partial^{\nu}K^{\mp\mu}_{a})+W^{\pm\mu}j^{\mp}_{\mu}\}}
\nonumber \\
&&-{g_{W}\over 4}sin\theta_{C}
\Delta m^{2}f_{a}W^{\pm}_{\mu}K^{\mp,\mu}
_{a}-{1\over4}\sin\theta_{C}
f_{K}W^{\pm}_{\mu}\partial^{\mu}K^{\mp},
\end{eqnarray}
where $j^{\pm}_{\mu}$ are obtained by substituting $K^{\pm}_{a\mu}
\rightarrow-{g_{W}\over4f_{a}}
sin\theta_{C}W^{\pm}_{\mu}$ into the vertex in
which $K_{a}$ fields are involved.
In the limit of \(m_{q}=0\), $f_{a}$, $\Delta m^{2}$ are the same
as Eqs.(23,25) and \(f_{K}=f_{\pi}\).
${\cal L}^{As}$(72) can be exploited to study
$\tau$ mesonic decays. On the other hand, $\tau$ mesonic decays
of \(\Delta s=1\) provide crucial test on ${\cal L}^{As}$.

\section{$K_{a}$ dominance in $\tau\rightarrow K\pi\pi\nu$
decay}
The processes $\tau\rightarrow K\pi\pi\nu$
have been studied by many authors. In Ref.[3]
a chiral Lagrangian of the psedoscalars with introduction of vector
resonances($\rho$ and $K^{*}$) has been used to calculate the branching
ratios of $\tau\rightarrow K\pi\pi$. In Ref.[4] a chiral
Lagrangian of pseudoscalars and $\rho$ mesons is exploited.
In Refs.[38] the the mixture of the two $K_{1}$ resonances are
phenomenologically taken
into account in
studying the decay $\tau\rightarrow K_{1}(1400)(K_{1}(1270))\nu$.
In Refs.[5] meson vertices(independent of momentum)
and normalized Breit-Wigner propergators of the resonances are
exploited.

In this theory, like $\tau\rightarrow 3\pi\nu$, the contribution of the
contact terms containing more than three mesons to the processes
$\tau\rightarrow K\pi\pi\nu$ is too small and the processes are
dominated by $\tau\rightarrow K^{*}\pi\nu$ and $K\rho\nu$.

\subsection{$\tau\rightarrow K^{*}\pi\nu$}
We study the decay $\tau\rightarrow K^{*}\pi\nu$ first.
Both the vector and axial-vector currents contribute to the decay
$\tau^{-}\rightarrow \bar{K}^{*0}\pi^{-}\nu$. The vertex ${\cal L}
^{\pi K^{*}\bar{K^{*}}}$ contributes to the vector part and
has abnormal parity. It is from
anomaly. This vertex is derived from
\[-i{2m\over f_{\pi}}\pi^{i}<\bar{\psi}\tau_{i}\gamma_{5}\psi>.\]
The method obtaining the vertex ${\cal L}^{\pi K^{*}\bar{K^{*}}}$
from this quantity is the same as the one
used to derive the vertices of $\eta vv$(\(v=\rho, \omega, \phi\))
in Ref.[11].
\begin{eqnarray}
\lefteqn{{\cal L}^{\pi K^{*}\bar{K}^{*}}=
-\frac{N_{C}}{\sqrt{2}\pi^{2}g^{2}
f_{\pi}}\varepsilon^{\mu\nu\alpha\beta}\{\partial_{\mu}K^{+}_{\nu}
\partial_{\alpha}\bar{K}^{0}_{\beta}\pi^{-}+\partial_{\mu}K^{-}
_{\nu}\partial_{\alpha}K^{0}_{\beta}\pi^{+}}\nonumber \\
&&+{1\over\sqrt{2}}\pi^{0}(\partial_{\mu}K^{+}_{\nu}
\partial_{\alpha}K^{-}_{\nu}-\partial_{\mu}K^{0}_{\nu}
\partial_{\alpha}\bar{K}^{0}_{\beta})\}.
\end{eqnarray}
The vertex ${\cal L}^{WK^{*0}\pi^{-}}$ is derived by using the
substitution. Using ${\cal L}^{Vs}$(68) and the vertex(73),
the vector matrix element is obtained
\begin{equation}
<\bar{K}^{*0}\pi^{-}|\bar{\psi}\lambda_{+}\gamma_{\mu}\psi|0>=
\frac{-1}{\sqrt{4\omega E}}{1\over\sqrt{2}}\frac{N_{C}}
{\pi^{2}g^{2}f_{\pi}}\frac{m^{2}_{K^{*}}-i\sqrt{q^{2}}
\Gamma_{K^{*}}
(q^{2})}{q^{2}-m^{2}_{
K^{*}}+i\sqrt{q^{2}}\Gamma_{K^{*}}(q^{2})}\varepsilon^{\mu\nu
\alpha\beta}k_{\nu}p_{\alpha}\epsilon^{*\sigma}_{\beta},
\end{equation}
where p and k are momentum of $K^{*}$ and pion respectively,
\(q=k+p\).

The axial-vector matrix element is obtained by using the vertices:
$K_{a}K^{*}\pi$, $KK^{*}\pi$ which are presented in Ref.[11]. In the
chiral limit, the expression of the matrix element of the axial-vector
current
is similar to Eq.(35)
\begin{eqnarray}
\lefteqn{<\bar{K}^{*0}\pi^{-}|\bar{\psi}\lambda_{+}\gamma_{\mu}
\gamma_{5}\psi|0>=
\frac{i}{\sqrt{4\omega E}}
{1\over\sqrt{2}}({q_{\mu}q_{\nu}\over q^{2}}-g_{\mu\nu})
\epsilon^{*\lambda}_{\sigma}}\nonumber \\
&&\{\frac{g^{2}f_{a}m^{2}_{K^{*}}
-iqf^{-1}_{a}\Gamma_{K_{1}(1400)}(q^{2})}{
q^{2}-m^{2}_{K_{1}(1400)}+iq\Gamma_{K_{1}(1400)}(q^{2})}cos\theta
(A_{K_{1}(1400)}(q^{2})_{K^{*}}g^{\nu\lambda}+B_{K_{1}(1400)}
k^{\nu}k^{\lambda}) \nonumber \\
&&+\frac{g^{2}f_{a}m^{2}_{K^{*}}
-iqf^{-1}_{a}\Gamma_{K_{1}(1270)}(q^{2})}{
q^{2}-m^{2}_{K_{1}(1270)}+iq\Gamma_{K_{1}(1270)}(q^{2})}sin\theta
(A_{K_{1}(1270)}(q^{2})_{K^{*}}g^{\nu\lambda}+B_{K_{1}(1270)}
k^{\nu}k^{\lambda})\}
\end{eqnarray}
Let's determine
the amplitudes $A_{K_{1}(1400)}$, $B_{K_{1}(1400)}$,
$A_{K_{1}(1270)}$, and $B_{K_{1}(1270)}$. Eqs.(71) are written as
\begin{eqnarray}
K_{1}(1400)=cos\theta K_{a}-sin\theta K_{b},\nonumber \\
K_{1}(1270)=sin\theta K_{a}+cos\theta K_{b}.
\end{eqnarray}
The vertex of $K_{1}VP$ is presented in Ref.[11]
\begin{equation}
{\cal L}^{K_{1}VP}=f_{abc}\{AK^{a}_{1\mu}V^{b\mu}P^{c}
-BK^{a\mu}V^{b\nu}\partial_{\mu}\partial_{\nu}P^{c}\}.
\end{equation}
It is similar to Eq.(27) the amplitude $A^{K^{*}}_{K_{a}}$ is
determined to be
\begin{eqnarray}
\lefteqn{A(q^{2})^{K^{*}}_{K_{a}}={2\over f_{\pi}}gf_{a}\{
{m^{2}_{K_{a}}\over g^{2}f^{2}_{a}}-m^{2}_{K^{*}}+m^{2}_{K^{*}}
[{2c\over g}+{3\over4
\pi^{2}g^{2}}(1-{2c\over g})]}\nonumber \\
&&+q^{2}[{1\over 2\pi^{2}g^{2}}-
{2c\over g}-{3\over4\pi^{2}g^{2}}(1-{2c\over g})]\}.
\end{eqnarray}
$B_{K_{a}}$ is the same as Eq.(29). The amplitudes $A^{K^{*}}_{
K_{b}}$ and $B^{K^{*}}_{K_{b}}$ are unknown and we take them as
parameters.
Both $K_{1}(1400)$ and $K_{1}(1270)$ decay to $K\rho$ and
$K\omega$. Using the $SU(3)$ coefficients, for both $K_{1}$, it
is determined
\[B(K\omega)={1\over3}B(K\rho).\]
This relation agrees with data[28] reasonably well.
For the $K\rho$ decay mode
$A^{\rho}_{K_{b}}$ and $B^{\rho}_{K_{b}}$ are other
two parameters. In the decays of the two $K_{1}$ mesons the
momentum of pion or kaon is low, therefore, the decay widths are
insensitive to the amplitude B. We take
\[B^{K^{*}}_{K_{b}}=B^{\rho}_{K_{b}}\equiv B_{b}.\]
The decay width of the $K_{1}$ meson is derived from Eq.(77)
\begin{equation}
\Gamma_{K_{1}}=\frac{k}{32\pi}\frac{1}{\sqrt{q^{2}}
m_{K_{1}}}\{(3+{k^{2}\over m^{2}_{v}})A^{2}(q^{2})
-A(q^{2})B(q^{2}+m^{2}_{v})\frac{k^{2}}{m^{2}_{v}}
+{q^{2}\over m^{2}_{v}}k^{4}B^{2}\},
\end{equation}
where \(q^{2}=m^{2}_{K_{1}}\), \(v=K^{*}, \rho\),
k is the momentum of pion or kaon
\[k=\{{1\over4m^{2}_{K_{1}}}(m^{2}_{K_{1}}+m^{2}_{v}-m^{2}_{P})
^{2}-4m^{2}_{v}\}^{{1\over2}},\]
$m_{P}$ is the mass of pion or kaon.

we choose the parameters as
\begin{equation}
\theta=30^{0},\;\;\;A^{K^{*}}_{b}=-4.5GeV,\;\;\;
A^{\rho}_{b}=5.0GeV,\;\;\;B_{b}=0.8GeV^{-1},
\end{equation}
from which the decay widths are obtained
\begin{eqnarray}
\Gamma(K_{1}(1400)\rightarrow K^{*}\pi)=159MeV,\;\;\;
\Gamma(K_{1}(1400)\rightarrow K\rho)=10.5MeV,\nonumber \\
\Gamma(K_{1}(1270)\rightarrow K^{*}\pi)=12.4MeV,\;\;\;
\Gamma(K_{1}(1270)\rightarrow K\rho)=26.8MeV.
\end{eqnarray}
The value of $\theta$ is about the same as the one determined in
Ref.[38].
The data[28] are 163.4$(1\pm0.13)$MeV, $5.22\pm 5.22$ MeV,
$37.8(1\pm0.28)$MeV, and 14.4$(1\pm 0.27)$MeV respectively.

Using the two matrix elements(74,75), the distribution of the decay
rate is derived
\begin{eqnarray}
\lefteqn{{d\Gamma\over dq^{2}}(\tau^{-}\rightarrow
\bar{K}^{*0}\pi^{-}\nu)={G^{2}\over
(2\pi)^{3}}\frac{sin^{2}\theta_{C}}{128m^{3}_{\tau}q^{4}}
(m^{2}_{\tau}-q^{2})^{2}(m^{2}_{\tau}+2q^{2})
\{(q^{2}+m^{2}_{K^{*}}
-m^{2}_{\pi})^{2}-4q^{2}m^{2}_{K^{*}}\}^{{1\over2}}}\nonumber \\
&&\{\frac{6}{\pi^{4}g^{2}f^{2}_{\pi}}\frac{m^{4}_{K^{*}}+q^{2}
\Gamma^{2}_{K^{*}}(q^{2})}{(q^{2}-m^{2}_{K^{*}})^{2}+q^{2}
\Gamma^{2}_{K^{*}}(q^{2})}[(p\cdot q)^{2}-q^{2}m^{2}_{K^{*}}]
\nonumber \\
&&+|A|^{2}[1+{1\over12m^{2}_{K^{*}}q^{2}}(q^{2}-m^{2}_{K^{*}})^{2}]
-(BA^{*}+B^{*}A){1\over24m^{2}_{K^{*}}
q^{2}}(q^{2}+m^{2}_{K^{*}})(q^{2}-m^{2}
_{K^{*}})^{2}\nonumber \\
&&+\frac{|B|^{2}}{48m^{2}_{K^{*}}q^{2}}(q^{2}
-m^{2}_{K^{*}})^{4}\}.
\end{eqnarray}
where p is the momentum of $K^{*}$, $q^{2}$ is the invariant mass
squared of $K^{*}\pi$, and
\begin{eqnarray}
\lefteqn{A=\frac{g^{2}f_{a}m^{2}_{K^{*}}
-i\sqrt{q^{2}}f^{-1}_{a}\Gamma_{K_{1}
(1400)}}{q^{2}-m^{2}_{K_{1}(1400)}+i\sqrt{q^{2}}\Gamma_{K_{1}(1400)}}
cos\theta A^{K^{*}}_{K_{1}(1400)}
+\frac{g^{2}f_{a}m^{2}_{K^{*}}-i\sqrt{q^{2}}f^{-1}_{a}\Gamma_{K_{1}
(1270)}}{q^{2}-m^{2}_{K_{1}(1270)}+i\sqrt{q^{2}}\Gamma_{K_{1}(1270)}}
sin\theta A^{K^{*}}_{K_{1}(1270)},}\nonumber \\
&&B=\frac{g^{2}f_{a}m^{2}_{K^{*}}-i\sqrt{q^{2}}f^{-1}_{a}
\Gamma_{K_{1}
(1400)}}{q^{2}-m^{2}_{K_{1}(1400)}+i\sqrt{q^{2}}\Gamma_{K_{1}(1400)}}
cos\theta B^{K^{*}}_{K_{1}(1400)}
+\frac{g^{2}f_{a}m^{2}_{K^{*}}-i\sqrt{q^{2}}f^{-1}_{a}\Gamma_{K_{1}
(1270)}}{q^{2}-m^{2}_{K_{1}(1270)}+i\sqrt{q^{2}}\Gamma_{K_{1}(1270)}}
sin\theta B^{K^{*}}_{K_{1}(1270)},
\end{eqnarray}
where
\begin{eqnarray}
A^{K^{*}}_{K_{1}(1400)}=cos\theta A^{K^{*}}_{a}-sin\theta A^{K^{*}}_{b}
,\;\;\;
B^{K^{*}}_{K_{1}(1400)}=cos\theta B_{a}-sin\theta B_{b},\nonumber \\
A^{K^{*}}_{K_{1}(1270)}=sin\theta A^{K^{*}}_{a}+cos\theta A^{K^{*}}_{b}
,\;\;\;
B^{K^{*}}_{K_{1}(1270)}=sin\theta B_{a}+cos\theta B_{b},
\end{eqnarray}
In the range of $q^{2}$ the main decay channels of $K^{*}$ are
$K\pi$ and $K\eta$(the vertex of $K^{*}K\eta$ is shown in (88)).
The decay width of $K^{*}$ is derived
\begin{eqnarray}
\lefteqn{\Gamma(q^{2})_{K^{*}}=\frac{f^{2}_{\rho\pi\pi}(q^{2})}
{8\pi}\frac{k^{3}}{\sqrt{q^{2}}m_{K^{*}}}+cos^{2}20^{0}
\frac{f^{2}_{\rho\pi\pi}(q^{2})}
{8\pi}\frac{k^{'3}}{\sqrt{q^{2}}m_{K^{*}}},}\nonumber \\
&&k=\{{1\over4q^{2}}(q^{2}+m^{2}_{K}-m^{2}_{\pi})^{2}-m^{2}_{K}
\}^{{1\over2}},\nonumber \\
&&k'=\{{1\over4q^{2}}(q^{2}+m^{2}_{K}-m^{2}_{\eta})^{2}-m^{2}_{K}
\}^{{1\over2}}.
\end{eqnarray}
In $\Gamma_{K_{1}}(q^{2})$ the
decay modes $K^{*}\pi$, $K\rho$ and
$K\omega$ are included.
\begin{eqnarray}
\lefteqn{\Gamma(q^{2})_{K_{1}}=\frac{k}{32\pi}\frac{1}{\sqrt{q^{2}}
m_{K_{1}}}\{(3+{k^{2}\over m^{2}_{K^{*}}})A^{2}(q^{2})_{K^{*}}
-A(q^{2})_{K^{*}}B(q^{2}+m^{2}_{K^{*}})\frac{k^{2}}{m^{2}_{K^{*}}}
+{q^{2}\over m^{2}_{K^{*}}}k^{4}B^{2}\}}\nonumber \\
&&+{4\over3}\frac{k'}{32\pi}\frac{1}{\sqrt{q^{2}}
m_{K_{1}}}\{(3+{k^{2}\over m^{2}_{\rho}})A^{2}(q^{2})
-A(q^{2})B(q^{2}+m^{2}_{\rho})\frac{k^{2}}{m^{2}_{\rho}}
+{q^{2}\over m^{2}_{\rho}}k^{4}B^{2}\},\nonumber \\
&&k=\{{1\over4q^{2}}(q^{2}+m^{2}_{K^{*}}-m^{2}_{\pi})^{2}
-m^{2}_{K^{*}}
\}^{{1\over2}},\nonumber \\
&&k'=\{{1\over4q^{2}}(q^{2}+m^{2}_{K}-m^{2}_{\rho})^{2}
-m^{2}_{K}\}^{{1\over2}}.
\end{eqnarray}
For $K_{1}(1270)$ \(\Gamma(K_{1}(1270)\rightarrow K^{*}_{0}(1430)\pi)
=25.2MeV\) is included.
The distribution is shown in Fig.7 and the branching ratio
is calculated

\begin{figure}
\begin{center}

\setlength{\unitlength}{0.240900pt}
\ifx\plotpoint\undefined\newsavebox{\plotpoint}\fi
\begin{picture}(1500,900)(0,0)
\font\gnuplot=cmr10 at 10pt
\gnuplot
\sbox{\plotpoint}{\rule[-0.500pt]{1.000pt}{1.000pt}}%
\put(220.0,113.0){\rule[-0.500pt]{292.934pt}{1.000pt}}
\put(220.0,113.0){\rule[-0.500pt]{4.818pt}{1.000pt}}
\put(198,113){\makebox(0,0)[r]{0}}
\put(1416.0,113.0){\rule[-0.500pt]{4.818pt}{1.000pt}}
\put(220.0,203.0){\rule[-0.500pt]{4.818pt}{1.000pt}}
\put(198,203){\makebox(0,0)[r]{2}}
\put(1416.0,203.0){\rule[-0.500pt]{4.818pt}{1.000pt}}
\put(220.0,293.0){\rule[-0.500pt]{4.818pt}{1.000pt}}
\put(198,293){\makebox(0,0)[r]{4}}
\put(1416.0,293.0){\rule[-0.500pt]{4.818pt}{1.000pt}}
\put(220.0,383.0){\rule[-0.500pt]{4.818pt}{1.000pt}}
\put(198,383){\makebox(0,0)[r]{6}}
\put(1416.0,383.0){\rule[-0.500pt]{4.818pt}{1.000pt}}
\put(220.0,473.0){\rule[-0.500pt]{4.818pt}{1.000pt}}
\put(198,473){\makebox(0,0)[r]{8}}
\put(1416.0,473.0){\rule[-0.500pt]{4.818pt}{1.000pt}}
\put(220.0,562.0){\rule[-0.500pt]{4.818pt}{1.000pt}}
\put(198,562){\makebox(0,0)[r]{10}}
\put(1416.0,562.0){\rule[-0.500pt]{4.818pt}{1.000pt}}
\put(220.0,652.0){\rule[-0.500pt]{4.818pt}{1.000pt}}
\put(198,652){\makebox(0,0)[r]{12}}
\put(1416.0,652.0){\rule[-0.500pt]{4.818pt}{1.000pt}}
\put(220.0,742.0){\rule[-0.500pt]{4.818pt}{1.000pt}}
\put(198,742){\makebox(0,0)[r]{14}}
\put(1416.0,742.0){\rule[-0.500pt]{4.818pt}{1.000pt}}
\put(220.0,832.0){\rule[-0.500pt]{4.818pt}{1.000pt}}
\put(198,832){\makebox(0,0)[r]{16}}
\put(1416.0,832.0){\rule[-0.500pt]{4.818pt}{1.000pt}}
\put(220.0,113.0){\rule[-0.500pt]{1.000pt}{4.818pt}}
\put(220,68){\makebox(0,0){1}}
\put(220.0,857.0){\rule[-0.500pt]{1.000pt}{4.818pt}}
\put(372.0,113.0){\rule[-0.500pt]{1.000pt}{4.818pt}}
\put(372,68){\makebox(0,0){1.1}}
\put(372.0,857.0){\rule[-0.500pt]{1.000pt}{4.818pt}}
\put(524.0,113.0){\rule[-0.500pt]{1.000pt}{4.818pt}}
\put(524,68){\makebox(0,0){1.2}}
\put(524.0,857.0){\rule[-0.500pt]{1.000pt}{4.818pt}}
\put(676.0,113.0){\rule[-0.500pt]{1.000pt}{4.818pt}}
\put(676,68){\makebox(0,0){1.3}}
\put(676.0,857.0){\rule[-0.500pt]{1.000pt}{4.818pt}}
\put(828.0,113.0){\rule[-0.500pt]{1.000pt}{4.818pt}}
\put(828,68){\makebox(0,0){1.4}}
\put(828.0,857.0){\rule[-0.500pt]{1.000pt}{4.818pt}}
\put(980.0,113.0){\rule[-0.500pt]{1.000pt}{4.818pt}}
\put(980,68){\makebox(0,0){1.5}}
\put(980.0,857.0){\rule[-0.500pt]{1.000pt}{4.818pt}}
\put(1132.0,113.0){\rule[-0.500pt]{1.000pt}{4.818pt}}
\put(1132,68){\makebox(0,0){1.6}}
\put(1132.0,857.0){\rule[-0.500pt]{1.000pt}{4.818pt}}
\put(1284.0,113.0){\rule[-0.500pt]{1.000pt}{4.818pt}}
\put(1284,68){\makebox(0,0){1.7}}
\put(1284.0,857.0){\rule[-0.500pt]{1.000pt}{4.818pt}}
\put(1436.0,113.0){\rule[-0.500pt]{1.000pt}{4.818pt}}
\put(1436,68){\makebox(0,0){1.8}}
\put(1436.0,857.0){\rule[-0.500pt]{1.000pt}{4.818pt}}
\put(220.0,113.0){\rule[-0.500pt]{292.934pt}{1.000pt}}
\put(1436.0,113.0){\rule[-0.500pt]{1.000pt}{184.048pt}}
\put(220.0,877.0){\rule[-0.500pt]{292.934pt}{1.000pt}}
\put(45,495){\makebox(0,0){${d\Gamma\over d\sqrt{q^{2}}}\times10^{15}$ }}
\put(828,23){\makebox(0,0){Fig.7   GeV}}
\put(220.0,113.0){\rule[-0.500pt]{1.000pt}{184.048pt}}
\put(306,230){\usebox{\plotpoint}}
\multiput(307.83,230.00)(0.493,0.571){22}{\rule{0.119pt}{1.450pt}}
\multiput(303.92,230.00)(15.000,14.990){2}{\rule{1.000pt}{0.725pt}}
\multiput(321.00,249.83)(0.467,0.493){22}{\rule{1.250pt}{0.119pt}}
\multiput(321.00,245.92)(12.406,15.000){2}{\rule{0.625pt}{1.000pt}}
\multiput(336.00,264.83)(0.500,0.492){18}{\rule{1.327pt}{0.118pt}}
\multiput(336.00,260.92)(11.246,13.000){2}{\rule{0.663pt}{1.000pt}}
\multiput(350.00,277.83)(0.591,0.489){14}{\rule{1.523pt}{0.118pt}}
\multiput(350.00,273.92)(10.840,11.000){2}{\rule{0.761pt}{1.000pt}}
\multiput(364.00,288.83)(0.723,0.485){10}{\rule{1.806pt}{0.117pt}}
\multiput(364.00,284.92)(10.252,9.000){2}{\rule{0.903pt}{1.000pt}}
\multiput(378.00,297.83)(0.814,0.481){8}{\rule{2.000pt}{0.116pt}}
\multiput(378.00,293.92)(9.849,8.000){2}{\rule{1.000pt}{1.000pt}}
\multiput(392.00,305.84)(1.092,0.462){4}{\rule{2.583pt}{0.111pt}}
\multiput(392.00,301.92)(8.638,6.000){2}{\rule{1.292pt}{1.000pt}}
\multiput(406.00,311.84)(1.092,0.462){4}{\rule{2.583pt}{0.111pt}}
\multiput(406.00,307.92)(8.638,6.000){2}{\rule{1.292pt}{1.000pt}}
\put(420,315.42){\rule{3.373pt}{1.000pt}}
\multiput(420.00,313.92)(7.000,3.000){2}{\rule{1.686pt}{1.000pt}}
\put(434,318.42){\rule{3.132pt}{1.000pt}}
\multiput(434.00,316.92)(6.500,3.000){2}{\rule{1.566pt}{1.000pt}}
\put(474,318.42){\rule{3.132pt}{1.000pt}}
\multiput(474.00,319.92)(6.500,-3.000){2}{\rule{1.566pt}{1.000pt}}
\multiput(487.00,316.69)(1.092,-0.462){4}{\rule{2.583pt}{0.111pt}}
\multiput(487.00,316.92)(8.638,-6.000){2}{\rule{1.292pt}{1.000pt}}
\multiput(501.00,310.68)(0.745,-0.481){8}{\rule{1.875pt}{0.116pt}}
\multiput(501.00,310.92)(9.108,-8.000){2}{\rule{0.937pt}{1.000pt}}
\multiput(514.00,302.68)(0.498,-0.491){16}{\rule{1.333pt}{0.118pt}}
\multiput(514.00,302.92)(10.233,-12.000){2}{\rule{0.667pt}{1.000pt}}
\multiput(528.83,286.43)(0.491,-0.628){16}{\rule{0.118pt}{1.583pt}}
\multiput(524.92,289.71)(12.000,-12.714){2}{\rule{1.000pt}{0.792pt}}
\multiput(540.83,268.94)(0.492,-0.820){18}{\rule{0.118pt}{1.942pt}}
\multiput(536.92,272.97)(13.000,-17.969){2}{\rule{1.000pt}{0.971pt}}
\multiput(553.83,244.70)(0.492,-1.100){18}{\rule{0.118pt}{2.481pt}}
\multiput(549.92,249.85)(13.000,-23.851){2}{\rule{1.000pt}{1.240pt}}
\multiput(566.83,213.79)(0.492,-1.340){18}{\rule{0.118pt}{2.942pt}}
\multiput(562.92,219.89)(13.000,-28.893){2}{\rule{1.000pt}{1.471pt}}
\multiput(579.83,178.55)(0.491,-1.370){16}{\rule{0.118pt}{3.000pt}}
\multiput(575.92,184.77)(12.000,-26.773){2}{\rule{1.000pt}{1.500pt}}
\multiput(591.83,158.00)(0.492,1.060){18}{\rule{0.118pt}{2.404pt}}
\multiput(587.92,158.00)(13.000,23.011){2}{\rule{1.000pt}{1.202pt}}
\multiput(604.83,186.00)(0.491,20.556){16}{\rule{0.118pt}{39.667pt}}
\multiput(600.92,186.00)(12.000,390.670){2}{\rule{1.000pt}{19.833pt}}
\multiput(616.83,592.93)(0.491,-8.129){16}{\rule{0.118pt}{15.917pt}}
\multiput(612.92,625.96)(12.000,-154.964){2}{\rule{1.000pt}{7.958pt}}
\multiput(628.83,471.00)(0.491,1.544){16}{\rule{0.118pt}{3.333pt}}
\multiput(624.92,471.00)(12.000,30.082){2}{\rule{1.000pt}{1.667pt}}
\multiput(640.83,508.00)(0.492,0.980){18}{\rule{0.118pt}{2.250pt}}
\multiput(636.92,508.00)(13.000,21.330){2}{\rule{1.000pt}{1.125pt}}
\multiput(653.83,534.00)(0.491,0.934){16}{\rule{0.118pt}{2.167pt}}
\multiput(649.92,534.00)(12.000,18.503){2}{\rule{1.000pt}{1.083pt}}
\multiput(665.83,557.00)(0.491,0.890){16}{\rule{0.118pt}{2.083pt}}
\multiput(661.92,557.00)(12.000,17.676){2}{\rule{1.000pt}{1.042pt}}
\multiput(677.83,579.00)(0.491,0.847){16}{\rule{0.118pt}{2.000pt}}
\multiput(673.92,579.00)(12.000,16.849){2}{\rule{1.000pt}{1.000pt}}
\multiput(689.83,600.00)(0.489,0.926){14}{\rule{0.118pt}{2.159pt}}
\multiput(685.92,600.00)(11.000,16.519){2}{\rule{1.000pt}{1.080pt}}
\multiput(700.83,621.00)(0.491,0.847){16}{\rule{0.118pt}{2.000pt}}
\multiput(696.92,621.00)(12.000,16.849){2}{\rule{1.000pt}{1.000pt}}
\multiput(712.83,642.00)(0.491,0.847){16}{\rule{0.118pt}{2.000pt}}
\multiput(708.92,642.00)(12.000,16.849){2}{\rule{1.000pt}{1.000pt}}
\multiput(724.83,663.00)(0.489,0.974){14}{\rule{0.118pt}{2.250pt}}
\multiput(720.92,663.00)(11.000,17.330){2}{\rule{1.000pt}{1.125pt}}
\multiput(735.83,685.00)(0.491,0.847){16}{\rule{0.118pt}{2.000pt}}
\multiput(731.92,685.00)(12.000,16.849){2}{\rule{1.000pt}{1.000pt}}
\multiput(747.83,706.00)(0.491,0.803){16}{\rule{0.118pt}{1.917pt}}
\multiput(743.92,706.00)(12.000,16.022){2}{\rule{1.000pt}{0.958pt}}
\multiput(759.83,726.00)(0.489,0.782){14}{\rule{0.118pt}{1.886pt}}
\multiput(755.92,726.00)(11.000,14.085){2}{\rule{1.000pt}{0.943pt}}
\multiput(770.83,744.00)(0.489,0.639){14}{\rule{0.118pt}{1.614pt}}
\multiput(766.92,744.00)(11.000,11.651){2}{\rule{1.000pt}{0.807pt}}
\multiput(780.00,760.83)(0.495,0.489){14}{\rule{1.341pt}{0.118pt}}
\multiput(780.00,756.92)(9.217,11.000){2}{\rule{0.670pt}{1.000pt}}
\multiput(792.00,771.86)(0.830,0.424){2}{\rule{2.450pt}{0.102pt}}
\multiput(792.00,767.92)(5.915,5.000){2}{\rule{1.225pt}{1.000pt}}
\put(803,771.42){\rule{2.650pt}{1.000pt}}
\multiput(803.00,772.92)(5.500,-3.000){2}{\rule{1.325pt}{1.000pt}}
\multiput(814.00,769.68)(0.447,-0.489){14}{\rule{1.250pt}{0.118pt}}
\multiput(814.00,769.92)(8.406,-11.000){2}{\rule{0.625pt}{1.000pt}}
\multiput(826.83,752.04)(0.489,-0.926){14}{\rule{0.118pt}{2.159pt}}
\multiput(822.92,756.52)(11.000,-16.519){2}{\rule{1.000pt}{1.080pt}}
\multiput(837.83,727.64)(0.489,-1.357){14}{\rule{0.118pt}{2.977pt}}
\multiput(833.92,733.82)(11.000,-23.821){2}{\rule{1.000pt}{1.489pt}}
\multiput(848.83,694.62)(0.489,-1.740){14}{\rule{0.118pt}{3.705pt}}
\multiput(844.92,702.31)(11.000,-30.311){2}{\rule{1.000pt}{1.852pt}}
\multiput(859.83,653.60)(0.489,-2.123){14}{\rule{0.118pt}{4.432pt}}
\multiput(855.92,662.80)(11.000,-36.802){2}{\rule{1.000pt}{2.216pt}}
\multiput(870.83,606.47)(0.489,-2.267){14}{\rule{0.118pt}{4.705pt}}
\multiput(866.92,616.24)(11.000,-39.235){2}{\rule{1.000pt}{2.352pt}}
\multiput(881.83,557.09)(0.489,-2.315){14}{\rule{0.118pt}{4.795pt}}
\multiput(877.92,567.05)(11.000,-40.047){2}{\rule{1.000pt}{2.398pt}}
\multiput(892.83,507.09)(0.489,-2.315){14}{\rule{0.118pt}{4.795pt}}
\multiput(888.92,517.05)(11.000,-40.047){2}{\rule{1.000pt}{2.398pt}}
\multiput(903.83,456.45)(0.487,-2.405){12}{\rule{0.117pt}{4.950pt}}
\multiput(899.92,466.73)(10.000,-36.726){2}{\rule{1.000pt}{2.475pt}}
\multiput(913.83,413.11)(0.489,-1.932){14}{\rule{0.118pt}{4.068pt}}
\multiput(909.92,421.56)(11.000,-33.556){2}{\rule{1.000pt}{2.034pt}}
\multiput(924.83,372.24)(0.489,-1.788){14}{\rule{0.118pt}{3.795pt}}
\multiput(920.92,380.12)(11.000,-31.122){2}{\rule{1.000pt}{1.898pt}}
\multiput(935.83,334.26)(0.487,-1.660){12}{\rule{0.117pt}{3.550pt}}
\multiput(931.92,341.63)(10.000,-25.632){2}{\rule{1.000pt}{1.775pt}}
\multiput(945.83,304.02)(0.489,-1.309){14}{\rule{0.118pt}{2.886pt}}
\multiput(941.92,310.01)(11.000,-23.009){2}{\rule{1.000pt}{1.443pt}}
\multiput(956.83,275.58)(0.487,-1.235){12}{\rule{0.117pt}{2.750pt}}
\multiput(952.92,281.29)(10.000,-19.292){2}{\rule{1.000pt}{1.375pt}}
\multiput(966.83,253.04)(0.489,-0.926){14}{\rule{0.118pt}{2.159pt}}
\multiput(962.92,257.52)(11.000,-16.519){2}{\rule{1.000pt}{1.080pt}}
\multiput(977.83,232.49)(0.487,-0.863){12}{\rule{0.117pt}{2.050pt}}
\multiput(973.92,236.75)(10.000,-13.745){2}{\rule{1.000pt}{1.025pt}}
\multiput(987.83,215.74)(0.487,-0.703){12}{\rule{0.117pt}{1.750pt}}
\multiput(983.92,219.37)(10.000,-11.368){2}{\rule{1.000pt}{0.875pt}}
\multiput(997.83,201.98)(0.487,-0.544){12}{\rule{0.117pt}{1.450pt}}
\multiput(993.92,204.99)(10.000,-8.990){2}{\rule{1.000pt}{0.725pt}}
\multiput(1006.00,193.68)(0.447,-0.489){14}{\rule{1.250pt}{0.118pt}}
\multiput(1006.00,193.92)(8.406,-11.000){2}{\rule{0.625pt}{1.000pt}}
\multiput(1017.00,182.68)(0.483,-0.485){10}{\rule{1.361pt}{0.117pt}}
\multiput(1017.00,182.92)(7.175,-9.000){2}{\rule{0.681pt}{1.000pt}}
\multiput(1027.00,173.69)(0.606,-0.475){6}{\rule{1.679pt}{0.114pt}}
\multiput(1027.00,173.92)(6.516,-7.000){2}{\rule{0.839pt}{1.000pt}}
\multiput(1037.00,166.69)(0.606,-0.475){6}{\rule{1.679pt}{0.114pt}}
\multiput(1037.00,166.92)(6.516,-7.000){2}{\rule{0.839pt}{1.000pt}}
\multiput(1047.00,159.71)(0.660,-0.424){2}{\rule{2.250pt}{0.102pt}}
\multiput(1047.00,159.92)(5.330,-5.000){2}{\rule{1.125pt}{1.000pt}}
\multiput(1057.00,154.71)(0.660,-0.424){2}{\rule{2.250pt}{0.102pt}}
\multiput(1057.00,154.92)(5.330,-5.000){2}{\rule{1.125pt}{1.000pt}}
\put(1067,148.42){\rule{2.409pt}{1.000pt}}
\multiput(1067.00,149.92)(5.000,-3.000){2}{\rule{1.204pt}{1.000pt}}
\put(1077,144.92){\rule{2.409pt}{1.000pt}}
\multiput(1077.00,146.92)(5.000,-4.000){2}{\rule{1.204pt}{1.000pt}}
\put(1087,141.42){\rule{2.409pt}{1.000pt}}
\multiput(1087.00,142.92)(5.000,-3.000){2}{\rule{1.204pt}{1.000pt}}
\put(1097,138.92){\rule{2.409pt}{1.000pt}}
\multiput(1097.00,139.92)(5.000,-2.000){2}{\rule{1.204pt}{1.000pt}}
\put(1107,136.92){\rule{2.168pt}{1.000pt}}
\multiput(1107.00,137.92)(4.500,-2.000){2}{\rule{1.084pt}{1.000pt}}
\put(1116,134.92){\rule{2.409pt}{1.000pt}}
\multiput(1116.00,135.92)(5.000,-2.000){2}{\rule{1.204pt}{1.000pt}}
\put(1126,132.92){\rule{2.409pt}{1.000pt}}
\multiput(1126.00,133.92)(5.000,-2.000){2}{\rule{1.204pt}{1.000pt}}
\put(1136,130.92){\rule{2.409pt}{1.000pt}}
\multiput(1136.00,131.92)(5.000,-2.000){2}{\rule{1.204pt}{1.000pt}}
\put(1146,129.42){\rule{2.168pt}{1.000pt}}
\multiput(1146.00,129.92)(4.500,-1.000){2}{\rule{1.084pt}{1.000pt}}
\put(1155,128.42){\rule{2.409pt}{1.000pt}}
\multiput(1155.00,128.92)(5.000,-1.000){2}{\rule{1.204pt}{1.000pt}}
\put(1165,126.92){\rule{2.168pt}{1.000pt}}
\multiput(1165.00,127.92)(4.500,-2.000){2}{\rule{1.084pt}{1.000pt}}
\put(1174,125.42){\rule{2.409pt}{1.000pt}}
\multiput(1174.00,125.92)(5.000,-1.000){2}{\rule{1.204pt}{1.000pt}}
\put(1184,124.42){\rule{2.168pt}{1.000pt}}
\multiput(1184.00,124.92)(4.500,-1.000){2}{\rule{1.084pt}{1.000pt}}
\put(1193,123.42){\rule{2.409pt}{1.000pt}}
\multiput(1193.00,123.92)(5.000,-1.000){2}{\rule{1.204pt}{1.000pt}}
\put(1203,122.42){\rule{2.168pt}{1.000pt}}
\multiput(1203.00,122.92)(4.500,-1.000){2}{\rule{1.084pt}{1.000pt}}
\put(447.0,322.0){\rule[-0.500pt]{6.504pt}{1.000pt}}
\put(1222,121.42){\rule{2.168pt}{1.000pt}}
\multiput(1222.00,121.92)(4.500,-1.000){2}{\rule{1.084pt}{1.000pt}}
\put(1231,120.42){\rule{2.168pt}{1.000pt}}
\multiput(1231.00,120.92)(4.500,-1.000){2}{\rule{1.084pt}{1.000pt}}
\put(1240,119.42){\rule{2.409pt}{1.000pt}}
\multiput(1240.00,119.92)(5.000,-1.000){2}{\rule{1.204pt}{1.000pt}}
\put(1250,118.42){\rule{2.168pt}{1.000pt}}
\multiput(1250.00,118.92)(4.500,-1.000){2}{\rule{1.084pt}{1.000pt}}
\put(1212.0,124.0){\rule[-0.500pt]{2.409pt}{1.000pt}}
\put(1268,117.42){\rule{2.168pt}{1.000pt}}
\multiput(1268.00,117.92)(4.500,-1.000){2}{\rule{1.084pt}{1.000pt}}
\put(1277,116.42){\rule{2.168pt}{1.000pt}}
\multiput(1277.00,116.92)(4.500,-1.000){2}{\rule{1.084pt}{1.000pt}}
\put(1259.0,120.0){\rule[-0.500pt]{2.168pt}{1.000pt}}
\put(1295,115.42){\rule{2.409pt}{1.000pt}}
\multiput(1295.00,115.92)(5.000,-1.000){2}{\rule{1.204pt}{1.000pt}}
\put(1305,114.42){\rule{2.168pt}{1.000pt}}
\multiput(1305.00,114.92)(4.500,-1.000){2}{\rule{1.084pt}{1.000pt}}
\put(1286.0,118.0){\rule[-0.500pt]{2.168pt}{1.000pt}}
\put(1323,113.42){\rule{2.168pt}{1.000pt}}
\multiput(1323.00,113.92)(4.500,-1.000){2}{\rule{1.084pt}{1.000pt}}
\put(1314.0,116.0){\rule[-0.500pt]{2.168pt}{1.000pt}}
\put(1341,112.42){\rule{1.927pt}{1.000pt}}
\multiput(1341.00,112.92)(4.000,-1.000){2}{\rule{0.964pt}{1.000pt}}
\put(1332.0,115.0){\rule[-0.500pt]{2.168pt}{1.000pt}}
\put(1367,111.42){\rule{2.168pt}{1.000pt}}
\multiput(1367.00,111.92)(4.500,-1.000){2}{\rule{1.084pt}{1.000pt}}
\put(1349.0,114.0){\rule[-0.500pt]{4.336pt}{1.000pt}}
\put(1376.0,113.0){\rule[-0.500pt]{4.336pt}{1.000pt}}

\end{picture}
\end{center}
\end{figure}

\[B(\tau^{-}\rightarrow \bar{K}^{*0}\pi^{-}\nu)=0.23\%,\]
The contribution of the vector current is about $7.4\%$. Therefore,
$K_{a}$ is dominant in this decay.
The data are\\
\(0.38\pm0.11\pm0.13\%\)(CLEO[39])\\
\(0.25\pm0.10\pm0.05\%\)(ARGUS[40])\\
There is another decay channel $\tau^{-}\rightarrow K^{*-}\pi^{0}\nu$
whose branching ratio is one half of $B(\tau^{-}\rightarrow\bar{K}^{*0}
\pi^{-}\nu$. The total branching ratio is
\[B(\tau^{-}\rightarrow \bar{K}\pi\nu)=0.35\%,\]
The narrow peak in Fig.7 is from $K_{1}(1270)$ and the wider peak comes
from $K_{1}(1400)$. The width is about 230MeV.

\subsection{$\tau\rightarrow K\rho\nu$ and $K\omega\nu$}
It is the same as $\tau\rightarrow K^{*}\pi\nu$, $K_{a}$
dominates the decay $\tau\rightarrow K\rho\nu$.
Both the vector and axial-vector currents contribute to this decay
mode. The matrix element of the vector current, $<\bar{K^{0}}\rho^{-}
|\bar{\psi}\lambda_{+}\gamma_{\mu}\psi|0>$
is determined by the
vertex ${\cal L}^{K^{*}K\rho}$(47) and is the same as Eq.(50).
The axial-vector matrix element $<\bar{K^{0}}\rho^{-}|\bar{\psi}
\lambda_{+}\gamma_{\mu}\gamma_{5}\psi|0>$ is obtained by
substituting
\[K^{*}\rightarrow \rho,\;\;\;K\rightarrow\pi\]
in Eq.(75). Using the same substitutions in Eq.(83),
the distribution of
the decay rate of $\tau\rightarrow K\rho\nu$ is found.
The branching ratio of $\tau\rightarrow K\rho\nu$(two modes
$\bar{K}^{0}\rho^{-}$ and $K^{-}\rho^{0}$)
is computed to be
\begin{equation}
B=0.75\times10^{-3}.
\end{equation}
It is about $18\%$ of $\tau\rightarrow K\pi\pi\nu$.
The vector current makes $8\%$ contribution.
The DELPHI[41] has reported that $\tau\rightarrow K^{*}\pi\nu$ is
dominant the decay $\tau\rightarrow K\pi\pi\nu$ and $K\rho\nu$ decay
mode has not been observed. The ALEPH[42] has reported the $K^{*}\pi$
dominance and a branching ratio of $30\pm 11\%$ for the $K\rho$ mode.

Due to the $SU(3)$ coefficient we expect
\[B(\tau\rightarrow K\omega\nu)={1\over3}B(\tau\rightarrow K\rho\nu).\]

The theoretical results are in reasonably agreement with data.
In this paper the
spontaneous chiral symmetry breaking effect(for the mass difference
between $K^{*}$ and $K_{a}$) is taken into account and the resonance
formula is(Eq.(83))
\[BW_{K_{1}}[s]\equiv\frac{-g^{2}f^{2}_{a}m^{2}_{K^{*}}
+i\sqrt{q^{2}}\Gamma_{K_{1}}}
{q^{2}-m^{2}_{K_{1}}+i\sqrt{q^{2}}\Gamma_{K_{1}}}.\]
Because the spontaneous chiral symmetry breaking effect doesn't
disappear in the limit of $q^{2}\rightarrow 0$, we have a different
low energy limit which is $g^{2}f^{2}_{a}{m^{2}_{K^{*}}\over
m^{2}_{K_{1}}}$.
On the other hand, in this theory the amplitude A
strongly depends on $q^{2}$ and this dependence plays important role
in understanding the large branching ratio of $K^{*}\pi$ mode and the
smaller one for $K\rho$ mode.

\section{$\tau\rightarrow K^{*}\eta$}
There are vector and axial-vector parts
in this decay. The
calculation of the decay rate is similar to the decay of $\tau
\rightarrow K^{*}\pi\nu$. The vertices ${\cal L}^{K^{*}\bar{K}^{*}
\eta}$ and ${\cal L}^{WK^{*}\eta}$ via the Lagrangian $L^{Vs}$(68)
contribute to the vector part and the vertices ${\cal L}^{K_{1}
K^{*}\eta}$, ${\cal L}^{KK^{*}\eta}$, and ${\cal L}^{WK^{*}\eta}$
via $L^{As}$ take the responsibility for the axial-vector part.
The vertex ${\cal L}^{K^{*}\bar{K}^{*}\eta}$ comes from anomaly.
Using the same method deriving the vertices $\eta vv$ (in Ref.[11])
, it is found
\begin{equation}
{\cal L}^{K^{*}\bar{K^{*}}\eta}=
-\frac{3a}{2\pi^{2}g^{2}f_{\pi}}
d_{ab8}\varepsilon^{\mu\nu\alpha\beta}\eta\partial_{\mu}K^{a}_{\nu}
\partial_{\alpha}K^{b}_{\beta}
-\frac{3b}{2\pi^{2}g^{2}f_{\pi}}
\varepsilon^{\mu\nu\alpha\beta}\eta\partial_{\mu}K^{a}_{\nu}
\partial_{\alpha}K^{a}_{\beta},
\end{equation}
where a and b are the octet and singlet component of $\eta$
respectively,
\(a=cos\theta\), \(b=\sqrt{{2\over3}}cos\theta\),
and \(\theta=-20^{0}\). Due to the cancellation
between the two components
the vector matrix element is very small and can be ignored.

The vertices ${\cal L}^{K_{1}K^{*}\eta}$ and ${\cal L}^{KK^{*}\eta}$
contribute to the axial-vector
matrix element and they are derived from the effective Lagrangian
presented in Ref.[11].
\begin{eqnarray}
\lefteqn{{\cal L}^{K_{1}K^{*}\eta}=af_{ab8}\{A(q^{2})_{K^{*}}
K^{a}_{\mu}
K^{b\nu}\eta-BK^{a}_{\mu}K^{b}_{\nu}\partial^{\mu\nu}\eta\}}
\\
&&{\cal L}^{K^{*}K\eta}=af_{K^{*}K\eta}f_{ab8}K^{a}_{\mu}(K^{b}
\partial^{\mu}\eta-\eta\partial^{\mu}K^{b}),
\end{eqnarray}
where $f_{K^{*}K\eta}$ is the same as $f_{\rho\pi\pi}$(37) in the limit
of \(m_{q}=0\).
The decay width is similar to the one of $\tau\rightarrow K^{*}\pi\nu$
\begin{eqnarray}
\lefteqn{{d\Gamma\over dq^{2}}(\tau^{-}\rightarrow
K^{*-}\eta\nu)={G^{2}\over
(2\pi)^{3}}cos^{2}20^{0}\frac{sin^{2}\theta_{C}}{64m^{3}_{\tau}q^{4}}
(m^{2}_{\tau}-q^{2})^{2}(m^{2}_{\tau}+2q^{2})}\nonumber \\
&&\{{3\over4}\{|A|^{2}[1+{1\over12m^{2}_{K^{*}}q^{2}}
(q^{2}-m^{2}_{K^{*}})^{2}]
-(BA^{*}+B^{*}A){1\over24m^{2}_{K^{*}}
q^{2}}(q^{2}+m^{2}_{K^{*}})(q^{2}-m^{2}
_{K^{*}})^{2}\nonumber \\
&&+\frac{|B|^{2}}{48m^{2}_{K^{*}}q^{2}}
(q^{2}-m^{2}_{K^{*}})
^{4}]\}\}.
\end{eqnarray}
The distribution is shown in Fig.8. The branching ratio is computed
to be
\[B=1.01\times10^{-4}.\]
The axial-vector current is dominant.

\begin{figure}
\begin{center}

\setlength{\unitlength}{0.240900pt}
\ifx\plotpoint\undefined\newsavebox{\plotpoint}\fi
\sbox{\plotpoint}{\rule[-0.500pt]{1.000pt}{1.000pt}}%
\begin{picture}(1500,900)(0,0)
\font\gnuplot=cmr10 at 10pt
\gnuplot
\sbox{\plotpoint}{\rule[-0.500pt]{1.000pt}{1.000pt}}%
\put(220.0,113.0){\rule[-0.500pt]{292.934pt}{1.000pt}}
\put(220.0,113.0){\rule[-0.500pt]{4.818pt}{1.000pt}}
\put(198,113){\makebox(0,0)[r]{0}}
\put(1416.0,113.0){\rule[-0.500pt]{4.818pt}{1.000pt}}
\put(220.0,209.0){\rule[-0.500pt]{4.818pt}{1.000pt}}
\put(198,209){\makebox(0,0)[r]{2}}
\put(1416.0,209.0){\rule[-0.500pt]{4.818pt}{1.000pt}}
\put(220.0,304.0){\rule[-0.500pt]{4.818pt}{1.000pt}}
\put(198,304){\makebox(0,0)[r]{4}}
\put(1416.0,304.0){\rule[-0.500pt]{4.818pt}{1.000pt}}
\put(220.0,400.0){\rule[-0.500pt]{4.818pt}{1.000pt}}
\put(198,400){\makebox(0,0)[r]{6}}
\put(1416.0,400.0){\rule[-0.500pt]{4.818pt}{1.000pt}}
\put(220.0,495.0){\rule[-0.500pt]{4.818pt}{1.000pt}}
\put(198,495){\makebox(0,0)[r]{8}}
\put(1416.0,495.0){\rule[-0.500pt]{4.818pt}{1.000pt}}
\put(220.0,591.0){\rule[-0.500pt]{4.818pt}{1.000pt}}
\put(198,591){\makebox(0,0)[r]{10}}
\put(1416.0,591.0){\rule[-0.500pt]{4.818pt}{1.000pt}}
\put(220.0,686.0){\rule[-0.500pt]{4.818pt}{1.000pt}}
\put(198,686){\makebox(0,0)[r]{12}}
\put(1416.0,686.0){\rule[-0.500pt]{4.818pt}{1.000pt}}
\put(220.0,782.0){\rule[-0.500pt]{4.818pt}{1.000pt}}
\put(198,782){\makebox(0,0)[r]{14}}
\put(1416.0,782.0){\rule[-0.500pt]{4.818pt}{1.000pt}}
\put(220.0,877.0){\rule[-0.500pt]{4.818pt}{1.000pt}}
\put(198,877){\makebox(0,0)[r]{16}}
\put(1416.0,877.0){\rule[-0.500pt]{4.818pt}{1.000pt}}
\put(220.0,113.0){\rule[-0.500pt]{1.000pt}{4.818pt}}
\put(220,68){\makebox(0,0){1.4}}
\put(220.0,857.0){\rule[-0.500pt]{1.000pt}{4.818pt}}
\put(372.0,113.0){\rule[-0.500pt]{1.000pt}{4.818pt}}
\put(372,68){\makebox(0,0){1.45}}
\put(372.0,857.0){\rule[-0.500pt]{1.000pt}{4.818pt}}
\put(524.0,113.0){\rule[-0.500pt]{1.000pt}{4.818pt}}
\put(524,68){\makebox(0,0){1.5}}
\put(524.0,857.0){\rule[-0.500pt]{1.000pt}{4.818pt}}
\put(676.0,113.0){\rule[-0.500pt]{1.000pt}{4.818pt}}
\put(676,68){\makebox(0,0){1.55}}
\put(676.0,857.0){\rule[-0.500pt]{1.000pt}{4.818pt}}
\put(828.0,113.0){\rule[-0.500pt]{1.000pt}{4.818pt}}
\put(828,68){\makebox(0,0){1.6}}
\put(828.0,857.0){\rule[-0.500pt]{1.000pt}{4.818pt}}
\put(980.0,113.0){\rule[-0.500pt]{1.000pt}{4.818pt}}
\put(980,68){\makebox(0,0){1.65}}
\put(980.0,857.0){\rule[-0.500pt]{1.000pt}{4.818pt}}
\put(1132.0,113.0){\rule[-0.500pt]{1.000pt}{4.818pt}}
\put(1132,68){\makebox(0,0){1.7}}
\put(1132.0,857.0){\rule[-0.500pt]{1.000pt}{4.818pt}}
\put(1284.0,113.0){\rule[-0.500pt]{1.000pt}{4.818pt}}
\put(1284,68){\makebox(0,0){1.75}}
\put(1284.0,857.0){\rule[-0.500pt]{1.000pt}{4.818pt}}
\put(1436.0,113.0){\rule[-0.500pt]{1.000pt}{4.818pt}}
\put(1436,68){\makebox(0,0){1.8}}
\put(1436.0,857.0){\rule[-0.500pt]{1.000pt}{4.818pt}}
\put(220.0,113.0){\rule[-0.500pt]{292.934pt}{1.000pt}}
\put(1436.0,113.0){\rule[-0.500pt]{1.000pt}{184.048pt}}
\put(220.0,877.0){\rule[-0.500pt]{292.934pt}{1.000pt}}
\put(45,495){\makebox(0,0){${d\Gamma\over d\sqrt{q^{2}}}\times10^{16}$ }}
\put(828,23){\makebox(0,0){Fig.8   GeV}}
\put(220.0,113.0){\rule[-0.500pt]{1.000pt}{184.048pt}}
\put(394,649){\usebox{\plotpoint}}
\multiput(395.83,649.00)(0.496,1.525){34}{\rule{0.119pt}{3.298pt}}
\multiput(391.92,649.00)(21.000,57.156){2}{\rule{1.000pt}{1.649pt}}
\multiput(416.83,713.00)(0.496,0.967){36}{\rule{0.119pt}{2.205pt}}
\multiput(412.92,713.00)(22.000,38.424){2}{\rule{1.000pt}{1.102pt}}
\multiput(438.83,756.00)(0.496,0.649){34}{\rule{0.119pt}{1.583pt}}
\multiput(434.92,756.00)(21.000,24.714){2}{\rule{1.000pt}{0.792pt}}
\multiput(458.00,785.83)(0.593,0.494){26}{\rule{1.485pt}{0.119pt}}
\multiput(458.00,781.92)(17.917,17.000){2}{\rule{0.743pt}{1.000pt}}
\multiput(479.00,802.83)(1.364,0.481){8}{\rule{3.000pt}{0.116pt}}
\multiput(479.00,798.92)(15.773,8.000){2}{\rule{1.500pt}{1.000pt}}
\put(501,807.42){\rule{5.059pt}{1.000pt}}
\multiput(501.00,806.92)(10.500,1.000){2}{\rule{2.529pt}{1.000pt}}
\put(522,805.92){\rule{5.059pt}{1.000pt}}
\multiput(522.00,807.92)(10.500,-4.000){2}{\rule{2.529pt}{1.000pt}}
\multiput(543.00,803.68)(1.142,-0.485){10}{\rule{2.583pt}{0.117pt}}
\multiput(543.00,803.92)(15.638,-9.000){2}{\rule{1.292pt}{1.000pt}}
\multiput(564.00,794.68)(0.803,-0.491){16}{\rule{1.917pt}{0.118pt}}
\multiput(564.00,794.92)(16.022,-12.000){2}{\rule{0.958pt}{1.000pt}}
\multiput(584.00,782.68)(0.674,-0.493){22}{\rule{1.650pt}{0.119pt}}
\multiput(584.00,782.92)(17.575,-15.000){2}{\rule{0.825pt}{1.000pt}}
\multiput(605.00,767.68)(0.593,-0.494){26}{\rule{1.485pt}{0.119pt}}
\multiput(605.00,767.92)(17.917,-17.000){2}{\rule{0.743pt}{1.000pt}}
\multiput(626.00,750.68)(0.503,-0.495){30}{\rule{1.303pt}{0.119pt}}
\multiput(626.00,750.92)(17.296,-19.000){2}{\rule{0.651pt}{1.000pt}}
\multiput(646.00,731.68)(0.479,-0.496){34}{\rule{1.250pt}{0.119pt}}
\multiput(646.00,731.92)(18.406,-21.000){2}{\rule{0.625pt}{1.000pt}}
\multiput(668.83,707.40)(0.495,-0.529){32}{\rule{0.119pt}{1.350pt}}
\multiput(664.92,710.20)(20.000,-19.198){2}{\rule{1.000pt}{0.675pt}}
\multiput(688.83,684.98)(0.495,-0.580){32}{\rule{0.119pt}{1.450pt}}
\multiput(684.92,687.99)(20.000,-20.990){2}{\rule{1.000pt}{0.725pt}}
\multiput(708.83,661.42)(0.496,-0.528){34}{\rule{0.119pt}{1.345pt}}
\multiput(704.92,664.21)(21.000,-20.208){2}{\rule{1.000pt}{0.673pt}}
\multiput(729.83,637.77)(0.495,-0.606){32}{\rule{0.119pt}{1.500pt}}
\multiput(725.92,640.89)(20.000,-21.887){2}{\rule{1.000pt}{0.750pt}}
\multiput(749.83,612.98)(0.495,-0.580){32}{\rule{0.119pt}{1.450pt}}
\multiput(745.92,615.99)(20.000,-20.990){2}{\rule{1.000pt}{0.725pt}}
\multiput(769.83,588.77)(0.495,-0.606){32}{\rule{0.119pt}{1.500pt}}
\multiput(765.92,591.89)(20.000,-21.887){2}{\rule{1.000pt}{0.750pt}}
\multiput(789.83,563.50)(0.495,-0.638){30}{\rule{0.119pt}{1.566pt}}
\multiput(785.92,566.75)(19.000,-21.750){2}{\rule{1.000pt}{0.783pt}}
\multiput(808.83,538.77)(0.495,-0.606){32}{\rule{0.119pt}{1.500pt}}
\multiput(804.92,541.89)(20.000,-21.887){2}{\rule{1.000pt}{0.750pt}}
\multiput(828.83,513.98)(0.495,-0.580){32}{\rule{0.119pt}{1.450pt}}
\multiput(824.92,516.99)(20.000,-20.990){2}{\rule{1.000pt}{0.725pt}}
\multiput(848.83,489.72)(0.495,-0.611){30}{\rule{0.119pt}{1.513pt}}
\multiput(844.92,492.86)(19.000,-20.859){2}{\rule{1.000pt}{0.757pt}}
\multiput(867.83,465.98)(0.495,-0.580){32}{\rule{0.119pt}{1.450pt}}
\multiput(863.92,468.99)(20.000,-20.990){2}{\rule{1.000pt}{0.725pt}}
\multiput(887.83,441.94)(0.495,-0.584){30}{\rule{0.119pt}{1.461pt}}
\multiput(883.92,444.97)(19.000,-19.969){2}{\rule{1.000pt}{0.730pt}}
\multiput(906.83,419.19)(0.495,-0.554){32}{\rule{0.119pt}{1.400pt}}
\multiput(902.92,422.09)(20.000,-20.094){2}{\rule{1.000pt}{0.700pt}}
\multiput(926.83,396.16)(0.495,-0.557){30}{\rule{0.119pt}{1.408pt}}
\multiput(922.92,399.08)(19.000,-19.078){2}{\rule{1.000pt}{0.704pt}}
\multiput(945.83,374.16)(0.495,-0.557){30}{\rule{0.119pt}{1.408pt}}
\multiput(941.92,377.08)(19.000,-19.078){2}{\rule{1.000pt}{0.704pt}}
\multiput(963.00,355.68)(0.478,-0.495){32}{\rule{1.250pt}{0.119pt}}
\multiput(963.00,355.92)(17.406,-20.000){2}{\rule{0.625pt}{1.000pt}}
\multiput(984.83,332.59)(0.495,-0.503){30}{\rule{0.119pt}{1.303pt}}
\multiput(980.92,335.30)(19.000,-17.296){2}{\rule{1.000pt}{0.651pt}}
\multiput(1002.00,315.68)(0.476,-0.495){30}{\rule{1.250pt}{0.119pt}}
\multiput(1002.00,315.92)(16.406,-19.000){2}{\rule{0.625pt}{1.000pt}}
\multiput(1021.00,296.68)(0.476,-0.495){30}{\rule{1.250pt}{0.119pt}}
\multiput(1021.00,296.92)(16.406,-19.000){2}{\rule{0.625pt}{1.000pt}}
\multiput(1040.00,277.68)(0.503,-0.494){26}{\rule{1.309pt}{0.119pt}}
\multiput(1040.00,277.92)(15.283,-17.000){2}{\rule{0.654pt}{1.000pt}}
\multiput(1058.00,260.68)(0.533,-0.494){26}{\rule{1.368pt}{0.119pt}}
\multiput(1058.00,260.92)(16.161,-17.000){2}{\rule{0.684pt}{1.000pt}}
\multiput(1077.00,243.68)(0.567,-0.494){24}{\rule{1.438pt}{0.119pt}}
\multiput(1077.00,243.92)(16.016,-16.000){2}{\rule{0.719pt}{1.000pt}}
\multiput(1096.00,227.68)(0.605,-0.493){22}{\rule{1.517pt}{0.119pt}}
\multiput(1096.00,227.92)(15.852,-15.000){2}{\rule{0.758pt}{1.000pt}}
\multiput(1115.00,212.68)(0.660,-0.492){18}{\rule{1.635pt}{0.118pt}}
\multiput(1115.00,212.92)(14.607,-13.000){2}{\rule{0.817pt}{1.000pt}}
\multiput(1133.00,199.68)(0.700,-0.492){18}{\rule{1.712pt}{0.118pt}}
\multiput(1133.00,199.92)(15.448,-13.000){2}{\rule{0.856pt}{1.000pt}}
\multiput(1152.00,186.68)(0.716,-0.491){16}{\rule{1.750pt}{0.118pt}}
\multiput(1152.00,186.92)(14.368,-12.000){2}{\rule{0.875pt}{1.000pt}}
\multiput(1170.00,174.68)(0.830,-0.489){14}{\rule{1.977pt}{0.118pt}}
\multiput(1170.00,174.92)(14.896,-11.000){2}{\rule{0.989pt}{1.000pt}}
\multiput(1189.00,163.68)(0.863,-0.487){12}{\rule{2.050pt}{0.117pt}}
\multiput(1189.00,163.92)(13.745,-10.000){2}{\rule{1.025pt}{1.000pt}}
\multiput(1207.00,153.68)(0.962,-0.485){10}{\rule{2.250pt}{0.117pt}}
\multiput(1207.00,153.92)(13.330,-9.000){2}{\rule{1.125pt}{1.000pt}}
\multiput(1225.00,144.68)(1.089,-0.481){8}{\rule{2.500pt}{0.116pt}}
\multiput(1225.00,144.92)(12.811,-8.000){2}{\rule{1.250pt}{1.000pt}}
\multiput(1243.00,136.69)(1.258,-0.475){6}{\rule{2.821pt}{0.114pt}}
\multiput(1243.00,136.92)(12.144,-7.000){2}{\rule{1.411pt}{1.000pt}}
\multiput(1261.00,129.69)(1.606,-0.462){4}{\rule{3.417pt}{0.111pt}}
\multiput(1261.00,129.92)(11.909,-6.000){2}{\rule{1.708pt}{1.000pt}}
\put(1280,121.92){\rule{4.336pt}{1.000pt}}
\multiput(1280.00,123.92)(9.000,-4.000){2}{\rule{2.168pt}{1.000pt}}
\put(1298,117.92){\rule{4.336pt}{1.000pt}}
\multiput(1298.00,119.92)(9.000,-4.000){2}{\rule{2.168pt}{1.000pt}}
\put(1316,114.42){\rule{4.095pt}{1.000pt}}
\multiput(1316.00,115.92)(8.500,-3.000){2}{\rule{2.048pt}{1.000pt}}
\put(1333,112.42){\rule{4.336pt}{1.000pt}}
\multiput(1333.00,112.92)(9.000,-1.000){2}{\rule{2.168pt}{1.000pt}}
\end{picture}

\end{center}
\end{figure}

\section{$\tau\rightarrow K\eta\nu$}
The decay $\tau\rightarrow\eta K\nu$ has been studied in terms a chiral
Lagrangian[3,7] and only the vector current contributes.
The prediction is $1.2\times10^{-4}$.
The experiments are \\
CLEO[43]: $(2.6\pm0.5)\times10^{-4}$, \\
ALEPH[44]: $(2.9^{+1.3}_{-1.2}\pm0.7)\times10^{-4}$.

In the effective chiral theory the vertex $K_{1}K\eta$ doesn't exist.
The reason is that if it exists it has abnormal parity and comes
from the anomaly in which there is antisymmetric tensor. It is
impossible to construct a vertex with a antisymmetric tensor by using
$K_{1}$, K, and $\eta$ fields. Therefore,
only the vector current contributes to this process and $K^{*}$ is
dominant in this process.
The vertex
${\cal L}^{K^{*}K\eta}$ is shown in Eq.(90).
The decay width is found
\begin{eqnarray}
\lefteqn{{d\Gamma\over dq^{2}}={3\over4}{G^{2}\over(2\pi)
^{3}}sin^{2}\theta_{C}cos^{2}20^{0}
{1\over384m^{3}_{\tau}}{1\over q^{2}}(m^{2}_{\tau}-q^{2})^{2}
(m^{2}_{\tau}+2q^{2})}\nonumber \\
&&[(q^{2}+m^{2}_{\eta}-m^{2}_{K})^{2}-4q^{2}
m^{2}_{\eta}]^{{1\over2}}g^{2}
f^{2}_{\rho\pi\pi}(q^{2})\frac{m^{4}_{K^{*}}+q^{2}\Gamma^{2}
_{K^{*}}(q^{2})}{(q^{2}-m^{2}_{K^{*}})^{2}+q^{2}
\Gamma^{2}_{K^{*}}(q^{2})}.
\end{eqnarray}
The branching ratio is computed to be
\[B(\tau^{-}\rightarrow\eta K^{-}\nu)=2.22\times10^{-4}.\]

The distribution of the invariant mass of $\eta$ and K is shown in
Fig.(9). The figure indicates a peak at 1.086GeV which is slightly
above
the threshold. The peak is resulted by the effects of the threshold
and the resonance.

The branching ratio of $\tau\rightarrow\eta'K\nu$ is 200 times smaller.
\begin{figure}
\begin{center}

\setlength{\unitlength}{0.240900pt}
\ifx\plotpoint\undefined\newsavebox{\plotpoint}\fi
\begin{picture}(1500,900)(0,0)
\font\gnuplot=cmr10 at 10pt
\gnuplot
\sbox{\plotpoint}{\rule[-0.500pt]{1.000pt}{1.000pt}}%
\put(220.0,113.0){\rule[-0.500pt]{292.934pt}{1.000pt}}
\put(220.0,113.0){\rule[-0.500pt]{4.818pt}{1.000pt}}
\put(198,113){\makebox(0,0)[r]{0}}
\put(1416.0,113.0){\rule[-0.500pt]{4.818pt}{1.000pt}}
\put(220.0,252.0){\rule[-0.500pt]{4.818pt}{1.000pt}}
\put(198,252){\makebox(0,0)[r]{50}}
\put(1416.0,252.0){\rule[-0.500pt]{4.818pt}{1.000pt}}
\put(220.0,391.0){\rule[-0.500pt]{4.818pt}{1.000pt}}
\put(198,391){\makebox(0,0)[r]{100}}
\put(1416.0,391.0){\rule[-0.500pt]{4.818pt}{1.000pt}}
\put(220.0,530.0){\rule[-0.500pt]{4.818pt}{1.000pt}}
\put(198,530){\makebox(0,0)[r]{150}}
\put(1416.0,530.0){\rule[-0.500pt]{4.818pt}{1.000pt}}
\put(220.0,669.0){\rule[-0.500pt]{4.818pt}{1.000pt}}
\put(198,669){\makebox(0,0)[r]{200}}
\put(1416.0,669.0){\rule[-0.500pt]{4.818pt}{1.000pt}}
\put(220.0,808.0){\rule[-0.500pt]{4.818pt}{1.000pt}}
\put(198,808){\makebox(0,0)[r]{250}}
\put(1416.0,808.0){\rule[-0.500pt]{4.818pt}{1.000pt}}
\put(220.0,113.0){\rule[-0.500pt]{1.000pt}{4.818pt}}
\put(220,68){\makebox(0,0){1}}
\put(220.0,857.0){\rule[-0.500pt]{1.000pt}{4.818pt}}
\put(372.0,113.0){\rule[-0.500pt]{1.000pt}{4.818pt}}
\put(372,68){\makebox(0,0){1.1}}
\put(372.0,857.0){\rule[-0.500pt]{1.000pt}{4.818pt}}
\put(524.0,113.0){\rule[-0.500pt]{1.000pt}{4.818pt}}
\put(524,68){\makebox(0,0){1.2}}
\put(524.0,857.0){\rule[-0.500pt]{1.000pt}{4.818pt}}
\put(676.0,113.0){\rule[-0.500pt]{1.000pt}{4.818pt}}
\put(676,68){\makebox(0,0){1.3}}
\put(676.0,857.0){\rule[-0.500pt]{1.000pt}{4.818pt}}
\put(828.0,113.0){\rule[-0.500pt]{1.000pt}{4.818pt}}
\put(828,68){\makebox(0,0){1.4}}
\put(828.0,857.0){\rule[-0.500pt]{1.000pt}{4.818pt}}
\put(980.0,113.0){\rule[-0.500pt]{1.000pt}{4.818pt}}
\put(980,68){\makebox(0,0){1.5}}
\put(980.0,857.0){\rule[-0.500pt]{1.000pt}{4.818pt}}
\put(1132.0,113.0){\rule[-0.500pt]{1.000pt}{4.818pt}}
\put(1132,68){\makebox(0,0){1.6}}
\put(1132.0,857.0){\rule[-0.500pt]{1.000pt}{4.818pt}}
\put(1284.0,113.0){\rule[-0.500pt]{1.000pt}{4.818pt}}
\put(1284,68){\makebox(0,0){1.7}}
\put(1284.0,857.0){\rule[-0.500pt]{1.000pt}{4.818pt}}
\put(1436.0,113.0){\rule[-0.500pt]{1.000pt}{4.818pt}}
\put(1436,68){\makebox(0,0){1.8}}
\put(1436.0,857.0){\rule[-0.500pt]{1.000pt}{4.818pt}}
\put(220.0,113.0){\rule[-0.500pt]{292.934pt}{1.000pt}}
\put(1436.0,113.0){\rule[-0.500pt]{1.000pt}{184.048pt}}
\put(220.0,877.0){\rule[-0.500pt]{292.934pt}{1.000pt}}
\put(45,495){\makebox(0,0){${d\Gamma\over d\sqrt{q^{2}}}\times 10^{17}$ }}
\put(828,23){\makebox(0,0){FiG.9       $\sqrt{q^{2}}$        GeV}}
\put(220.0,113.0){\rule[-0.500pt]{1.000pt}{184.048pt}}
\put(285,117){\usebox{\plotpoint}}
\multiput(286.83,117.00)(0.494,16.738){26}{\rule{0.119pt}{32.897pt}}
\multiput(282.92,117.00)(17.000,486.721){2}{\rule{1.000pt}{16.449pt}}
\multiput(303.83,672.00)(0.494,4.076){24}{\rule{0.119pt}{8.250pt}}
\multiput(299.92,672.00)(16.000,110.877){2}{\rule{1.000pt}{4.125pt}}
\multiput(319.83,800.00)(0.494,1.565){24}{\rule{0.119pt}{3.375pt}}
\multiput(315.92,800.00)(16.000,42.995){2}{\rule{1.000pt}{1.688pt}}
\multiput(334.00,851.83)(0.538,0.492){20}{\rule{1.393pt}{0.119pt}}
\multiput(334.00,847.92)(13.109,14.000){2}{\rule{0.696pt}{1.000pt}}
\multiput(350.00,861.69)(1.298,-0.462){4}{\rule{2.917pt}{0.111pt}}
\multiput(350.00,861.92)(9.946,-6.000){2}{\rule{1.458pt}{1.000pt}}
\multiput(366.00,855.68)(0.470,-0.494){24}{\rule{1.250pt}{0.119pt}}
\multiput(366.00,855.92)(13.406,-16.000){2}{\rule{0.625pt}{1.000pt}}
\multiput(383.83,834.87)(0.493,-0.708){22}{\rule{0.119pt}{1.717pt}}
\multiput(379.92,838.44)(15.000,-18.437){2}{\rule{1.000pt}{0.858pt}}
\multiput(398.83,812.48)(0.494,-0.760){24}{\rule{0.119pt}{1.812pt}}
\multiput(394.92,816.24)(16.000,-21.238){2}{\rule{1.000pt}{0.906pt}}
\multiput(414.83,786.77)(0.493,-0.846){22}{\rule{0.119pt}{1.983pt}}
\multiput(410.92,790.88)(15.000,-21.883){2}{\rule{1.000pt}{0.992pt}}
\multiput(429.83,760.49)(0.493,-0.880){22}{\rule{0.119pt}{2.050pt}}
\multiput(425.92,764.75)(15.000,-22.745){2}{\rule{1.000pt}{1.025pt}}
\multiput(444.83,733.77)(0.493,-0.846){22}{\rule{0.119pt}{1.983pt}}
\multiput(440.92,737.88)(15.000,-21.883){2}{\rule{1.000pt}{0.992pt}}
\multiput(459.83,707.77)(0.493,-0.846){22}{\rule{0.119pt}{1.983pt}}
\multiput(455.92,711.88)(15.000,-21.883){2}{\rule{1.000pt}{0.992pt}}
\multiput(474.83,682.04)(0.493,-0.812){22}{\rule{0.119pt}{1.917pt}}
\multiput(470.92,686.02)(15.000,-21.022){2}{\rule{1.000pt}{0.958pt}}
\multiput(489.83,656.85)(0.492,-0.834){20}{\rule{0.119pt}{1.964pt}}
\multiput(485.92,660.92)(14.000,-19.923){2}{\rule{1.000pt}{0.982pt}}
\multiput(503.83,633.87)(0.493,-0.708){22}{\rule{0.119pt}{1.717pt}}
\multiput(499.92,637.44)(15.000,-18.437){2}{\rule{1.000pt}{0.858pt}}
\multiput(518.83,611.44)(0.492,-0.760){20}{\rule{0.119pt}{1.821pt}}
\multiput(514.92,615.22)(14.000,-18.220){2}{\rule{1.000pt}{0.911pt}}
\multiput(532.83,590.03)(0.492,-0.686){20}{\rule{0.119pt}{1.679pt}}
\multiput(528.92,593.52)(14.000,-16.516){2}{\rule{1.000pt}{0.839pt}}
\multiput(546.83,570.33)(0.492,-0.649){20}{\rule{0.119pt}{1.607pt}}
\multiput(542.92,573.66)(14.000,-15.664){2}{\rule{1.000pt}{0.804pt}}
\multiput(560.83,551.63)(0.492,-0.612){20}{\rule{0.119pt}{1.536pt}}
\multiput(556.92,554.81)(14.000,-14.813){2}{\rule{1.000pt}{0.768pt}}
\multiput(574.83,533.92)(0.492,-0.575){20}{\rule{0.119pt}{1.464pt}}
\multiput(570.92,536.96)(14.000,-13.961){2}{\rule{1.000pt}{0.732pt}}
\multiput(588.83,517.22)(0.492,-0.538){20}{\rule{0.119pt}{1.393pt}}
\multiput(584.92,520.11)(14.000,-13.109){2}{\rule{1.000pt}{0.696pt}}
\multiput(602.83,501.22)(0.492,-0.538){20}{\rule{0.119pt}{1.393pt}}
\multiput(598.92,504.11)(14.000,-13.109){2}{\rule{1.000pt}{0.696pt}}
\multiput(615.00,488.68)(0.464,-0.492){20}{\rule{1.250pt}{0.119pt}}
\multiput(615.00,488.92)(11.406,-14.000){2}{\rule{0.625pt}{1.000pt}}
\multiput(630.83,471.49)(0.492,-0.500){18}{\rule{0.118pt}{1.327pt}}
\multiput(626.92,474.25)(13.000,-11.246){2}{\rule{1.000pt}{0.663pt}}
\multiput(642.00,460.68)(0.541,-0.491){16}{\rule{1.417pt}{0.118pt}}
\multiput(642.00,460.92)(11.060,-12.000){2}{\rule{0.708pt}{1.000pt}}
\multiput(656.00,448.68)(0.498,-0.491){16}{\rule{1.333pt}{0.118pt}}
\multiput(656.00,448.92)(10.233,-12.000){2}{\rule{0.667pt}{1.000pt}}
\multiput(669.00,436.68)(0.498,-0.491){16}{\rule{1.333pt}{0.118pt}}
\multiput(669.00,436.92)(10.233,-12.000){2}{\rule{0.667pt}{1.000pt}}
\multiput(682.00,424.68)(0.543,-0.489){14}{\rule{1.432pt}{0.118pt}}
\multiput(682.00,424.92)(10.028,-11.000){2}{\rule{0.716pt}{1.000pt}}
\multiput(695.00,413.68)(0.650,-0.487){12}{\rule{1.650pt}{0.117pt}}
\multiput(695.00,413.92)(10.575,-10.000){2}{\rule{0.825pt}{1.000pt}}
\multiput(709.00,403.68)(0.597,-0.487){12}{\rule{1.550pt}{0.117pt}}
\multiput(709.00,403.92)(9.783,-10.000){2}{\rule{0.775pt}{1.000pt}}
\multiput(722.00,393.68)(0.663,-0.485){10}{\rule{1.694pt}{0.117pt}}
\multiput(722.00,393.92)(9.483,-9.000){2}{\rule{0.847pt}{1.000pt}}
\multiput(735.00,384.68)(0.603,-0.485){10}{\rule{1.583pt}{0.117pt}}
\multiput(735.00,384.92)(8.714,-9.000){2}{\rule{0.792pt}{1.000pt}}
\multiput(747.00,375.68)(0.745,-0.481){8}{\rule{1.875pt}{0.116pt}}
\multiput(747.00,375.92)(9.108,-8.000){2}{\rule{0.937pt}{1.000pt}}
\multiput(760.00,367.68)(0.745,-0.481){8}{\rule{1.875pt}{0.116pt}}
\multiput(760.00,367.92)(9.108,-8.000){2}{\rule{0.937pt}{1.000pt}}
\multiput(773.00,359.68)(0.677,-0.481){8}{\rule{1.750pt}{0.116pt}}
\multiput(773.00,359.92)(8.368,-8.000){2}{\rule{0.875pt}{1.000pt}}
\multiput(785.00,351.69)(0.851,-0.475){6}{\rule{2.107pt}{0.114pt}}
\multiput(785.00,351.92)(8.627,-7.000){2}{\rule{1.054pt}{1.000pt}}
\multiput(798.00,344.69)(0.769,-0.475){6}{\rule{1.964pt}{0.114pt}}
\multiput(798.00,344.92)(7.923,-7.000){2}{\rule{0.982pt}{1.000pt}}
\multiput(810.00,337.69)(0.851,-0.475){6}{\rule{2.107pt}{0.114pt}}
\multiput(810.00,337.92)(8.627,-7.000){2}{\rule{1.054pt}{1.000pt}}
\multiput(823.00,330.69)(0.887,-0.462){4}{\rule{2.250pt}{0.111pt}}
\multiput(823.00,330.92)(7.330,-6.000){2}{\rule{1.125pt}{1.000pt}}
\multiput(835.00,324.69)(0.887,-0.462){4}{\rule{2.250pt}{0.111pt}}
\multiput(835.00,324.92)(7.330,-6.000){2}{\rule{1.125pt}{1.000pt}}
\multiput(847.00,318.69)(0.989,-0.462){4}{\rule{2.417pt}{0.111pt}}
\multiput(847.00,318.92)(7.984,-6.000){2}{\rule{1.208pt}{1.000pt}}
\multiput(860.00,312.69)(0.887,-0.462){4}{\rule{2.250pt}{0.111pt}}
\multiput(860.00,312.92)(7.330,-6.000){2}{\rule{1.125pt}{1.000pt}}
\multiput(872.00,306.69)(0.887,-0.462){4}{\rule{2.250pt}{0.111pt}}
\multiput(872.00,306.92)(7.330,-6.000){2}{\rule{1.125pt}{1.000pt}}
\multiput(884.00,300.71)(1.000,-0.424){2}{\rule{2.650pt}{0.102pt}}
\multiput(884.00,300.92)(6.500,-5.000){2}{\rule{1.325pt}{1.000pt}}
\multiput(896.00,295.69)(0.887,-0.462){4}{\rule{2.250pt}{0.111pt}}
\multiput(896.00,295.92)(7.330,-6.000){2}{\rule{1.125pt}{1.000pt}}
\multiput(908.00,289.71)(0.830,-0.424){2}{\rule{2.450pt}{0.102pt}}
\multiput(908.00,289.92)(5.915,-5.000){2}{\rule{1.225pt}{1.000pt}}
\multiput(919.00,284.71)(1.000,-0.424){2}{\rule{2.650pt}{0.102pt}}
\multiput(919.00,284.92)(6.500,-5.000){2}{\rule{1.325pt}{1.000pt}}
\multiput(931.00,279.71)(1.000,-0.424){2}{\rule{2.650pt}{0.102pt}}
\multiput(931.00,279.92)(6.500,-5.000){2}{\rule{1.325pt}{1.000pt}}
\multiput(943.00,274.71)(1.000,-0.424){2}{\rule{2.650pt}{0.102pt}}
\multiput(943.00,274.92)(6.500,-5.000){2}{\rule{1.325pt}{1.000pt}}
\multiput(955.00,269.71)(0.830,-0.424){2}{\rule{2.450pt}{0.102pt}}
\multiput(955.00,269.92)(5.915,-5.000){2}{\rule{1.225pt}{1.000pt}}
\multiput(966.00,264.71)(1.000,-0.424){2}{\rule{2.650pt}{0.102pt}}
\multiput(966.00,264.92)(6.500,-5.000){2}{\rule{1.325pt}{1.000pt}}
\multiput(978.00,259.71)(0.830,-0.424){2}{\rule{2.450pt}{0.102pt}}
\multiput(978.00,259.92)(5.915,-5.000){2}{\rule{1.225pt}{1.000pt}}
\put(989,252.92){\rule{2.891pt}{1.000pt}}
\multiput(989.00,254.92)(6.000,-4.000){2}{\rule{1.445pt}{1.000pt}}
\multiput(1001.00,250.71)(0.830,-0.424){2}{\rule{2.450pt}{0.102pt}}
\multiput(1001.00,250.92)(5.915,-5.000){2}{\rule{1.225pt}{1.000pt}}
\multiput(1012.00,245.71)(0.830,-0.424){2}{\rule{2.450pt}{0.102pt}}
\multiput(1012.00,245.92)(5.915,-5.000){2}{\rule{1.225pt}{1.000pt}}
\put(1023,238.92){\rule{2.891pt}{1.000pt}}
\multiput(1023.00,240.92)(6.000,-4.000){2}{\rule{1.445pt}{1.000pt}}
\multiput(1035.00,236.71)(0.830,-0.424){2}{\rule{2.450pt}{0.102pt}}
\multiput(1035.00,236.92)(5.915,-5.000){2}{\rule{1.225pt}{1.000pt}}
\multiput(1046.00,231.71)(0.830,-0.424){2}{\rule{2.450pt}{0.102pt}}
\multiput(1046.00,231.92)(5.915,-5.000){2}{\rule{1.225pt}{1.000pt}}
\put(1057,224.92){\rule{2.650pt}{1.000pt}}
\multiput(1057.00,226.92)(5.500,-4.000){2}{\rule{1.325pt}{1.000pt}}
\multiput(1068.00,222.71)(0.830,-0.424){2}{\rule{2.450pt}{0.102pt}}
\multiput(1068.00,222.92)(5.915,-5.000){2}{\rule{1.225pt}{1.000pt}}
\multiput(1079.00,217.71)(0.830,-0.424){2}{\rule{2.450pt}{0.102pt}}
\multiput(1079.00,217.92)(5.915,-5.000){2}{\rule{1.225pt}{1.000pt}}
\put(1090,210.92){\rule{2.650pt}{1.000pt}}
\multiput(1090.00,212.92)(5.500,-4.000){2}{\rule{1.325pt}{1.000pt}}
\multiput(1101.00,208.71)(0.830,-0.424){2}{\rule{2.450pt}{0.102pt}}
\multiput(1101.00,208.92)(5.915,-5.000){2}{\rule{1.225pt}{1.000pt}}
\put(1112,201.92){\rule{2.650pt}{1.000pt}}
\multiput(1112.00,203.92)(5.500,-4.000){2}{\rule{1.325pt}{1.000pt}}
\multiput(1123.00,199.71)(0.830,-0.424){2}{\rule{2.450pt}{0.102pt}}
\multiput(1123.00,199.92)(5.915,-5.000){2}{\rule{1.225pt}{1.000pt}}
\put(1134,192.92){\rule{2.409pt}{1.000pt}}
\multiput(1134.00,194.92)(5.000,-4.000){2}{\rule{1.204pt}{1.000pt}}
\multiput(1144.00,190.71)(0.830,-0.424){2}{\rule{2.450pt}{0.102pt}}
\multiput(1144.00,190.92)(5.915,-5.000){2}{\rule{1.225pt}{1.000pt}}
\put(1155,183.92){\rule{2.650pt}{1.000pt}}
\multiput(1155.00,185.92)(5.500,-4.000){2}{\rule{1.325pt}{1.000pt}}
\multiput(1166.00,181.71)(0.660,-0.424){2}{\rule{2.250pt}{0.102pt}}
\multiput(1166.00,181.92)(5.330,-5.000){2}{\rule{1.125pt}{1.000pt}}
\put(1176,174.92){\rule{2.650pt}{1.000pt}}
\multiput(1176.00,176.92)(5.500,-4.000){2}{\rule{1.325pt}{1.000pt}}
\put(1187,170.92){\rule{2.409pt}{1.000pt}}
\multiput(1187.00,172.92)(5.000,-4.000){2}{\rule{1.204pt}{1.000pt}}
\multiput(1197.00,168.71)(0.830,-0.424){2}{\rule{2.450pt}{0.102pt}}
\multiput(1197.00,168.92)(5.915,-5.000){2}{\rule{1.225pt}{1.000pt}}
\put(1208,161.92){\rule{2.409pt}{1.000pt}}
\multiput(1208.00,163.92)(5.000,-4.000){2}{\rule{1.204pt}{1.000pt}}
\put(1218,157.92){\rule{2.650pt}{1.000pt}}
\multiput(1218.00,159.92)(5.500,-4.000){2}{\rule{1.325pt}{1.000pt}}
\put(1229,153.92){\rule{2.409pt}{1.000pt}}
\multiput(1229.00,155.92)(5.000,-4.000){2}{\rule{1.204pt}{1.000pt}}
\put(1239,149.92){\rule{2.409pt}{1.000pt}}
\multiput(1239.00,151.92)(5.000,-4.000){2}{\rule{1.204pt}{1.000pt}}
\put(1249,145.92){\rule{2.650pt}{1.000pt}}
\multiput(1249.00,147.92)(5.500,-4.000){2}{\rule{1.325pt}{1.000pt}}
\put(1260,141.92){\rule{2.409pt}{1.000pt}}
\multiput(1260.00,143.92)(5.000,-4.000){2}{\rule{1.204pt}{1.000pt}}
\put(1270,137.92){\rule{2.409pt}{1.000pt}}
\multiput(1270.00,139.92)(5.000,-4.000){2}{\rule{1.204pt}{1.000pt}}
\put(1280,134.42){\rule{2.409pt}{1.000pt}}
\multiput(1280.00,135.92)(5.000,-3.000){2}{\rule{1.204pt}{1.000pt}}
\put(1290,131.42){\rule{2.409pt}{1.000pt}}
\multiput(1290.00,132.92)(5.000,-3.000){2}{\rule{1.204pt}{1.000pt}}
\put(1300,128.42){\rule{2.409pt}{1.000pt}}
\multiput(1300.00,129.92)(5.000,-3.000){2}{\rule{1.204pt}{1.000pt}}
\put(1310,125.42){\rule{2.409pt}{1.000pt}}
\multiput(1310.00,126.92)(5.000,-3.000){2}{\rule{1.204pt}{1.000pt}}
\put(1320,122.42){\rule{2.409pt}{1.000pt}}
\multiput(1320.00,123.92)(5.000,-3.000){2}{\rule{1.204pt}{1.000pt}}
\put(1330,119.92){\rule{2.409pt}{1.000pt}}
\multiput(1330.00,120.92)(5.000,-2.000){2}{\rule{1.204pt}{1.000pt}}
\put(1340,117.92){\rule{2.409pt}{1.000pt}}
\multiput(1340.00,118.92)(5.000,-2.000){2}{\rule{1.204pt}{1.000pt}}
\put(1350,115.92){\rule{2.409pt}{1.000pt}}
\multiput(1350.00,116.92)(5.000,-2.000){2}{\rule{1.204pt}{1.000pt}}
\put(1360,113.92){\rule{2.409pt}{1.000pt}}
\multiput(1360.00,114.92)(5.000,-2.000){2}{\rule{1.204pt}{1.000pt}}
\put(1370,112.42){\rule{2.409pt}{1.000pt}}
\multiput(1370.00,112.92)(5.000,-1.000){2}{\rule{1.204pt}{1.000pt}}
\put(1380,111.42){\rule{2.409pt}{1.000pt}}
\multiput(1380.00,111.92)(5.000,-1.000){2}{\rule{1.204pt}{1.000pt}}
\put(1409,111.42){\rule{2.409pt}{1.000pt}}
\multiput(1409.00,110.92)(5.000,1.000){2}{\rule{1.204pt}{1.000pt}}
\put(1419,112.42){\rule{2.168pt}{1.000pt}}
\multiput(1419.00,111.92)(4.500,1.000){2}{\rule{1.084pt}{1.000pt}}
\put(1428,113.42){\rule{1.927pt}{1.000pt}}
\multiput(1428.00,112.92)(4.000,1.000){2}{\rule{0.964pt}{1.000pt}}
\put(1390.0,113.0){\rule[-0.500pt]{4.577pt}{1.000pt}}
\end{picture}

\end{center}
\end{figure}

\section{Form factors of $K^{+}\rightarrow\gamma l\nu$}
As a test of ${\cal L}^{Vs,As}$(68,72)
we study the form factors of $K^{-}\rightarrow e\gamma\nu$.
The form factors of $K^{+}\rightarrow\gamma l\nu$ are calculated in
the chiral limit.
The vector form factor is
determined by the vertex which comes from the anomaly[11]
\begin{equation}
{\cal L}^{K^{*+}K^{-}\gamma}=-\frac{e}{2\pi^{2}gf_{\pi}}\varepsilon
^{\mu\nu\alpha\beta}K^{+}_{\mu}\partial_{\beta}K^{-}
\partial_{\nu}A_{\alpha}.
\end{equation}
The vector form factor is determined
\begin{equation}
F^{V}={1\over2\sqrt{2}\pi^{2}}{m_{K}\over f_{\pi}}=0.095.
\end{equation}
The vertices ${\cal L}^{K_{1}K\gamma}$ can be
found from Ref.[11].
\begin{equation}
{\cal L}^{K_{1}K\gamma}={i\over2}eg(Ag_{\nu\lambda}+Bp_{\nu}p_
{\lambda}+Dk_{\nu}k_{\lambda})K^{-\nu}_{1}K^{+}A^{\lambda}.
\end{equation}
${\cal L}^{KK\gamma}$ is[11]
\begin{equation}
{\cal L}^{K^{+}K^{-}\gamma}=ie(K^{+}\partial_{\mu}K^{-}-K^{-}
\partial_{\mu}K^{+})A^{\mu}.
\end{equation}
The axial-vector form factors are derived
\begin{eqnarray}
\lefteqn{F^{A}=\frac{1}{2\sqrt{2}\pi^{2}}\frac{m_{K}}{f_{\pi}}
\frac{m^{2}_{K^{*}}}{m^{2}_{K_{1}}}(1-{2c\over g})(1-{1\over2\pi^{2}
g^{2}})^{-1}=0.04,\;\;\;m^{2}_{K_{1}}=(1.32GeV)^{2}}\\
&&R=\frac{g^{2}}{2\sqrt{2}}\frac{m_{K}}{f_{\pi}}\frac{m^{2}_{K^{*}}}
{m^{2}_{K_{1}}}\{{2c\over g}+{1\over\pi^{2}g^{2}}(1-{2c\over g})\}
(1-{1\over2\pi^{2}g^{2}})^{-1}=0.078.
\end{eqnarray}
We obtain
\(F^{A}+F^{V}=0.135\), \(F^{A}-F^{V}=-0.055\).
The data[28] are \\
\(F^{A}+F^{V}=0.147\pm 0.011\), \(0.150^{+0.018}_{-0.023}\),
\(F^{A}-F^{V}=< 0.49\).
In the calculation of $F_{A}$ the $K_{a}$ is used, we obtain the
\(F_{A}=0.032\).
It is necessary to point out that the factor $m_{K}$ in
Eqs.(88,91,92)
comes from the definitions of the form factors.

\section{Conclusions}
The Lagrangian of weak interaction of mesons consists of vector part
and axial-vector part. In the chiral limit,
the VMD takes responsibility for the vector part.
Based on chiral symmetry and spontaneous chiral symmetry breaking the
Lagrangian of the axial-vector part of weak interactions of mesons
is determined. The whole Lagrangian is derived from the effective chiral
theory of mesons. All the vertices of mesons are obtained from the same
theory. This theory provides a unified study for $\tau$ mesonic decays.
The $a_{1}$ dominance in the
matrix elements of the \(\Delta s=0\) axial-vector currents and $K_{a}$
dominance in the ones of \(\Delta s=1\) axial-vector currents in $\tau$
decays are found.
All theoretical studies are done in the limit of \(m_{q}=0\) and
the results are in reasonable agreements with data.
There are many other $\tau$ mesonic decay modes can be studied by this
theory.

\section{Acknowledgment}
The author wishes to thank E.Braaten, Jin Li, J.Smith,
N.D.Qin, M.L.Yan,
and Z.P.Zheng
for discussion.
The author specially appreciates
R.Stroynowski for sending CLEO's data to the
author.
This research was partially
supported by DOE Grant No. DE-91ER75661.


\begin{thebibliography}{40}
\bibitem{} Y.S.Tsai, Phys.Rev. {\bf D4}, 2821(1971).
\bibitem{} R.Fischer, J.Wess, and F.Wagner, Z.Phys. {\bf C3},313(1980).
\bibitem{} G.Aubrecht II, N.Chahrouri, and K.Slanec, Phys.Rev.{\bf D24},
1318(1981).
\bibitem{} E.Braaten, R.J.Oakes, and S.M.Tse, Inter.Jour.Modern Phys.,
{\bf 5},2737(1990).
\bibitem{} R.Decker, E.Mirkes, R.Sauer and Z.Was, Z.Phys.,
{\bf C58},445(1993);
M.Finkemeier and E.Mirkes, Z.Phys., {\bf C69}, 243(1996).
\bibitem{} G.Kramer and W.F.Palmer, Z.Phys., {\bf C25},195(1984) and
ibid, {\bf 39},423(1988).
\bibitem{} A.Pich, Phys.Lett., {\bf 196},561(1987).
\bibitem{} R.Decker, Z.Phys.,{\bf C36},487(1987).
\bibitem{} E.Braaten,R.oakes, and S.Z.Tse, Phys.Rev., {\bf D36},2187
(1987).
\bibitem{} R.Decker and E.Mirkes, Phys.Rev., {\bf D47},4012(1993).
\bibitem{} B.A.Li, Phys.Rev., {\bf D52}, 5165(1995), 5184(1995).
\bibitem{} J.J.Sakurai, Currents and Mesons, Univ. of Chicago Press,
1969.
\bibitem{} S.Weinberg, Phys.Rev.Lett., {\bf 18},507(1967).
\bibitem{} J. Schwinger, Phys. Lett., {\bf B24},473(1967);J. Wess and
B. Zumino, Phys. Rev., {\bf 163},1727(1967);
S. Weinberg, Phys. Rev.,{\bf 166},1568(1968);
B. W . Lee and H. T. Nieh, Phys. Rev., {\bf 166},1507(1968).
\bibitem{} J.Wess and B.Zumino, Phys.Lett., {\bf B37},95(1971),
E.Witten, Nucl.Phys., {\bf B223},422(1983).
\bibitem{} S.I.Eidelman and V.N.Ivanchenko, Proc. of the Third Workshop
on Tau Lepton Physics, p.131,
Montreux, Switzerland, 19-22 September 1994, ed.
by L.Rolandi.
\bibitem{} J.J.Gomez-Cadennas,M.C.Gonzalez-Garcia and A.Pich,
Phys.Rev., {\bf D42},3093(1990).
\bibitem{} M.G.Bowler, Phy.Lett., {\bf 182B}, 400(1986), N.A.Tornquist,
Z.Phys., {\bf C36}, 695(1987), J.H.Kuhn and A.Santamaria, Z.Phys.,
{\bf C48},445(1990), M.K.Volkov, Yu.P.Ivanov, A.A.Osipov, Z.Phys., {\bf
C49}, 563(1991).
\bibitem{} N.Isgur, C.Mornongstar, and C.Reader, Phys.Rev., {\bf D39},
1357(1989).
\bibitem{} H.J.Behrend et al., CELLO Collaboration, Z.Phys., {\bf C46},
537(1990).
\bibitem{} D.Decamp et al., ALEPH Collaboration, Z.Phys., {\bf C54}, 211
(1992).
\bibitem{} P.Abreu et al., DELPHI Collaboration, Z.Phys., {\bf C55}, 555
(1992).
\bibitem{} G.Crawford  CLEO Collaboration, Proc. of the Second Workshop
on Tau Lepton Physics, Columbus, Ohio, ed. by K.K.Ga, World Scientific,
p.183, 1992.
\bibitem{} D.Antreasyan et al., Crystalball Collaboration, Phys.Lett.,
{\bf B259},216(1991).
\bibitem{} H.Evans, ``Charged Current Measurements a Tau96 Overview``,
talk presented in Fourth International Workshop on Tau Lepton Physics,
Estes Park, Colorado, Sept.15-19,1996.
\bibitem{} BES Collaboration, High Energy Phys. and Nuclear Phys., {\bf
19},385(1994).
\bibitem{} H.Albrecht et al., ARGUS Collaboration, Z.Phys., {\bf C58}
, 61(1993).
\bibitem{} Particle data group, Phys.Rev., {\bf D50} No.3(1994).
\bibitem{} T.E.Coan et al., CLEO Collaboration, CLEO CONF 96-15,
ICHEP96 PA01-085.
\bibitem{} M.Goldberg et al., CLEO Collaboration, Phys.Lett.,
{\bf B251}, 223(1990).
\bibitem{} H.Albrecht et al., ARGUS Collaboration, Z.Phys., {\bf C68},
215(1995).
\bibitem{} see paper by M.Benayoun et al., Z.Phys., {\bf C58},31(1993).
\bibitem{} E.B.Dally et al., Phys.Rev.Lett., {\bf 48},375(1982).
\bibitem{} S.I.Eidenlman and V.N.Ivanchenko, Phys.Let., {\bf B257},437(
1991), S.Nelson and A.Pich, Phys.Lett., {\bf B304},359(1993).
\bibitem{} H.Aihara et al., TPC/2$\gamma$ Collaboration, Phys.Rev.Lett.,
{\bf 59},751(1987).
\bibitem{} D.Buskulic et al., ALEPH Collaboration, CERN-PPE/95-140.
\bibitem{} J.Barlet et al., CLEO Collaboration, CLNS 96/1391, CLEO 96-3.
\bibitem{} H.Lipkin, Phys Lett., {\bf B303},119(1993),
M.Suzuki, Phys.Rev., {\bf D47},1252(1993).
\bibitem{} M.Goldberg et al., CLEO collaboration, Phys.Lett., {\bf B251}
223(1992).
\bibitem{} H.Albrecht et al., ARGUS Collaboration, Z.Phys., {\bf C68},
215(1995).
\bibitem{} W.Hao(DELPHI Collaboration), "Study of charged kaon
production in three Prong tau decays", Doctoral thesis.
\bibitem{} M.Davier(ALEPH Collaboration), talk presented in Fourth
Inter. Workshop on Tau-Lepton Physics, Estes Park, Colorado, Sept.
15-19, 96.
\bibitem{} J.Barlet et al., CLEO Collaboration, CLNS 96/1395,
CLEO 96-5.
\bibitem{} D.Buskulic et al., ALEPH Collaboration, CERN-PDE/96-103.
\end{thebibliography}
\end{document}